\documentclass[useAMS,usenatbib,usegraphicx]{mn2e}
\usepackage{times}
\usepackage{amssymb,amsmath,color,soul}
\usepackage[colorlinks = true,linkcolor = blue,urlcolor  = blue,citecolor = blue,anchorcolor = blue]{hyperref}
\usepackage[caption = false]{subfig}
\usepackage{breqn}
\usepackage{float}

\DeclareRobustCommand{\ion}[2]{%
\relax\ifmmode
\ifx\testbx\f@series
{\mathbf{#1\,\mathsc{#2}}}\else
{\mathrm{#1\,\mathsc{#2}}}\fi
\else\textup{#1\,{\mdseries\textsc{#2}}}%
\fi}

\newcommand{\msun}{$M_\odot$}

\newcommand\aj{AJ}
\newcommand\an{AN}
\newcommand\araa{ARA\&A}
\newcommand\apj{ApJ}
\newcommand\apjl{ApJ}
\newcommand\apjs{ApJS}
\newcommand\aap{A\&A}
\newcommand\mnras{MNRAS}
\newcommand\pasp{PASP}

\title[The chemical evolution of local star forming galaxies]{The chemical evolution of local star forming galaxies: Radial profiles of ISM metallicity, gas mass, and stellar mass and constraints on galactic accretion and winds}

\author[Kudritzki et al.]
{Rolf-Peter Kudritzki$^{1,2}$\thanks{E-mail: \href{kud@ifa.hawaii.edu}{kud@ifa.hawaii.edu}}, I-Ting Ho$^1$, Andreas Schruba$^3$, Andreas Burkert$^2$, 
\newauthor H. Jabran Zahid$^4$, Fabio Bresolin$^1$, and Gabriel I. Dima$^1$\\ 
$^1$Institute for Astronomy, University of Hawaii, 2680 Woodlawn Drive, Honolulu, HI 96822, USA \\
$^2$University Observatory Munich, Scheinerstr. 1, D-81679 Munich, Germany\\
$^3$Max-Planck-Institute for Extraterrestrial Physics, Giessenbachstr., D-85741 Garching, Germany\\
$^4$Harvard-Smithsonian Center for Astrophysics, 60 Garden Street MS-20, Cambridge, MA 02138, USA \\
}

\begin{document}
\date{Accepted ??. Received ??; in original form ??}
\label{firstpage}
\maketitle

\begin{abstract}
The radially averaged metallicity distribution of the ISM and the young stellar population of a sample of 20 disk galaxies is investigated by means of an analytical chemical evolution model which assumes constant ratios of galactic wind mass loss and accretion mass gain to star formation rate. Based on this model the observed metallicities and their gradients can be described surprisingly well by the radially averaged distribution of the ratio of stellar mass to ISM gas mass. The comparison between observed and model predicted metallicity is used to constrain the rate of mass loss through galactic wind and accretion gain in units of the star formation rate. Three groups of galaxies are found: galaxies with either mostly winds and only weak accretion, or mostly accretion and only weak winds, and galaxies where winds are roughly balanced by accretion. The three groups are distinct in the properties of their gas disks. Galaxies with approximately equal rates of mass-loss and accretion gain have low metallicity, atomic hydrogen dominated gas disks with a flat spatial profile. The other two groups have gas disks dominated by molecular hydrogen out to 0.5 to 0.7 isophotal radii and show a radial exponential decline, which is on average steeper for the galaxies with small accretion rates. The rates of accretion ($\lesssim1.0\times\rm SFR$) and outflow ($\lesssim2.4\times\rm SFR$) are relatively low. The latter depend on the calibration of the zero point of the metallicity determination from the use of H\textsc{ii} region strong emission lines.
\end{abstract}


\section{Introduction}

The chemical composition of stars and the interstellar medium (ISM) reflect the evolutionary history of a galaxy from its formation to the present stage. While the physical processes leading to chemical evolution and enrichment with heavy elements such as star formation, nucleosynthesis, stellar winds, supernova explosions, galactic winds and accretion are complicated and coupled in a complex way, an intriguingly simple relationship between galactic stellar mass and the gas-phase oxygen abundance of the ISM or the metallicity of the young stellar population of star forming galaxies has been observed, the mass-metallicity relationship \citep[MZR;][]{lequeux79,tremonti04,kud12}. This indicates that there might be a straightforward simplified way to describe the key aspects of chemical evolution. For instance, \citet{lequeux79} already argued that a simple closed-box chemical evolution model, where galaxies are not affected by mass accretion or mass loss and which relates metallicity to the ratio of stellar to ISM gas mass, could explain the MZR of their sample of irregular galaxies well. Since then it has become clear that in a $\Lambda$CDM dominated universe processes of galaxy merging and accretion from the intergalactic medium of the cosmic web and galactic winds may play a crucial role in galaxy formation and evolution. Consequently, the situation is far more complex than in a simple closed-box model (see, for instance, \citealt{dave11a,dave11b,dave12,yates12,dayal13}). Moreover, comprehensive spectroscopic surveys from low to high redshifts (see \citealt{zahid14} and references therein) clearly indicate that there is an empirical upper limit in galaxy metallicity for the most massive galaxies \citep {zahid13}. This is in contradiction to the classical closed-box model, for which the metallicity increases logarithmically with increasing ratio of stellar mass to gas mass. On the other hand, as has been shown recently by \citet{zahid14}, when allowing for galactic winds and accretion the observed MZR can still be explained by a model where the observed metallicity is a simple analytical function of the stellar to gas mass ratio. \citet{ascasibar14} find a very similar result. They develop a new approach, in which even a closed box model would eventually reach a maximum metallicity, equal to the oxygen abundance of the supernova ejecta. They also point out that their approach should apply to spatially resolved observations.

The study of global galaxy metallicity as a function of global stellar and gas mass is only a first step to constrain galaxy evolution. An important extension of these investigations, as already noted by \citet{ascasibar14}, is to use the information of the observed resolved spatial distribution of metallicity in the disks of star forming galaxies. As is known since a long time \citep[e.g.,][]{searle71,garnett87,garnett97,skillman98,vilacostas92,zaritsky94}, disk galaxies show negative abundance gradients with increased metallicity towards the centres and lower metallicity in the outskirts. Since then a considerable effort has been made to model the spatially resolved chemical evolution of galactic disks \citep[e.g.,][]{molla97,prantzos00,chiappini01,fu09,pilkington12,mott13}.  

This is the starting point for the study presented here. With the indication that the global metallicity of galaxies can be related to the global ratio of stellar to gas mass by relatively simple chemical evolution models \citep{zahid14,ascasibar14,yabe15}, we investigate whether the spatially resolved radial distribution of metals in disk galaxies can be explained in a similar way. We use the radial distributions of stellar mass and ISM H\textsc{i} and H$_{2}$ gas mass column densities together with the observed metallicities and metallicity gradients of 20 well observed disk galaxies to find out whether the analytical chemical evolution models introduced below can explain the observed radial metallicity distribution, and whether the influence of galactic winds and accretion can be constrained in this way.

This work has been triggered by the work by \citet{ho15} who investigated IFU observations of 49 local field star-forming galaxies and found a narrow Gaussian distribution of metallicity gradients characterised by a mean $\pm$ standard deviation of $-0.39\pm0.18~{\rm dex}~R_{25}^{-1}$ ($R_{25}$ is the {\it B}-band iso-photal radius). \citet{ho15} compared this observed distribution of metallicity gradients with that predicted by chemical evolution models from a set of stellar and gas mass profiles of 13 nearby galaxies. The latter were taken as benchmark galaxies representing local galaxies. They concluded from the comparison of the two statistical distributions that the effects of galactic winds and accretion must be very small. \citet{ho15} used a statistical approach to compare the observed distribution of metallicity gradients of their sample of IFU-studied galaxies with a model-generated sample based on the observed mass profiles of a second set of galaxies. We will apply a different approach and will investigate one single sample of galaxies with a homogeneous set of metallicity and stellar and gas mass observations. In addition, we will not only use the metallicity gradients but also the absolute level of metallicity to constrain the effects of winds and infall for each galaxy in detail.

We describe the sample of local star-forming galaxies and their observations in Section~\ref{sec:galaxy-sample-and-observations}. The chemical evolution model is introduced in Section~\ref{sec:the-chemical evolution-model} and then applied in a first step in Section~\ref{sec:the-radial-metallicity-distribution-of-the-milky-way} to the Milky Way disk, where we use a fit of metallicity and metallicity gradient observed with Cepheids and B-stars to empirically constrain the metallicity yield. The next step is then to apply our models for a fit of the observed gas-phase ISM oxygen abundances of our star-forming galaxies. This is done in Section~\ref{sec:analysis-of-the-observed-metallicity-distribution} and the results are discussed and summarised in Section~\ref{sec:discussion-and-summary}.  

\section{Galaxy sample and observations}\label{sec:galaxy-sample-and-observations}

The sample investigated in this work consists of 19 nearby spiral galaxies selected from the study by \citet{schruba11} and \citet{leroy08}, who published carefully measured radial profiles of H\textsc{i} and H$_{2}$ gas mass column densities. We combine these observations with measurements of oxygen abundances and abundance gradients published in the comprehensive study by \citet{pilyugin14a}. The study by \citet{schruba11} contains 33 spirals, however, of those only 21 overlap with \citet{pilyugin14a}. Of those 21, NGC\,2841 and NGC\,4725 were not included into our final sample, because the oxygen abundance observations covered only a small radial range of the galaxies and were also very uncertain. The basic properties of our sample are summarised in Table~\ref{table:sample}. We note that while the majority of our sample consists of isolated field galaxies, NGC\,5194 (M51) is a merger and NGC\,4625 is a pair with NGC\,4618 (the latter is not included in our sample). For further information see \citet{schruba11} and \citet{leroy08}. In the following, we use the term ``SL-sample'' for this sample of 19 spiral galaxies.

We include the Milky Way as the 20th galaxy in our investigation. We will use the Milky way as test galaxy for the chemical evolution models and for the empirical calibration of our yields. 

\subsection{HI and H${2}$ mass column density profiles}

\subsubsection{The SL-sample}

The radial column densities of atomic hydrogen, $\Sigma_{H\textsc{i}},$ were obtained from VLA observations of the 21 cm hydrogen line carried out in the The H\textsc{i} Nearby Galaxy Survey survey \citep[THINGS;][]{walter08}. A factor 1.36 was included to account for helium and heavier elements (see \citealt{schruba11}). The molecular gas column densities, $\Sigma_{H_{2}}$, result from CO(2 $\rightarrow$ 1) line emission observations with the IRAM 30m telescope obtained as part of the HERA CO Line Extragalactic Survey survey \citep[HERACLES;][]{leroy09}. For all details, such as the conversion from CO to H$_{2}$, the azimuthal averaging and construction of deprojected radial profiles, the conversion of emission line intensities into mass column densities, etc., the readers are referred to \citet{schruba11}, who present the radial profiles of H\textsc{i} and H$_{2}$ gas mass column densities. The total ISM gas mass column density is then $\Sigma_{g}$ = $\Sigma_{H\textsc{i}}$ + $\Sigma_{H_{2}}$.

\subsubsection{The Milky Way}

For gas mass column densities of the Milky Way disk, we use the fit formulae provided by \cite{wolfire03}. We restrict ourselves to the range of galactocentric distances between 4.5 to 9 kpc for which a good description of the stellar mass profile and accurate measurements of the metallicity and metallicity gradient are available (see below). In this range, \cite{wolfire03} suggest a constant mass column density of the atomic gas, $\Sigma_{H\textsc{i}}$ = 6.8 \msun~pc$^{-2}$. A factor 1.36 is also included here to account for helium and heavier elements consistent with the SL-sample. 
For the radial distribution of the molecular gas, $\Sigma_{H_{2}}$, \cite{wolfire03} apply a Gaussian distribution in the galactocentric range between 3 to 6.97 kpc with a peak at 4.85 kpc and a FWHM value of 4.42 kpc. Beyond 6.97 kpc they use a radial exponential for $\Sigma_{H_{2}}$ 
with a scale length of 2.89 kpc. Same as the SL-sample, our total ISM gas mass column density in the Milky Way disk is then $\Sigma_{g}$ = $\Sigma_{H\textsc{i}}$ + $\Sigma_{H_{2}}$.

\subsection{Stellar mass column density profiles}
 
\subsubsection{The SL-sample}

Column densities of stellar mass of all galaxies of the SL-sample were obtained by surface photometry of infrared images observed by the Wide-field Infrared Survey Explorer \citep[WISE;][]{wright10} in the W1 band at 3.4~$\mu$m, the SIRTF Nearby Galaxies Survey survey \citep[SINGS;][]{kennicutt03} with Spitzer/IRAC at 3.6~$\mu$m, and the 2MASS survey \citep{jarrett03} in the $K_{s}$-band at 2.2~$\mu$m. A procedure similar as described in \citet{leroy08} is applied. We measure the median intensities in 10 arcsec wide tilted rings at different galactocentric distances. We find a tight relation between the WISE/W1 and SPITZER/IRAC intensities, $I_{3.6}=1.03I_{3.4}$, and the shape of the radial surface brightness distributions typically agrees very well. The $K_{s}$-band images are not deep enough to reach the outer rings, as opposed to the 3.4 and 3.6~$\mu$m observations, but for the inner and brighter parts of the galaxies where the data overlap, we find a relation $I_{K_{s}}=1.8I_{3.4}$. We use this relation and the median deprojected intensities $I_{3.4}(r)$ together with a fixed {\it K}-band mass-to-light ratio, $\Upsilon_{\ast}^{K}$ = 0.5 \msun/L$_{\odot,K}$ (see \citealt{leroy08}), to convert the surface brightness profiles to mass column density profiles $\Sigma_{\ast}(r)$.

In most cases the central parts of the mass column density are contaminated by the contribution of galactic bulges, which have a different star formation history and chemical evolution than disks. Since our goal is to investigate the chemical evolution of the ISM and the young stellar population in the disks of these galaxies, we subtract the bulge contribution by a applying a bulge-disk decomposition algorithm. For this purpose, the bulge contribution is represented by a general Sersic profile 
\begin{equation}
\Sigma_{bulge}(r) = \Sigma_{b}\exp[-b_{n}((r/r_{e})^{1/n}-1)]. 
\end{equation}
The disk contribution is represented by either a single exponential profile or two exponential profiles broken into inner and outer parts so that the total mass column density, $\Sigma_{t}$, is given by
\begin{equation}
\Sigma_{t}(r) = \Sigma_{bulge}(r) + \Sigma_{d,i}\exp(-r/h_{i}) {\rm\ at\ } r\leq r_{break}, 
\end{equation}
and
\begin{equation}
\Sigma_{t}(r) = \Sigma_{bulge}(r) + \Sigma_{d,o}\exp(-r/h_{o}) {\rm\ at\ } r\geq r_{break}.
\end{equation}

Note that we measure the galactocentric radius $r$, the effective radius $r_{e}$, the break radius r$_{break}$, and the the disk scale lengths $h_{i}$, $h_{o}$ in units of the iso-photal radius $R_{25}$. The mass column densities $\Sigma_{b}$, $\Sigma_{d,i}$, $\Sigma_{d,o}$ are given in \msun ~pc$^{-2}$, and $b_n$ is calculated as a function of the bulge shape parameter $n$ through the condition $\Gamma(2n)=2\gamma(2n,b_n)$, where $\Gamma$ and $\gamma$ are the complete and incomplete gamma functions, respectively \citep{graham01}. All the parameters are determined through fitting the observed radial column densities using the above equations and are summarised in Table~\ref{table:sample}. 


\begin{table*}
 \caption{The SL galaxy sample}
 \label{table:sample}
 \begin{tabular}{crrcccccccccc}
\hline
\hline
Name &  D  & $R_{25}$ & $({\rm O/H})_0$ & ${d({\rm O/H}) \over dr}$ & ${\log}\Sigma_b$ &  $n$   &  $r_e$  &${\log}\Sigma_{d,i}$ &$h_i$&${\log}\Sigma_{d,o}$&$h_o$& $r_{break}$\\
NGC  & [Mpc] &  [kpc]    &  [dex]      &  [dex~kpc$^{-1}$] & [\msun~pc$^{-2}$]&   & [$R_{25}$]& [\msun~pc$^{-2}$]&[$R_{25}$]& [\msun~pc$^{-2}$]&[$R_{25}$]&[$R_{25}$] \\
\hline
0628& 7.3& 10.40& 8.93$\pm$0.01& 0.051$\pm$0.0024& 2.449& 0.741& 0.044& 2.596& 0.206&  --    &   --   &   --    \\
0925& 9.2& 14.20& 8.63$\pm$0.02& 0.034$\pm$0.0026& 1.577& 0.665& 0.087& 1.796& 0.282&  --    &   --   &  --     \\
2403& 3.2&  7.30& 8.63$\pm$0.02& 0.051$\pm$0.0042& 1.770& 0.419& 0.063& 2.383& 0.228&  --    &   --   &   --    \\
2903& 8.9& 15.30& 8.97$\pm$0.03& 0.032$\pm$0.0044& 2.964& 0.817& 0.030& 2.901& 0.154&  --    & --     &   --    \\
3184&11.1& 11.90& 8.81$\pm$0.02& 0.015$\pm$0.0031& 1.597& 2.114& 0.169& 2.092& 0.645& 3.085& 0.166& 0.477 \\
3198&13.8& 13.00& 8.75$\pm$0.04& 0.025$\pm$0.0044& 2.087& 0.157& 0.045& 2.275& 0.249&  --    &   --   &    --   \\
3351&10.1& 10.60& 8.97$\pm$0.01& 0.019$\pm$0.0028& 3.131& 0.674& 0.052& 2.660& 0.240&  --    &   --   &    --   \\
3521&10.7& 12.90& 8.98$\pm$0.04& 0.039$\pm$0.0056& 2.522& 2.705& 0.082& 2.921& 0.207&  --    &   --   &    --   \\ 
3938&12.2&  6.28& 8.94$\pm$0.05& 0.089$\pm$0.0097& 0.727& 0.272& 0.189& 2.749& 0.295&  --    &  --    &    --   \\
4254&20.0& 14.60& 8.69$\pm$0.03& 0.004$\pm$0.0026& 2.626& 0.711& 0.076& 3.045& 0.210&  --    &   --   &    --   \\
4321&14.3& 12.52& 8.89$\pm$0.03& 0.017$\pm$0.0031& 3.079& 0.500& 0.049& 2.661& 0.331&  --    &   --   &   --    \\
4559&11.6& 17.68& 8.68$\pm$0.03& 0.021$\pm$0.0031& 1.361& 1.236& 0.089& 2.313& 0.166& 2.120& 0.196& 0.473 \\
4625& 9.5&  1.91& 8.73$\pm$0.01& 0.041$\pm$0.0146&   --   &  --    &  --    & 2.564& 0.347& --     &  --    &   --    \\
4736& 4.7&  5.30& 8.72$\pm$0.03& 0.009$\pm$0.0222& 3.586& 1.843& 0.083& 2.830& 0.270&   --   &   --   &  --     \\
5055&10.1& 17.40& 9.02$\pm$0.02& 0.026$\pm$0.0017& 2.648& 1.110& 0.081& 2.814& 0.199&  --    &  --    &   --    \\ 
5194& 8.0&  9.00& 9.03$\pm$0.03& 0.025$\pm$0.0041& 3.132& 0.748& 0.075& 2.811& 0.414& 3.624& 0.186& 0.701 \\
5457& 7.4& 25.81& 8.86$\pm$0.01& 0.025$\pm$0.0009& 2.402& 1.697& 0.021& 2.487& 0.176&  --    &  --    &  --     \\
6946& 5.9&  9.80& 8.87$\pm$0.06& 0.034$\pm$0.0101& 2.855& 0.584& 0.040& 2.801& 0.275&  --    &  --    &   --    \\
7331&14.1& 19.60& 8.82$\pm$0.08& 0.011$\pm$0.0090& 3.003& 1.460& 0.026& 3.086& 0.136& 2.036& 0.366& 0.490 \\
\hline
\end{tabular}
\end{table*}


\subsubsection{The Milky Way}

The radial stellar mass column density profile of the Milky Way disk  has been studied recently by \citet{bovy13}. From their dynamical measurement they derive a surface column density of the stellar disk of $\Sigma_{\ast}$(R$_{0}$=8~kpc) = 38 \msun~pc$^{-2}$, and show that an exponential radial distribution of $\Sigma_{\ast}$ around this galactocentric distance with a scale length of 2.1~kpc is a good fit to their observations. We adopt this fit for the stellar mass in the Milky Way disk.

\subsection{Metallicities and metallicity gradients}

\subsubsection{The SL-sample}

The standard method to determine metallicities and metallicity gradients of the ISM or the young stellar population in the disks of star forming spiral galaxies is the analysis of strong emission lines from H\textsc{ii} regions using empirical calibrations which turn flux ratios of particular emission lines into oxygen abundances. The calibrations of these ``strong line methods'' are subject to large systematic errors and are heavily debated (see, for instance \citealt{kewley08} or \citealt{bresolin09a}). An alternative method is the quantitative analysis of absorption line spectra of individual blue supergiant stars (BSGs) which has been carefully tested in the Milky Way and and has now been applied to galaxies as distant as 7~Mpc \citep{kud08,kud12,kud13,kud14}. While this method is much less affected by systematic uncertainties, it has so far been applied to only a handful of galaxies beyond the Local Group. Thus, for the investigation of our SL-sample we decide to use the homogeneous and comprehensive study by \citet{pilyugin14a}, who applied a new carefully tested strong-line calibration \citep{pilyugin12} on observations of H\textsc{ii} regions of a sample of 130 late type galaxies providing central oxygen abundances and abundance gradients. Since linear radial gradients depend on the distances to the galaxies and in many cases the distances used by \citet{schruba11} and \citet{leroy08}, which are also what we adopted, are different from the distances used by \citet{pilyugin14a}, we have scaled the metallicity gradients by \citet{pilyugin14a} to the distances by Schruba et al. and Leroy et al. 

In order to assess the zero point of the strong line calibration adopted by \citet{pilyugin14a}, we compare the H\textsc{ii} metallicities with BSG metallicities for the galaxies NGC 300 \citep{kud08}, M33 \citep{u09}, M81 \citep{kud12}, and NGC 3621 \citep{kud14}. We proceed with the comparison in the following way. The BSG metallicities represent a full range of heavy elements including the iron group and oxygen, and are measured relative to the sun on a logarithmic scale with a metallicity defined as $[Z] = \log(Z/Z_{\odot})$. Since the BSG metallicities including the contribution from oxygen are measured relative to the sun, we also relate the H\textsc{ii} oxygen abundances, defined as $\rm (O/H) = 12 + \log[N(O)/N(H)]$, to the sun by introducing the difference $\Delta (O/H) = (O/H) - (O/H)_{\odot}$, where $\rm (O/H)_{\odot} = 8.69$ \citep{asplund09}. We then calculate $\Delta [Z] = [Z]_{\rm BSG} - \Delta (O/H)$ at three galactocentric distances $r/R_{25} = 0.3, 0.6, 0.9$ and obtain $\Delta [Z] = 0.13, 0.08, 0.23, 0.15$ dex for NGC 300, M33, M81, NGC 3621, respectively, as the average over the galactocentric distances in each galaxy. The average zero point difference is $\langle \Delta [Z] \rangle$ = $0.15\pm0.06$ dex. For NGC 3621 we have only $r/R_{25} = 0.65, 0.93$ available for the BSGs. For the distance to these galaxies and $R_{25}$ we use the values given in the BSG papers. We apply this zero point shift to the Pilyugin et al. central galactic metallicities. We will also investigate the effect of the zero point shift on our results later. 

The radial dependence of the logarithmic oxygen abundance is then given by 
\begin{equation}\label{pili}
({\rm O/H})(r) \equiv 12 + \log{\rm N(O)\over N(H)} = ({\rm O/H})_0 - {d({\rm O/H}) \over dr}r.
\end{equation} 
The values of the zero points $({\rm O/H})_0$ and linear gradients ${d({\rm O/H}) \over dr}$ for each galaxy are given in Table~\ref{table:sample}.

\subsubsection{The Milky Way}

The investigation of the radial metallicity distribution of the ISM or the young stellar population in the Milky Way disk requires accurate knowledge of the distances of the individual objects investigated. In this regard, Cepheid stars are by far the best tracers of metallicity as a function of galactocentric distance because their distances can be inferred from the application of period luminosity relationship. Very recently, \citet{genovali14} published a comprehensive spectroscopic study of 450 Cepheids and determined the metallicity and metallicity gradient in the 
same range of galactocentric distances for which we have information about the gas and stellar mass column density distributions. They obtain ${\rm [Z]} = (0.57\pm0.02) - (0.06\pm{0.002})r$, where $r$ is in kpc. We note that \citet{genovali14} actually measured the spatial distribution of $\rm Fe/H$ but we use this as a proxy for metallicity. The metallicity gradient of 0.06$\pm{0.002}\rm~dex~kpc^{-1}$ is very well constrained by this study, while the zero point could still be systematically affected because non-LTE effects were neglected. On the other hand, \citet{nieva12} have provided an accurate determination of cosmic abundance standards by a careful non-LTE investigation of B-Stars in the solar neighbourhood. They obtain a total metallicity mass fraction of Z = 0.014 and an iron abundunce of 12 + log N(Fe)/N(H) = 7.52, which both are very similar to the sun and the protosolar nebula \citep{asplund09}. We use this result to apply a correction to the zero point of the Genovali et al. relation by $-0.09$ dex, resulting in a relation for the Milky Way disk metallicity of the young stellar population, ${\rm [Z]} = (0.48\pm0.02) - (0.06\pm{0.002})r$, which yields $\rm [Z] =  [Z]_{B-stars}$ = 0 at the galactocentric distance of $r = 8$ kpc.

\section{The chemical evolution model}\label{sec:the-chemical evolution-model}

There is a large variety of model approaches to describe the circulation of matter from the birth of stars in the ISM, through their life with stellar winds and mass-loss to their late phases as planetary nebulae or supernovae ranging from the simple closed-box model \citep{searle72} to models with radial gas flows \citep{edmunds95} or radial stellar migration \citep{sellwood02,schoenrich09a,schoenrich09b,minchev13} and models with a chemo-dynamical description (\citealt{burkert92}; see also \citealt{matteucci12} or \citealt{pagel09} for an overview). Since the goal of this study is not to investigate star formation history or the chemical enrichment of the older population but solely to focus on the ISM and the young population, we will use a simple analytical model. The model includes the effects of galactic winds and accretion and allows us to obtain constraints on galactic mass-loss and accretion mass-gain inferred from the ISM and young stars. In the following, we derive the model. We stress that this model is not new and a special case of the more general models introduced by \citet{recchi08}. Similar models were also used to study the global metallicities of galaxies (\citealt{spitoni10}, \citealt{dayal13}, \citealt{lilly13}, \citealt{pipino14}, \citealt{yabe15}, \citealt{zahid14}). We also refer to the pioneering work by \citet{edmunds90}.

We consider a volume element in a galactic disk with stellar mass $M_*$ and gas mass $M_g$ and start from the mass balance equation
\begin{equation}\label{mass-balance}
dM_g = -dM_* - (\dot{M}_{loss} - \dot{M}_{accr}) dt.
\end{equation}
The first term on the right-hand side is the gas consumed by star formation. We note that  $M_*$\footnote{Note that the $M_*$ here is equivalent to $M_{*o}$ in \citet{ho15}, where the subscript ``o'' denotes the ``observed'' stellar mass.} is the presently observed stellar mass and that its increase through the formation of new stars with a star formation rate $\psi$(t) is $dM_* = (1-R)\psi dt$, where $R$ is the fraction of stellar mass returned to the ISM through stellar winds and other processes. Note also that we make use of the instantaneous recycling approximation assuming that this return happens on a short time scale compared with galactic evolution. The second term describes the mass-loss and 
mass-gain through galactic winds and accretion, respectively; in this formulation both $\dot{M}_{loss}$ and $\dot{M}_{accr}$ are required to be positive or zero. 

Expressing mass-gain and mass-loss in units of the star formation rate by introducing the mass loading factor
\begin{equation}\label{eta}
\eta \equiv {\dot{M}_{loss} \over \psi}
\end{equation}
and the mass accretion factor 
\begin{equation}\label{lam}
\Lambda \equiv {\dot{M}_{accr} \over \psi},
\end{equation}
and with the definition
\begin{equation}\label{alpha}
\alpha \equiv { \dot{M}_{loss} - \dot{M}_{accr} \over (1-R) \psi} = {\eta- \Lambda \over 1-R},
\end{equation}
one can rewrite Equation~\ref{mass-balance} as
 \begin{equation}\label{mg-alpha}
dM_g =  -(1 + \alpha) dM_*.
\end{equation}

With the definition of metallicity $Z\equiv M_Z / M_g$ ($M_Z$ is the mass of metals in the ISM), one can calculate the derivative of $Z$ with respect to $M_*$ 
\begin{eqnarray}\label{dz_dm*}
{dZ \over dM_*} & = &  {1 \over M_g} ({dM_Z\over dM_*} - Z {dM_g\over dM_*}) \nonumber\\
                          & = &  {1 \over M_g} \left[{dM_Z\over dM_*} + Z (1+\alpha) \right].
\end{eqnarray}

The first term on the right hand side of Equation~\ref{dz_dm*} can be expressed as
\begin{equation}\label{dmz_dm*-1}
{dM_Z \over dM_*} = {y_Z \over (1-R)} - Z - {\zeta \over (1-R)},
\end{equation}
where $\zeta$ is defined as
\begin{eqnarray}\label{zeta}
\zeta & \equiv & (1-R)\left(Z_{loss} {dM_{loss} \over dM_*} - Z_{accr} {dM_{accr} \over dM_*}\right) \nonumber\\
         & = & Z_{loss} {\dot{M}_{loss} \over \psi} - Z_{accr} {\dot{M}_{accr} \over \psi} \nonumber\\
         & = & Z_{loss} \eta - Z_{accr} \Lambda.
\end{eqnarray}
There are three terms in the right-hand side of Equation~\ref{dmz_dm*-1}, and each term describes a different mechanism of increasing or decreasing the metal mass $dM_Z$ under the formation of some stars $dM_*$. The first term represents the metal production by star formation. $y_Z$ is the nucleosynthetic yield defined as the fraction of metal mass per stellar mass which is produced by star formation. The second term is the fraction of metals locked up in each generation of stars. The last terms describes the gain and loss of metals due to inflows and outflows. Here, we implicitly assume that the metals are recycled instantaneously and the gas and metals are well mixed.

If we assume that the accreted gas is pristine ($Z_{accr}\approx 0$), and the gas-loss through galactic winds predominately driven by supernovae has the same metallicity as the ISM ($Z_{loss}\approx Z$), we can express $\zeta$ in terms of the mass loading factor $\eta$:
\begin{equation}\label{zeta-1}
\zeta \approx  Z {\dot{M}_{loss} \over \psi} = Z\eta,
\end{equation}
which in combination with Equation~\ref{dz_dm*} and \ref{dmz_dm*-1} yields
\begin{equation}\label{dz_2}
{dZ \over dM_*} = {1 \over M_g (1-R)} (y_Z - Z\Lambda).
\end{equation}

The assumption $Z_{loss}\approx Z$ has already been discussed by \citet{ho15}. It seems valid if the energy and momentum of the sources generating the outflows (for instance, supernovae or stellar winds) is transfered to the neighboring ISM so that a fraction of the local ISM is lost in galactic winds, a scenario which is supported by observations \citep{rupke13}, although counter examples like the case of the dwarf galaxy NGC\,1569 \citep{martin02} have also been detected. A more general model, where $Z_{accr}$ and $Z_{loss}$ are arbitrary, could of course also be applied (see \citealt{dalcanton07}) but would introduce additional parameters.

It is worth pointing out that by definition $\eta$ and $\Lambda$ must be positive or zero
\begin{equation}\label{positive-eta}
\eta \geq 0 \ {\rm and}\ \Lambda \geq 0, 
\end{equation}
because of the ways $\dot{M}_{loss}$ and $\dot{M}_{accr}$ were defined. However, $\alpha$ defined in Equation~\ref{alpha} can be both positive (outflow dominates), negative (inflow dominates), 
or zero (balance between inflows and outflows). 

To solve the relation between $Z$, $M_*$ and $M_g$, we further assume that $R$, $y_Z$, $\eta$, and $\Lambda$ (and therefore $\alpha$) are all constant as a function of time and space. While constant $R$ and $y_Z$ is a standard assumption in many chemical evolution models, the assumption of constant mass-loading and mass accretion factors requires an explanation. First of all, this assumption allows to obtain an analytical solution for metallicity as a function of the ratio of stellar mass to gas mass. We note that by this assumption we do not restrict the time dependence of star formation or mass infall or loss, we only assume that the ratios of wind loss or accretion gain to star formation rate are on average constant. Since very likely the energy and momentum driving the outflow is related to star formation, $\eta$ = const. is a reasonable zero order approximation. At the same time, inflow increases the reservoir of gas capable to form stars and, thus, a constant $\Lambda$ seems also a good starting approximation. The work by \citet{recchi08} investigating effects of non-constant $\Lambda$ confirmed this conclusion. Recent work by \citet{bouche10} and \citet{lilly13} also indicates that mass accretion gain is proportional to the star formation rate.

Given the relation between $dM_*$ and $dM_g$ in Equation~\ref{mg-alpha}, we find
\begin{equation}\label{dz-dmg}
{(1-R)(1+\alpha)\over Z\Lambda-y_Z} dZ = {1\over M_g} dM_g.
\end{equation}
We note that this solution is only valid for $\alpha$ different from $-1$. We will discuss the case $\alpha=-1$ separately below. 
Integrating this equation and combining the result with the integral of  Equation~\ref{mg-alpha}
\begin{equation}\label{mg0-mgt}
M_g(0)= M_g(t) + (1+\alpha) M_*(t),
\end{equation}
we finally obtain the solution  for the metallicity
\begin{equation}\label{z1}
Z(t) = {y_Z \over \Lambda}\left\{1-\left[1+(1+\alpha){M_{*}(t)\over M_g(t)}\right]^ {-w}\right\}, 
\end{equation}
where 
\begin{equation}
w = {\Lambda \over (1-R)(1+\alpha)},
\end{equation}
$\eta \geq 0$, $\Lambda > 0$, and $\alpha \neq -1$.

Equation~\ref{z1} describes the general case with both inflows and outflows. However, there is a restriction in this general case for models with very strong inflows where $\alpha \lneq -1$ or $\Lambda \gneq \eta +(1-R)$. In such a situation, there are still solutions possible but only for small stellar mass to gas mass ratios limited to $M_*/M_g \leq -1/(1+\alpha)$. We also note that models, which start with zero gas mass, M$_g$(0) = 0, and develop with strong inflow ($\alpha \lneq -1$) have the solution of constant metallicity Z(t) = $y_Z/\Lambda$. Models corresponding to this case have been discussed by \citet{bouche10} and \citet{lilly13}.

The case $\alpha=-1$ which was excluded before can be easily addressed. From Equation~\ref{mg0-mgt}, $M_{g}(t)=const.$ is obtained in this case and $\Lambda$ becomes
\begin{equation}\label{z2}
\Lambda = (1-R)+\eta.
\end{equation}
With $M_{g}(t)=const.$ we can then directly integrate Equation~\ref{dz_2} and obtain
\begin{equation}\label{z3}
Z(t) = {y_Z \over \Lambda} \left(1 - e^{-{\Lambda \over (1-R)}{M_{*}(t) \over M_{g}(t)}}\right).
\end{equation}

In the special case whith only outflows and virtually no inflows, i.e. $\eta \neq 0$ and $\Lambda = 0$, Equation~\ref{z1} is no longer valid. 
However, for this case one can repeat the derivation from Equation~\ref{dz_2} with $\Lambda=0$ and show that  
\begin{multline}\label{z5}
Z(t) = \left({y_Z \over 1-R }\right) \left({1 \over 1+{\eta \over 1-R}}\right) \\
\ln \left[ 1 + \left( 1+ {\eta\over1-R } \right) {M_{*}(t)\over M_{g}(t)}  \right],\\
{\rm when}\ \eta \neq 0 \ {\rm and}\ \Lambda = 0. 
\end{multline}

In the alternative special case with only inflows but no outflows ($\Lambda\neq0$ and $\eta=0$), Equations~\ref{z1} remains valid and the 
reader is also directed to the similar models by \citet{koeppen99}. 

In the special case where there are no inflows nor outflows Equation~\ref{z5} becomes:
\begin{equation}\label{zcb}
Z(t) = {y_Z \over 1-R } \ln \left[ 1 + {M_{*}(t)\over M_{g}(t)}  \right],\  {\rm when} \ \eta = 0\ {\rm and}\ \Lambda = 0.
\end{equation}
This is the classical closed-box solution \citep{searle72,pagel75}. We note that in the limit of ${M_{*}(t)\over M_{g}(t)} \ll 1$ all different cases of the models with inflow or outflow (Equations~\ref{z1},~\ref{z3},~\ref{z5}) approach the closed-box case $ Z(t) = {y_{Z} \over 1-R }{M_{*}(t)\over M_{g}(t)}$.

Equations~\ref{z1},~\ref{z3},~\ref{z5},~\ref{zcb} describe the evolution of total metalliticity with the change of the ratio of stellar mass to gas mass. 
While in recent years quantitative stellar spectroscopy has advanced beyond the Local Group and has been applied to measure total metallicity including iron group elements from the population of young massive stars in galaxies \citep{kud08,kud12,kud13,kud14,gazak14}, the standard method of metallicity determination in the disks of star forming galaxies is still the analysis of ISM H\textsc{ii} region emission lines, which yields the number fraction of oxygen to hydrogen or, more specifically, the oxygen abundance $\rm (O/H)=12+log[N(O)/N(H)]$. If in the equations above the metallicity yield $y_Z$ is replaced by the oxygen yield $y_O$, then they can be used to calculate the oxygen mass fraction $O_m$, which is then converted to the oxygen number fraction by ${\rm N(O)/N(H)} = O_m/(16X)$ ($X$ is the hydrogen mass fraction).

The formulae derived, if applied to the presently observed radial distribution of the ratio of stellar to gas masses in the disks of star forming galaxies, predict the metallicity mass fraction of the interstellar medium or the young stellar population. The free parameters of the model are the metallicity or oxygen yield $y_Z$ or $y_O$, the stellar mass return fraction $R$, and the mass loading and accretion factors $\eta$ and $\Lambda$. We will empirically calibrate $y_Z$ and  $y_O$ in the next section using Milky Way observations described in the previous sections. The value of $R$ depends crucially on assumptions for the initial mass function (IMF) and the strengths of stellar winds and mass loss in all phases of stellar evolution and is, thus, a rather uncertain number. \citet{leitner11} present calculations of $R$ as a function of the age of the stellar populations and for different IMFs and find a range of approximately 0.15 to 0.5. For our study presented here we adopt $R = 0.4$ throughout. However, we have also carried out all calculations and fits with a value of $R=0.2$ in parallel. The resulting changes are small and will be discussed in the forthcoming sections.
\floatplacement{figure}{!t}
\begin{figure}
\includegraphics[width=6.5cm,angle=90]{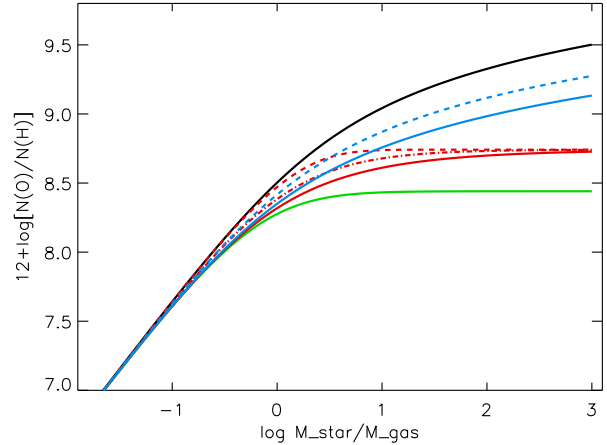}
\caption{Oxygen abundance as a function of the ratio of stellar to gas mass for a characteristic selection of the parameters $\eta$ and $\Lambda$. Black: closed-box
model with $\eta=0$, $\Lambda = 0$. Blue:  $\Lambda = 0$, $\eta = 0.5$ (dashed) and 1.0 (solid). Red: $\Lambda = 0.5$, $\eta = 0.0$ (dashed), 0.5 (dashed-dotted) and 1.0 (solid).
Green: $\Lambda = 1.0$, $\eta = 1.0$.}\label{figure1}
\end{figure}

The determination of the remaining two parameters $\eta$ and $\Lambda$ from the comparison of the observed metallicities and the model predictions is the major goal of this paper and will be described and discussed in the following sections. At this point, we discuss the major properties of the model. Fig.~\ref{figure1} shows how these parameters influence the metallicity as a function of the ratio of stellar to gas mass. The closed-box model has the largest metallicity. It does not show a saturation in metallicity but continuous with a logarithmical increase as a function of $M_{*}/M_{g}$. This is a consequence of the assumption of a constant yield during the course of chemical evolution. Models with $\Lambda = 0$ have a somewhat lower metallicity and the metallicity decreases with increasing $\eta$. These models do not saturate and increase logarithmically. However, models with $\Lambda$ larger than zero saturate at a metallicity $X_{\rm max} = y_Z/\Lambda$ as long as $\alpha \geq -1$. How fast this level of saturation is reached with increasing $M_{*}/M_{g}$ depends on $\eta$. Models with very small values of $\eta$ follow the closed-box model closely, but their metallicities suddenly turn into the maximum metallicity $Z_{\rm max}$, whereas models with larger $\eta$ depart early from the closed-box case and then approach $Z_{\rm max}$ gradually.

\floatplacement{figure}{!t}
\begin{figure}
\includegraphics[width=6.5cm,angle=90]{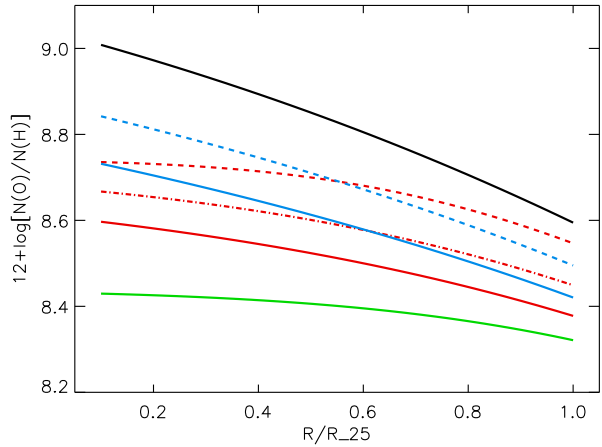}
\caption{Oxygen abundance as a function of galactocentric radius for a galactic disk with an exponential profile of the ratio of stellar to gas mass (see text) using the same models as in Fig.~\ref{figure1}.}\label{figure2}
\end{figure}

It is obvious that the systematically different behaviour of the curves in Fig.~\ref{figure1} as a function of the parameters $\eta$ and $\Lambda$ opens the possibility to determine these parameters for disk galaxies with well observed spatially resolved radial profiles of stellar and gas mass and metallicity. We demonstrate this in Fig.~\ref{figure2}, where we have adopted a simple exponential profile for the ratio of stellar to gas mass, $M_*/M_g = 5~\exp(-x/h)$ with $x=r/R_{25}$ and $h=0.5$ (see \citealt{ho15} for the choice of this exponential). The closed-box model produces the highest metallicity and the steepest gradient. Models with no inflow are slightly shallower and have lower metallicities which depend on the strength of the adopted outflows. Models with inflow produce even flatter gradients with increasing $\Lambda$.

In two recent publications, \citet{zahid14} and \citet{ascasibar14} have developed analytical chemical evolution models which relate the averaged metallicities of galaxies as observed in the large galaxy surveys to the ratio of total galactic stellar mass to total gas mass. These models reproduce the observed global MZR of galaxies very well. \citet{zahid14} use the observed scaling relation that the total ISM hydrogen mass in the disks of star forming galaxies relates to the total stellar disk mass by $M_g \propto M_*^g$ with an observed exponent $g \approx$ 0.5. However, in the spatially resolved case this relationship does not hold and, thus, the approach by \citet{zahid14} will require some modifications to predict the stratification of spatially resolved metallicity in galaxies.

\begin{figure}
\includegraphics[width=6.5cm,angle=90]{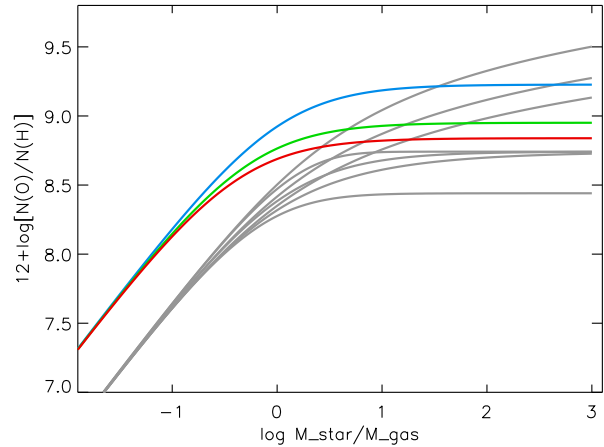}
\caption{Oxygen abundance as a function of the ratio of stellar to gas mass predicted by the global galaxy model by \citet{ascasibar14} and described by Equation~\ref{ascasibar}. Blue: $\Upsilon = 4.5$, green: $\Upsilon = 8.5$, red: $\Upsilon = 11.0$. The models in grey are the same as in Fig.~\ref{figure1}.}\label{figure3}
\end{figure}

\citet{ascasibar14} apply an analytical time averaging procedure of the general chemical evolution equations and show that their weak wind solution (the case where their $\tilde{\epsilon}_W$ = 0)
\begin{equation}\label{ascasibar}
12+\log{{\rm N(O) \over N(H)}} = 9.750 + \log({s \over 1+{1 \over 1.35}\Upsilon s})
\end{equation}
with $M_g = 1.35~M_H$, the ISM hydrogen mass, and
\begin{equation}\label{s}
s = {R \over 1-R}{M_* \over M_H}.
\end{equation}
leads to very good agreement with observed global galaxy oxygen abundances as a function of global gas and stellar mass. They argue that in the way as it was derived this relationship should also hold for the spatially resolved case in galaxy disks. \citet{ascasibar14} suggest a range of their parameter $\Upsilon$ from 4.5 to 11 and use a return fraction $R$ of $R=0.18$. Fig.~\ref{figure3} compares the \citet{ascasibar14} model with three values of $\Upsilon$ with our models. They all predict much higher metallicity at low ratios of stellar to gas mass. This is a consequence of the higher yield used by \citet{ascasibar14}, which leads to their constant 9.750. Modifying the yield to lower values would shift the curves downward. We will test this model in the next section.

\section{The radial metallicity distribution of the Milky Way disk and calibration of metallicity yield}\label{sec:the-radial-metallicity-distribution-of-the-milky-way}

With the well observed radial distributions of stellar mass, ISM gas mass and metallicity in the Milky Way disk as described in Section~\ref{sec:galaxy-sample-and-observations}, we can now empirically calibrate the metallicity yield $y_Z$ of our chemical evolution model of Section~\ref{sec:the-chemical evolution-model}. We are aware of the many theoretical model calculations of yields (see \citealt{zahid12}, \citealt{yabe15}, \citealt{pagel09}, or \citealt{matteucci12} for an overview), but with the apparent uncertainties introduced by the choices of IMF or assumptions about stellar winds and mass-loss, we regard a differential study of galaxies relative to the Milky Way with an empirically calibrated yield as more appropriate approach.

\begin{figure*}
\includegraphics[width=6.5cm,angle=90]{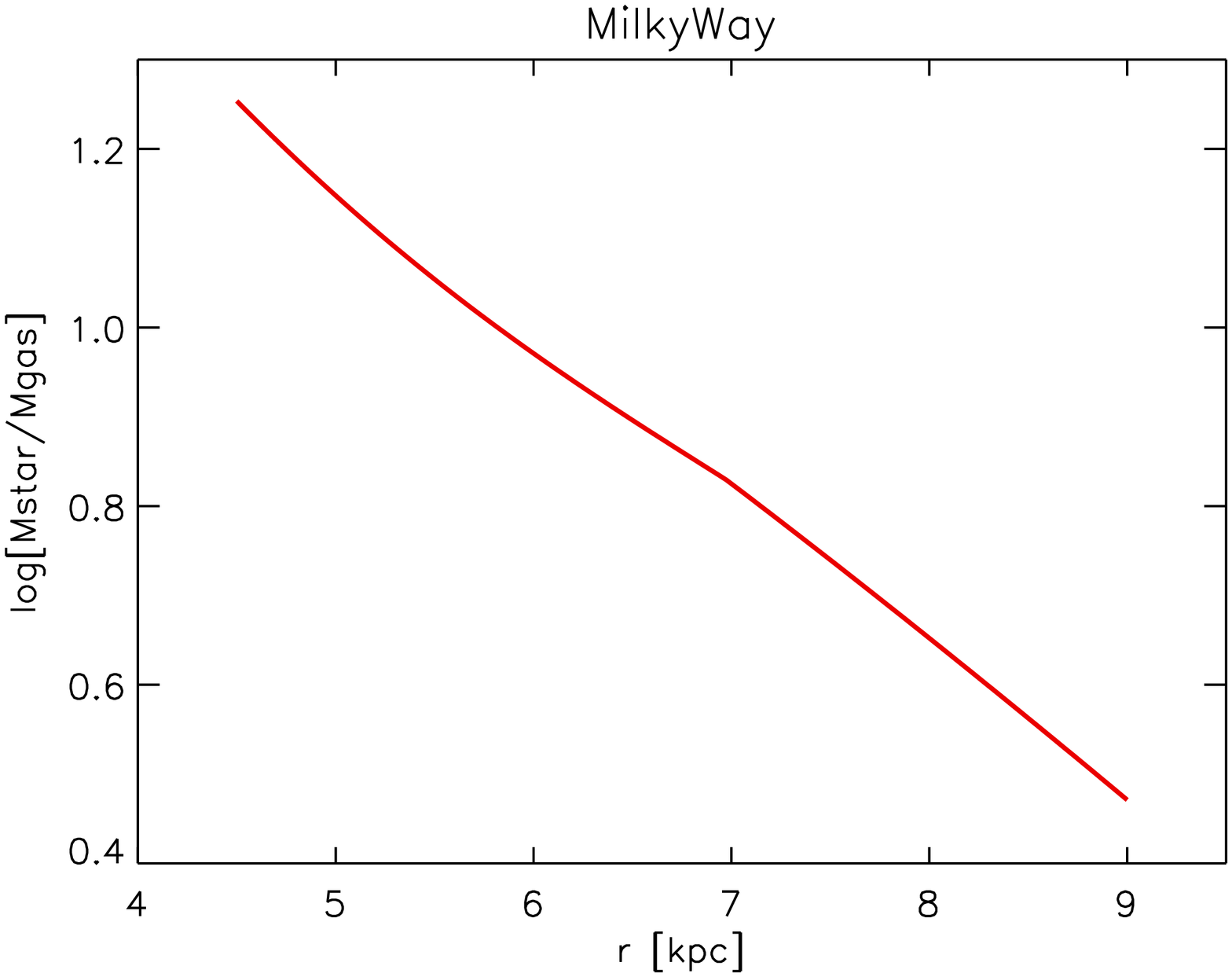}
\includegraphics[width=6.5cm,angle=90]{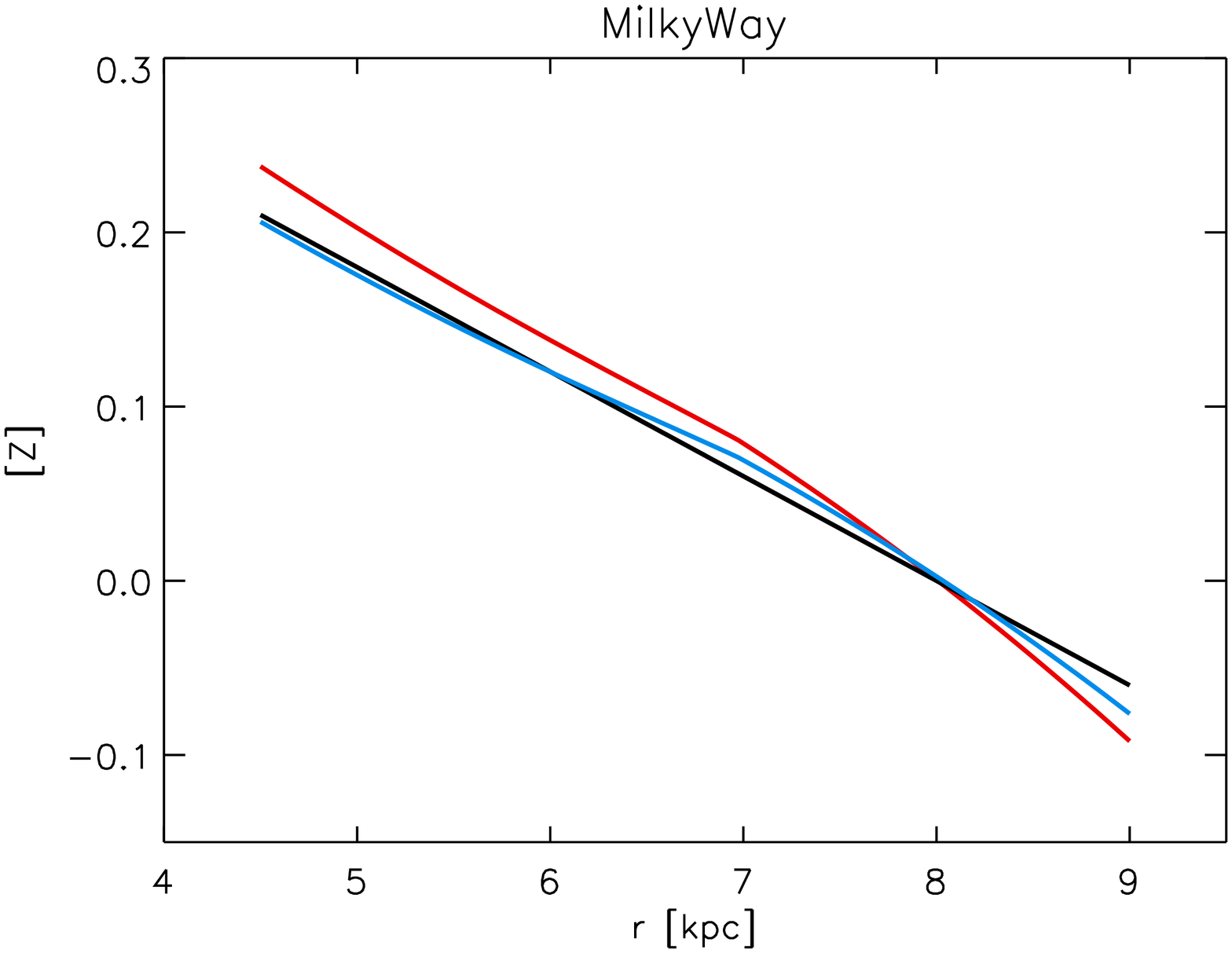}
\caption{Left: The radial profile of the observed ratio of stellar to gas mass column density of the Milky Way disk. Right: Observed Milky Way disk metallicity $[Z]$ (black) compared to the prediction of the closed-box model (red) and the best fitting chemical evolution model (blue) with $\eta = 0.432$ and $\Lambda = 0$.}\label{figure4}
\end{figure*}

We start with the assumption of the closed-box case described by Equation~\ref{zcb}. According to Section~\ref{sec:galaxy-sample-and-observations}, the stellar mass column density of the Milky Way disk is 38~\msun~pc$^{-2}$ and the total gas mass of the ISM (H\textsc{i} and H$_2$) is 8.465~\msun~pc$^{-2}$, so that the ratio of the two is 4.49. With a metallicity of the young stellar population $Z_{\rm B-stars} = 0.014$ (see Section~\ref{sec:galaxy-sample-and-observations}) and Equation~\ref{zcb}, we constrain the metallicity yield to $y_Z/(1-R) = 0.00822$. We can use this yield and apply Equation~\ref{zcb} on the observed radial profile of the ratio of stellar to gas mass (Fig.~\ref{figure4}) to predict the metallicity distribution of the ISM and the young stellar population in the Milky Way disk. The result is shown in Fig.~\ref{figure4} and compared to the observed metallicity gradient. The agreement, though not perfect, is surprisingly good.

In the next step, we relax the closed-box assumptions and use the full chemical evolution model with galactic winds and accretion as described in Section~\ref{sec:the-chemical evolution-model}. We adopt a stellar wind return fraction $R=0.4$ (see Section~\ref{sec:the-chemical evolution-model} for discussion) and a range of metallicity yields $y_Z$ from 0.004 to 0.012. For each yield $y_{Z,i}$ adopted in this range, we calculate chemical evolution model distributions $[Z]_{\rm mod}(r, \eta_j, \Lambda_k)$ for an array of discrete values of $\eta_j$ and $\Lambda_k$. We then use the difference between $[Z]_{\rm mod}(r, \eta_j, \Lambda_k)$ and the observed distribution of $[Z](r)$ at hundred radial points distributed equally between 4.5 to 9.0 kpc to calculate a $\chi^2$-matrix $\chi^2(y_{Z,i}, \eta_j, \Lambda_k)$ and search for the minimum. We find a well determined minimum $\chi^2_{\rm min}$ at $y_{Z} = 0.00671$, $\eta = 0.432$, $\Lambda = 0$. The corresponding $[Z](r)$ model distribution is shown in Fig.~\ref{figure4}. The fit is a clear improvement relative to the closed-box model. 

\begin{figure}
\includegraphics[width=6.5cm,angle=90]{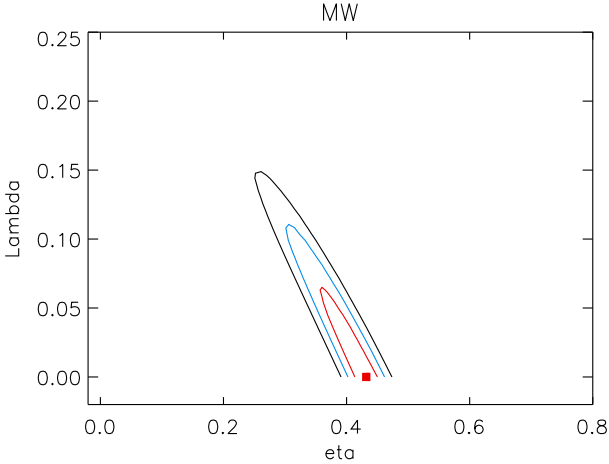}
\caption{$\Delta \chi^2$ isocontours in the $(\eta,\Lambda)$-plane of the observed Milky Way disk metallicity distribution $[Z](r)$ for a yield $y_Z= 0.00671$. The $\Delta \chi^2$ values adopted for the plot are 2.3 (red), 6.17 (blue), 11.8 (black).}\label{figure5}
\end{figure}

To assess the accuracy of the yield determination we repeat the procedure assuming metallicities at the error margins (see Section~\ref{sec:galaxy-sample-and-observations}), $[Z]_+(r)= 0.50 - 0.058r$ and $[Z]_-(r)= 0.46 - 0.062r $ (with the uncertainty $\sigma(r) = 0.02 +0.006r$ in the $\chi^2$ calculation). We obtain $y_{Z} = 0.00776$, $\eta=0.549$, $\Lambda = 0$ for $[Z]_+$ and $y_{Z} = 0.00572$, $\eta=0.306$, $\Lambda = 0$ for $[Z]_-$ and conclude that the yield is constrained through our procedure with an accuracy of $\Delta\log y_Z \approx 0.07$ dex. The value of $\eta$ has the uncertainty of $\Delta \eta \approx 0.12$. While all fits settle at values $\Lambda = 0$, the $\Delta \chi^2$ isocontours at a fixed yield as shown in Fig.~\ref{figure5} indicate an range of $\Delta \Lambda \approx$ 0.06 to 0.1. 

We also repeat the yield calibration procedure for a stellar mass return fraction of $R = 0.2$ and obtained $y_Z = 0.00895$, $\eta = 0.576$, $\Lambda = 0$ at the best fit solution. The fit is of similar quality as the one in Fig.~\ref{figure4}. We note that with the change from $R = 0.4$ to 0.2 the ratios $y_Z/(1-R)$ and $\eta/(1-R)$ have remained constant. This is a consequence of the fact that the best fits for both $R = 0.4$ and 0.2, respectively, have settled on a solution with $\Lambda = 0$ for which Equation~\ref{z5} holds. In the following we will use $R = 0.4$ and the correspondingly calibrated yield.

To apply our chemical evolution model to the observed ISM oxygen abundances of the SL-sample of galaxies in the forthcoming section, we need to turn the calibrated metallicity yield $y_Z$ into an oxygen yield $y_O$. For this purpose, we use the results of the work by \citet{nieva12} which we have discussed in Section~\ref{sec:galaxy-sample-and-observations}. For their sample of B-stars in the solar neighbourhood, they have found an oxygen abundance $\rm(O/H)_{B-stars} = 8.76$. With an average hydrogen mass fraction X=0.710, this corresponds to an oxygen mass fraction $O_{B} = 0.00654$ and an observed ratio of oxygen mass fraction to total metallicity mass fraction $O_B/Z_B = 0.467$. Thus, in order to obtain with our chemical evolution model the same ratio of oxygen mass fraction to total metallicity as observed in B-stars in the solar neighbourhood, we need the relation between oxygen and metallicity yield as $y_O = (O_{B}/Z_B)y_Z = 0.467 y_Z$. This means that for $R=0.4$ we will work with an oxygen yield of $y_O = 0.00313$.

\begin{figure}
\includegraphics[width=6.5cm,angle=90]{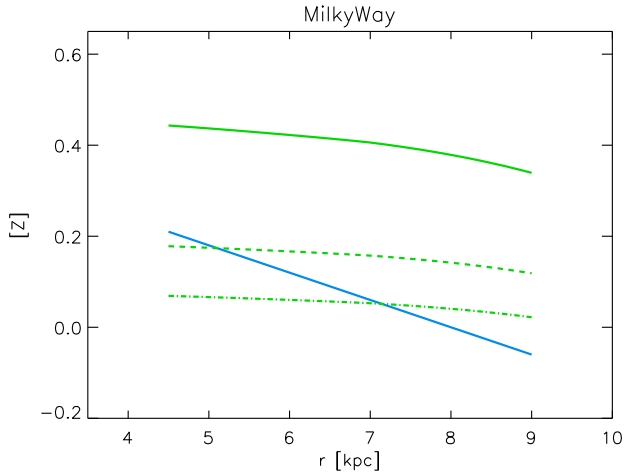}
\caption{Observed Milky Way disk metallicity $[Z]$ (blue solid) compared with the model by \citet{ascasibar14} (see Equation~\ref{ascasibar}) for $\Upsilon = 4.5$ (green solid), 8.5 (green dashed) 
and 11 (green dashed-dotted).}\label{figure6}
\end{figure}

With the good fit obtained in Fig.~\ref{figure4}, it is interesting to carry out a similar comparison with the chemical evolution model by \citet{ascasibar14} introduced in Section~\ref{sec:the-chemical evolution-model}. For this purpose we use the relation $[Z] = \rm (O/H) - (O/H)_{B-stars}$. The comparison shown in Fig.~\ref{figure6} demonstrates that the model which worked well for global models of total galactic metallicity using total gas and stellar masses does not fit well in the spatially resolved case of the Milky Way, because the metallicities saturate too early at still relatively low ratios of stellar to gas mass. This is evident from Equations~\ref{ascasibar} (see also Fig.~\ref{figure3}). The turnover from the linear unsaturated relation to maximum metallicity begins at $\Upsilon(R/(1-R))(M_*/M_g) \geq 1$ or  $M_*/M_g \geq 4.55/\Upsilon$ (for $R = 0.18$, as used by \citealt{ascasibar14}, larger values of $R$ make things worse). This means that very small values of $\Upsilon$ would be needed to avoid too early saturation. Since $\Upsilon$ in their model is proportional to the ratio of mass-weighted metallicity of all stars (not only the young population) to the gas-phase metallicity, such small values could be obtained by significant amounts of infall at very early times, which would strongly reduce this ratio. However, at the same time, the maximum metallicity in their model is given by $[Z]_{\rm max} = const. - \log \Upsilon$ and, thus, increases dramatically with decreasing $\Upsilon$. To compensate for that would require a strong reduction of the yield, as the constant in $[Z]_{\rm max}$ changes with $\Delta\log y_Z$.

\floatplacement{figure}{!t}
\begin{figure*}
\centering
\subfloat{\includegraphics[width = 6.5cm,angle=90]{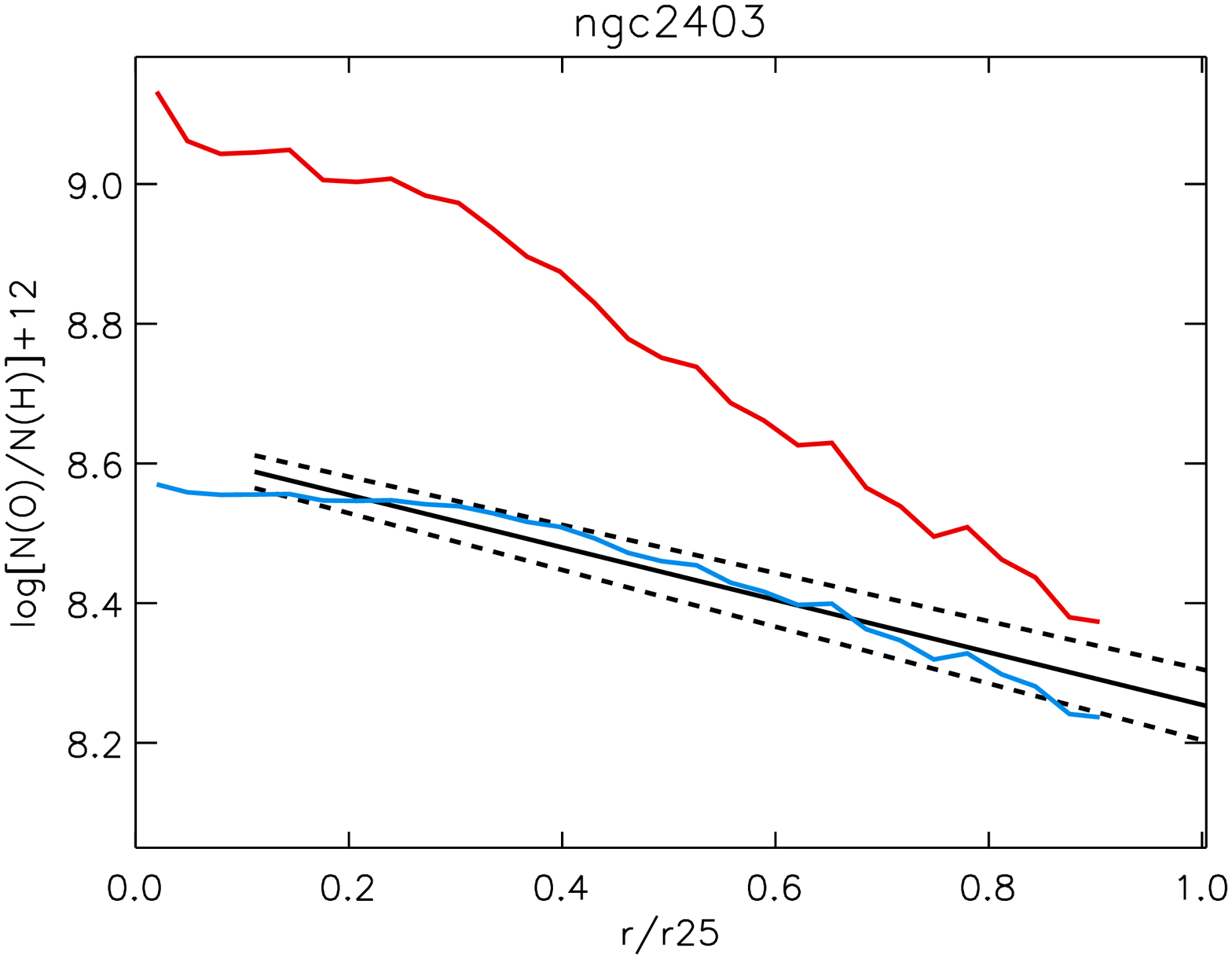}} \subfloat{\includegraphics[width = 6.5cm,angle=90]{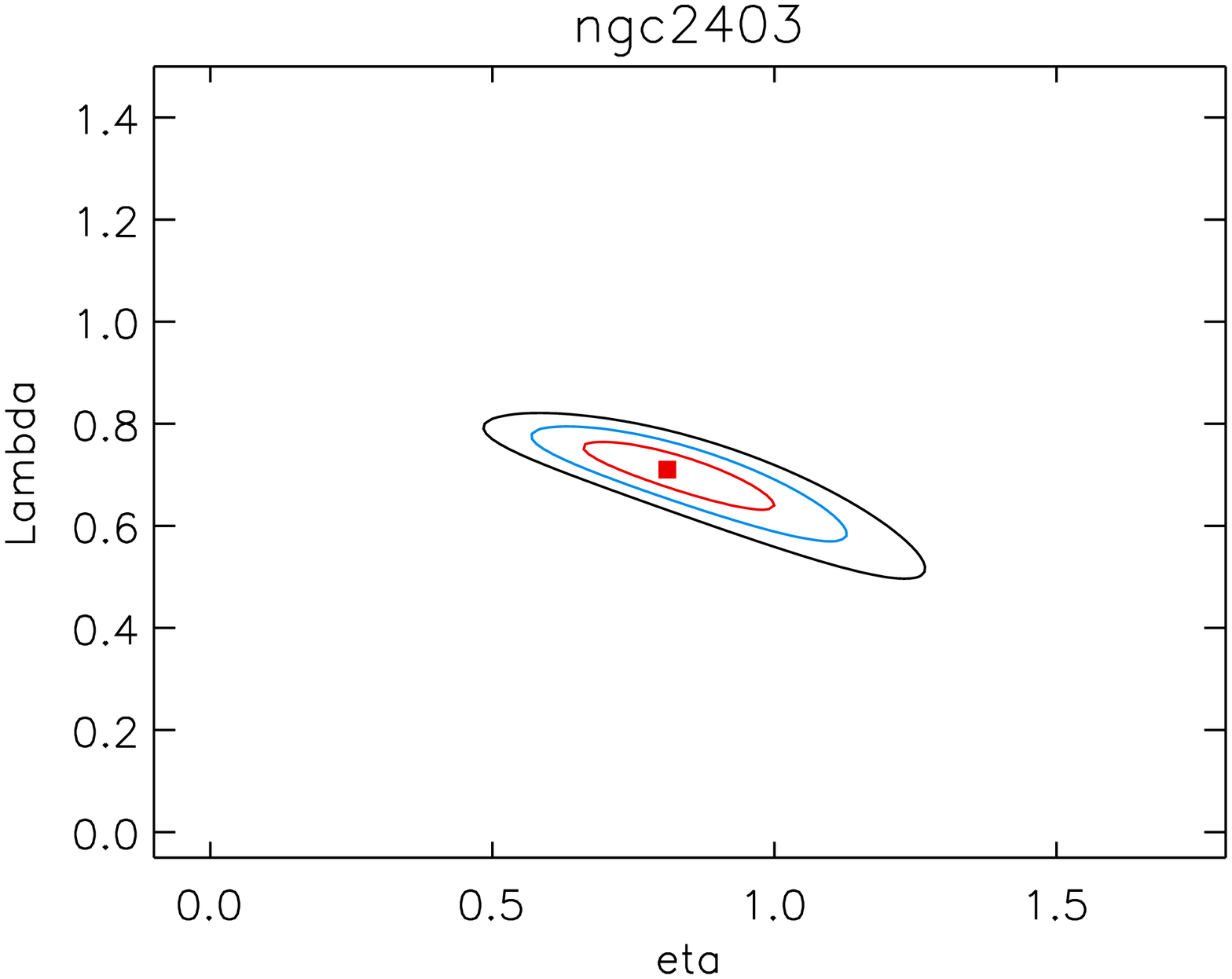}}\\
\subfloat{\includegraphics[width = 6.5cm,angle=90]{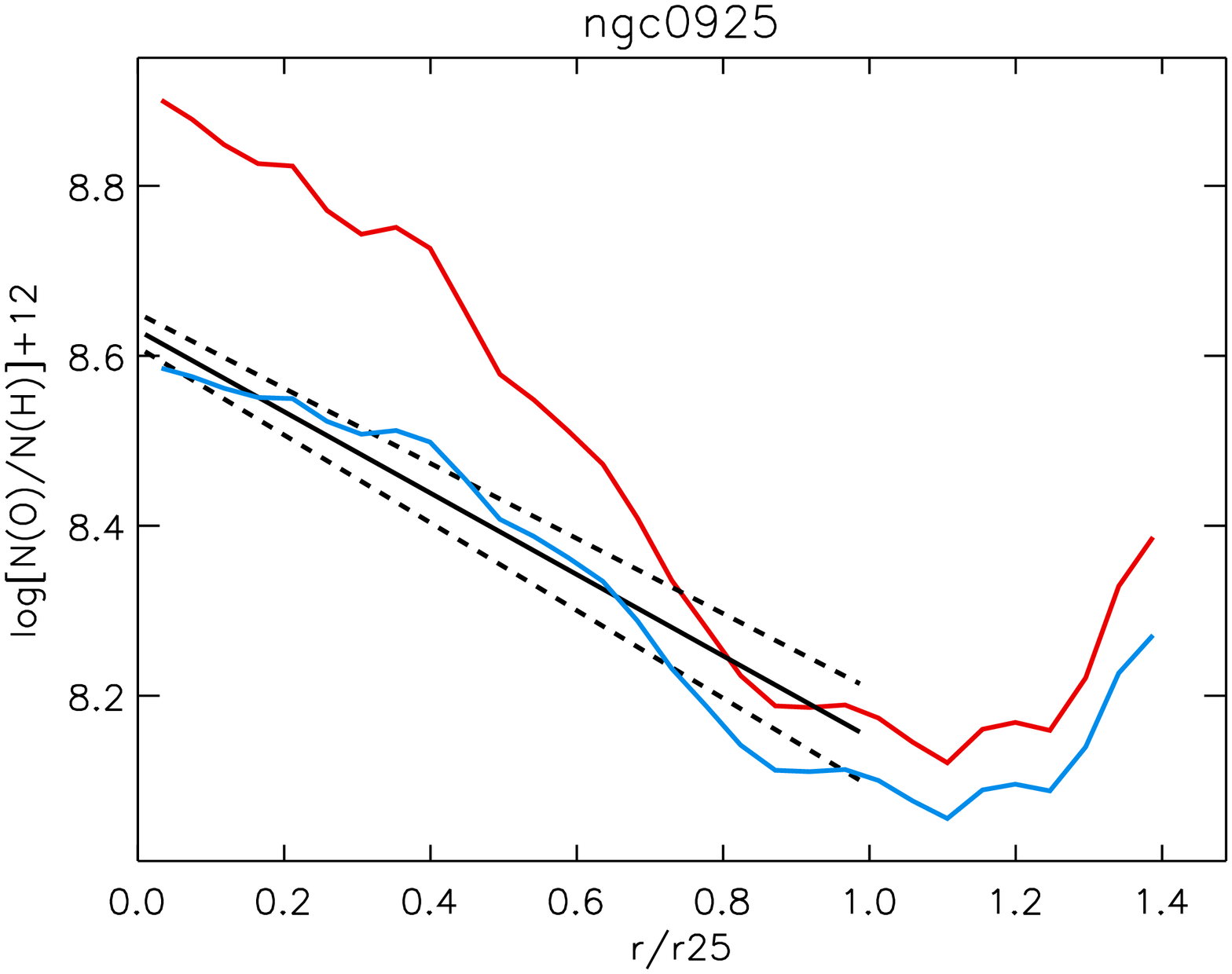}} \subfloat{\includegraphics[width = 6.5cm,angle=90]{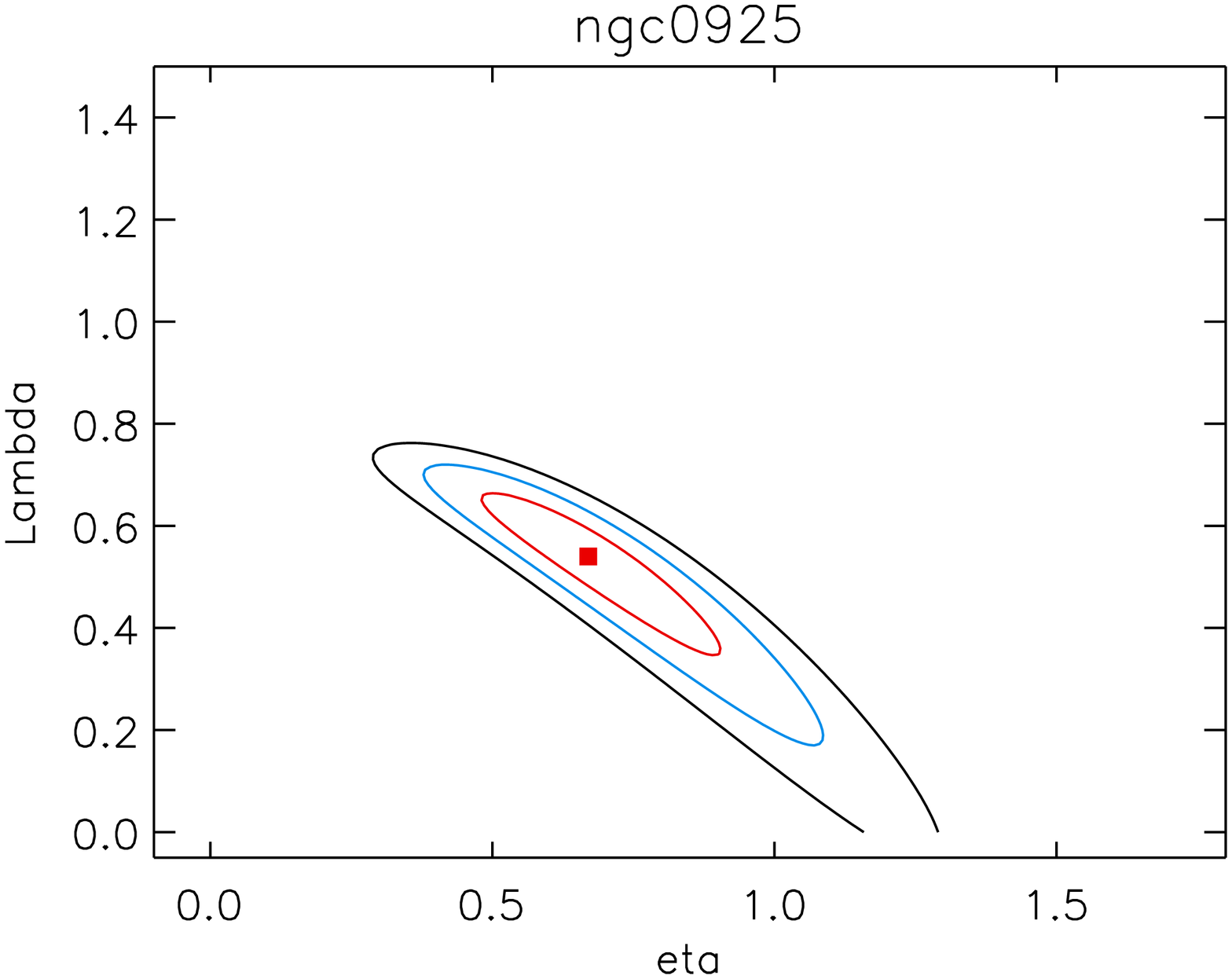}}\\
\subfloat{\includegraphics[width = 6.5cm,angle=90]{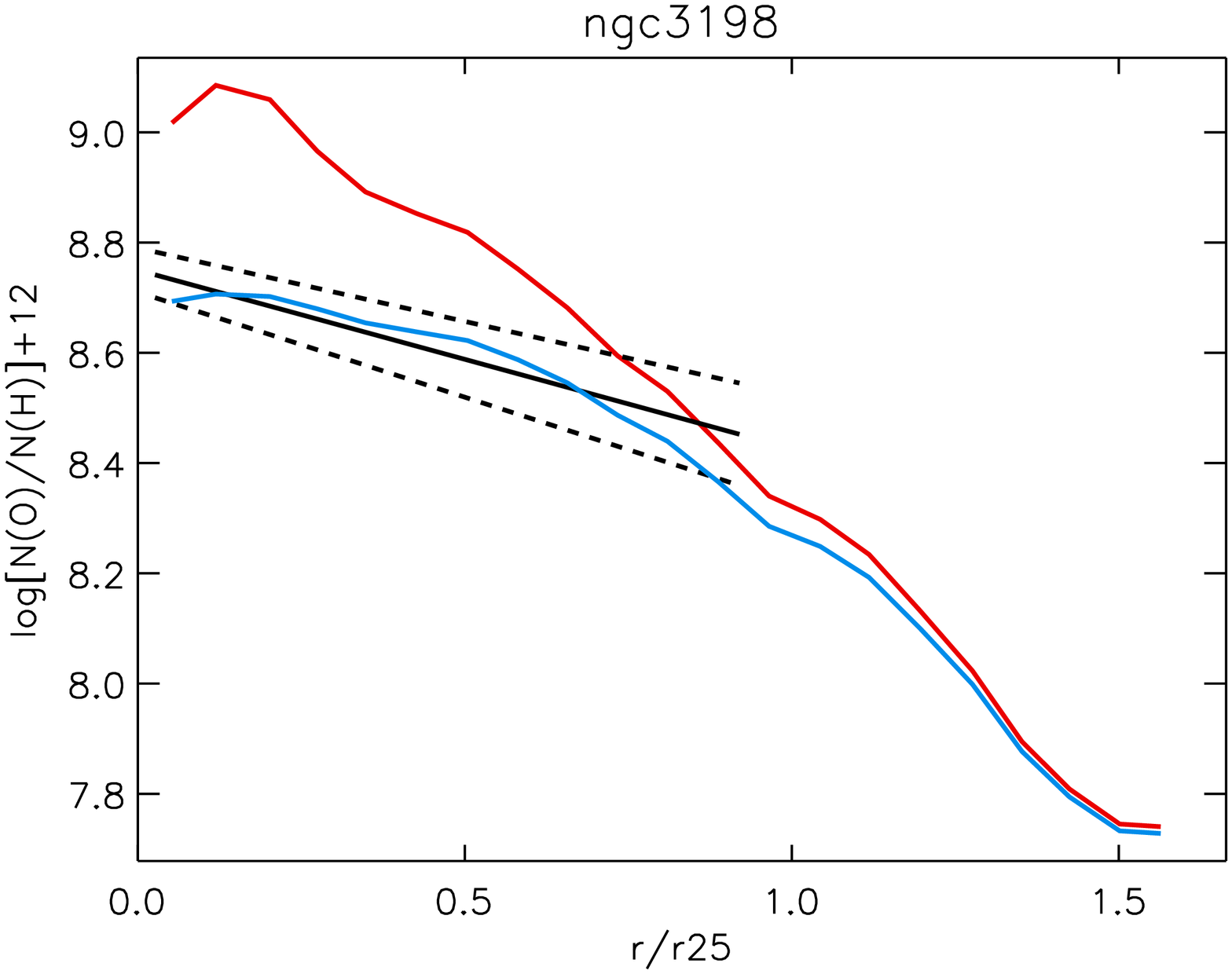}} \subfloat{\includegraphics[width = 6.5cm,angle=90]{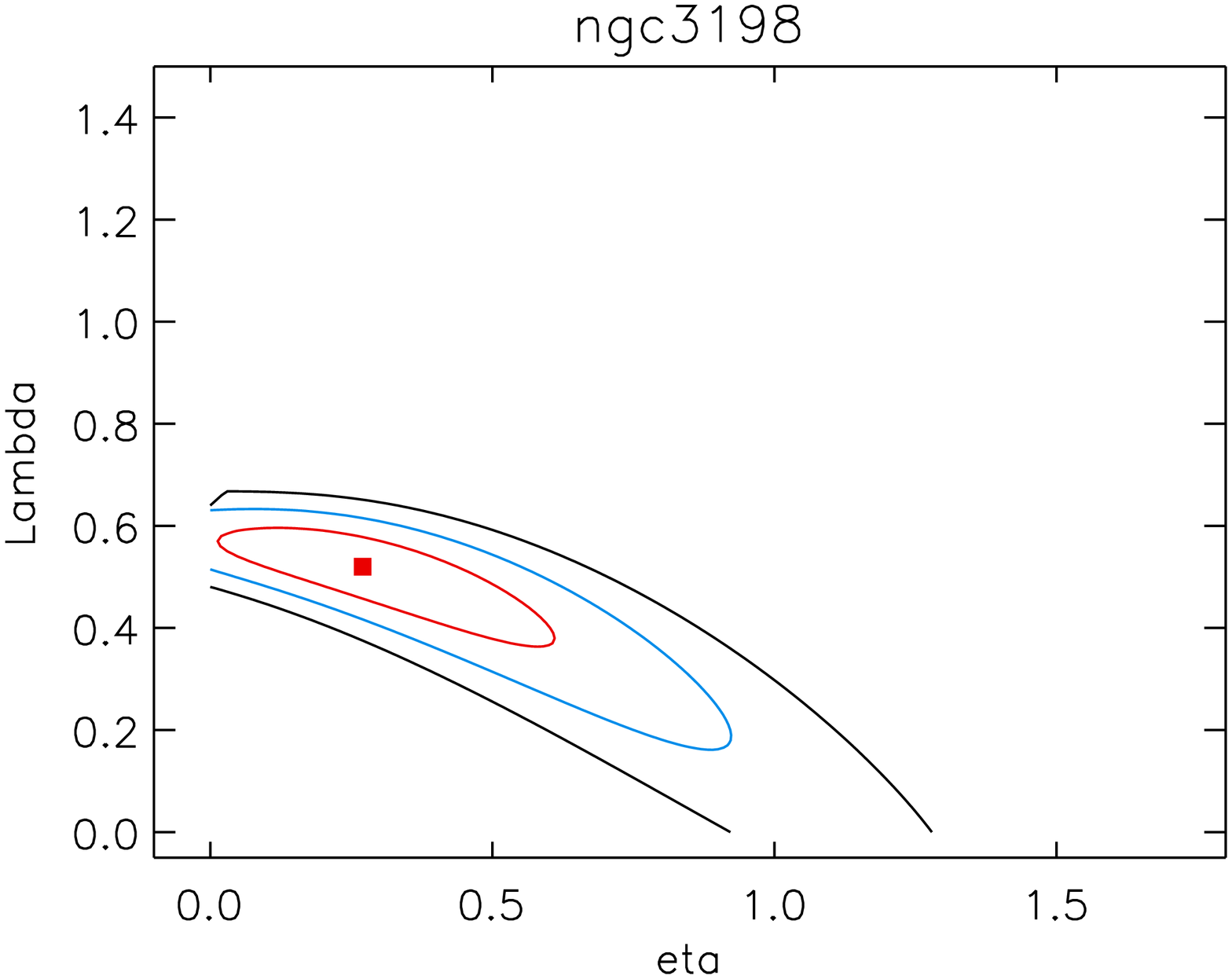}}\\
\caption{Chemical evolution fits of the galaxies NGC\,2403 (top), 925 (middle), 3198 (bottom). Left: comparison of the observed radial oxygen abundance distribution (black with the 1$\sigma$ 
uncertainty indicated by the dashed lines) with the final model (blue) and the closed-box model (red). Right: $\Delta \chi^2$ isocontours in the $(\eta,\Lambda)$-plane corresponding
to $\Delta \chi^2 = 2.3$ (red), 6.17 (blue) and 11.8 (black).}\label{figure7}
\end{figure*}
\floatplacement{figure}{!t}
\begin{figure*}
\centering
\subfloat{\includegraphics[width = 6.5cm,angle=90]{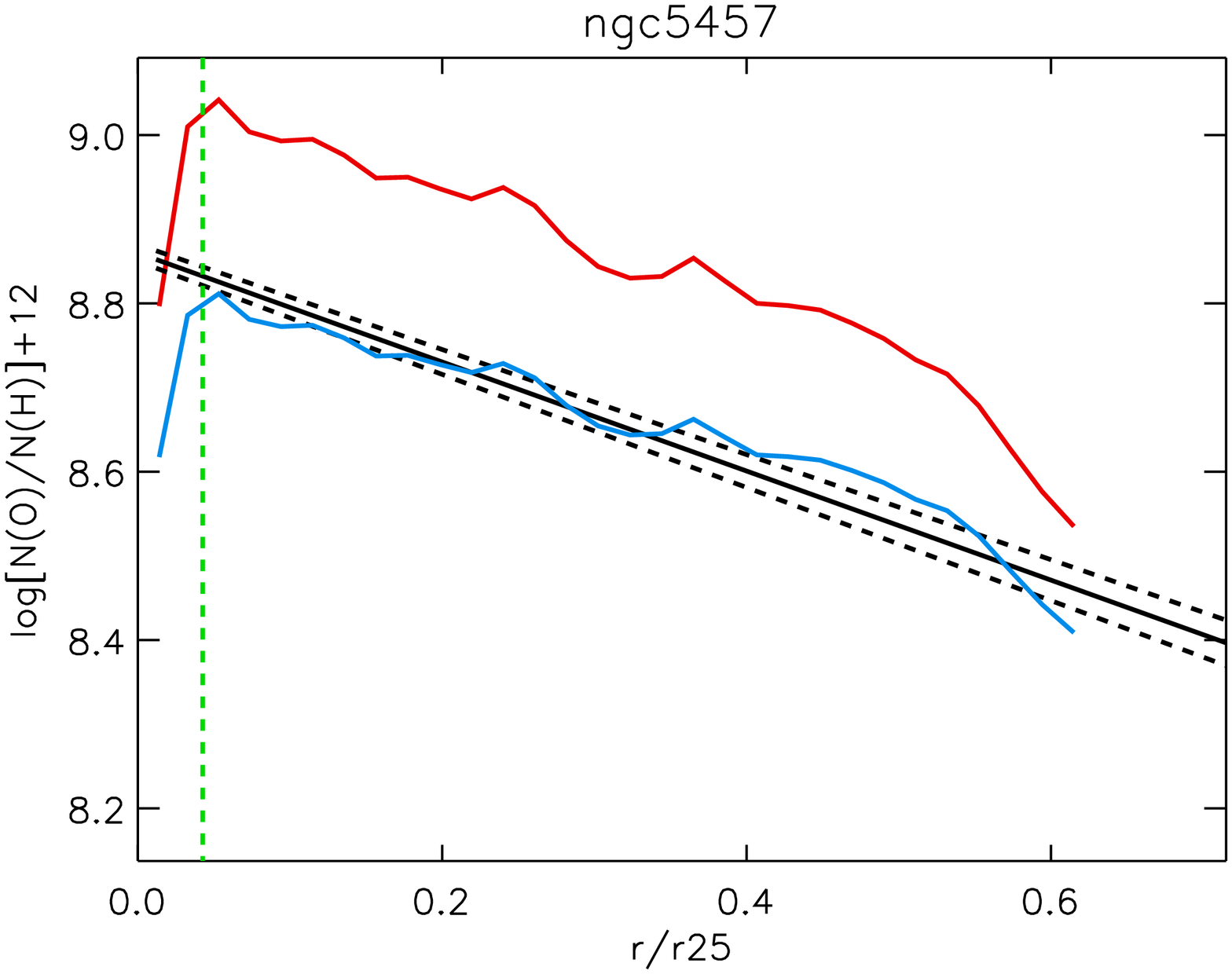}} \subfloat{\includegraphics[width = 6.5cm,angle=90]{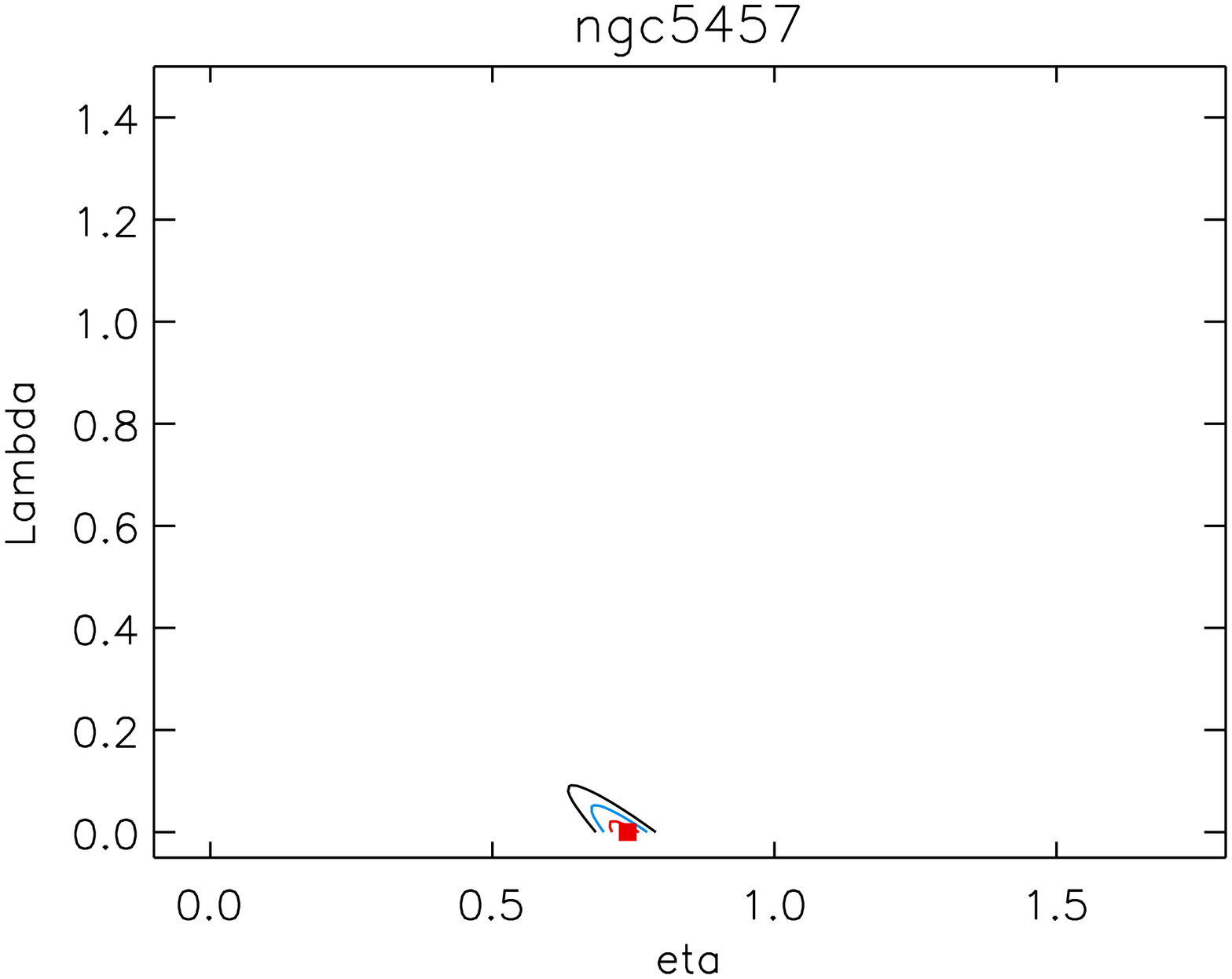}}\\
\subfloat{\includegraphics[width = 6.5cm,angle=90]{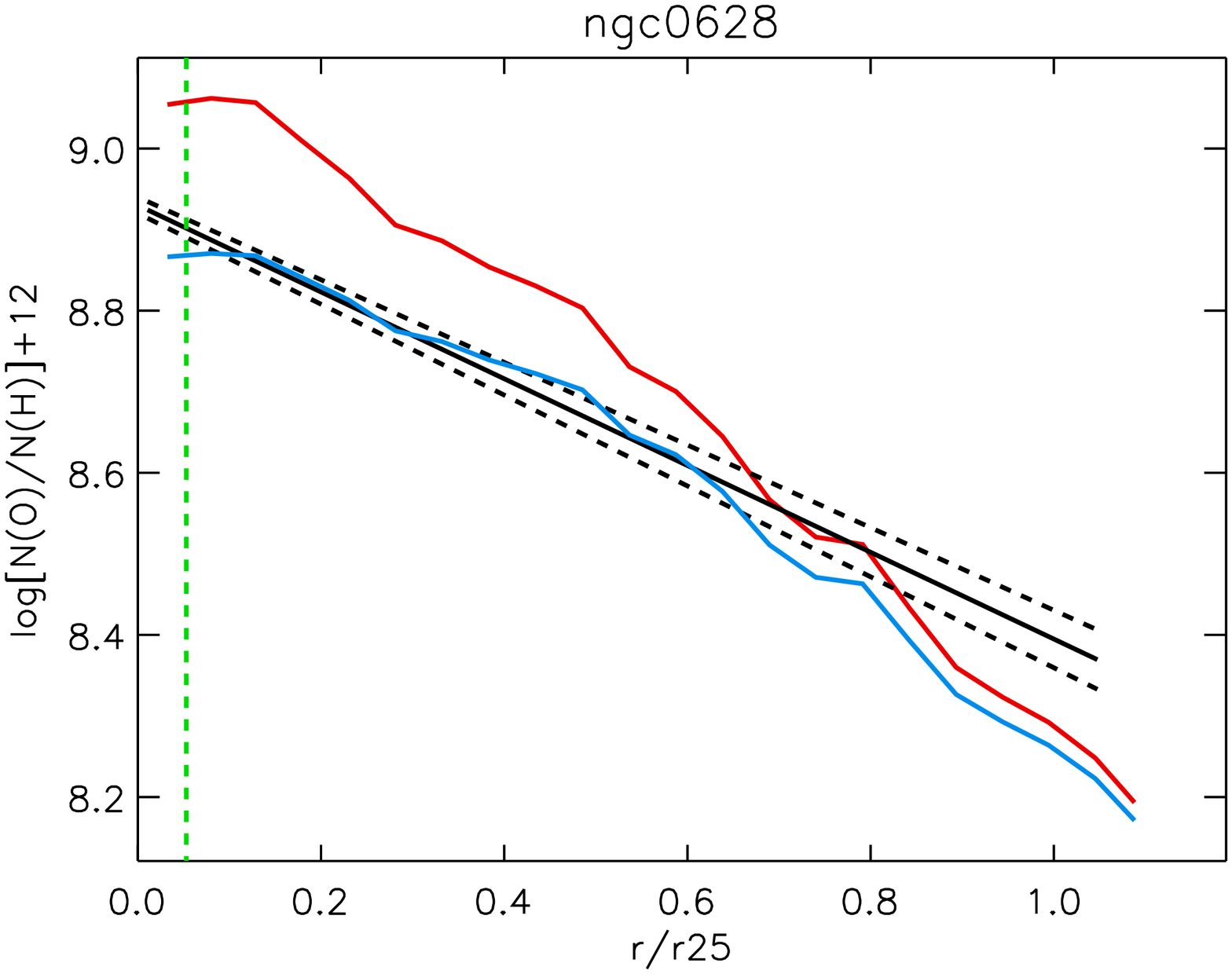}} \subfloat{\includegraphics[width = 6.5cm,angle=90]{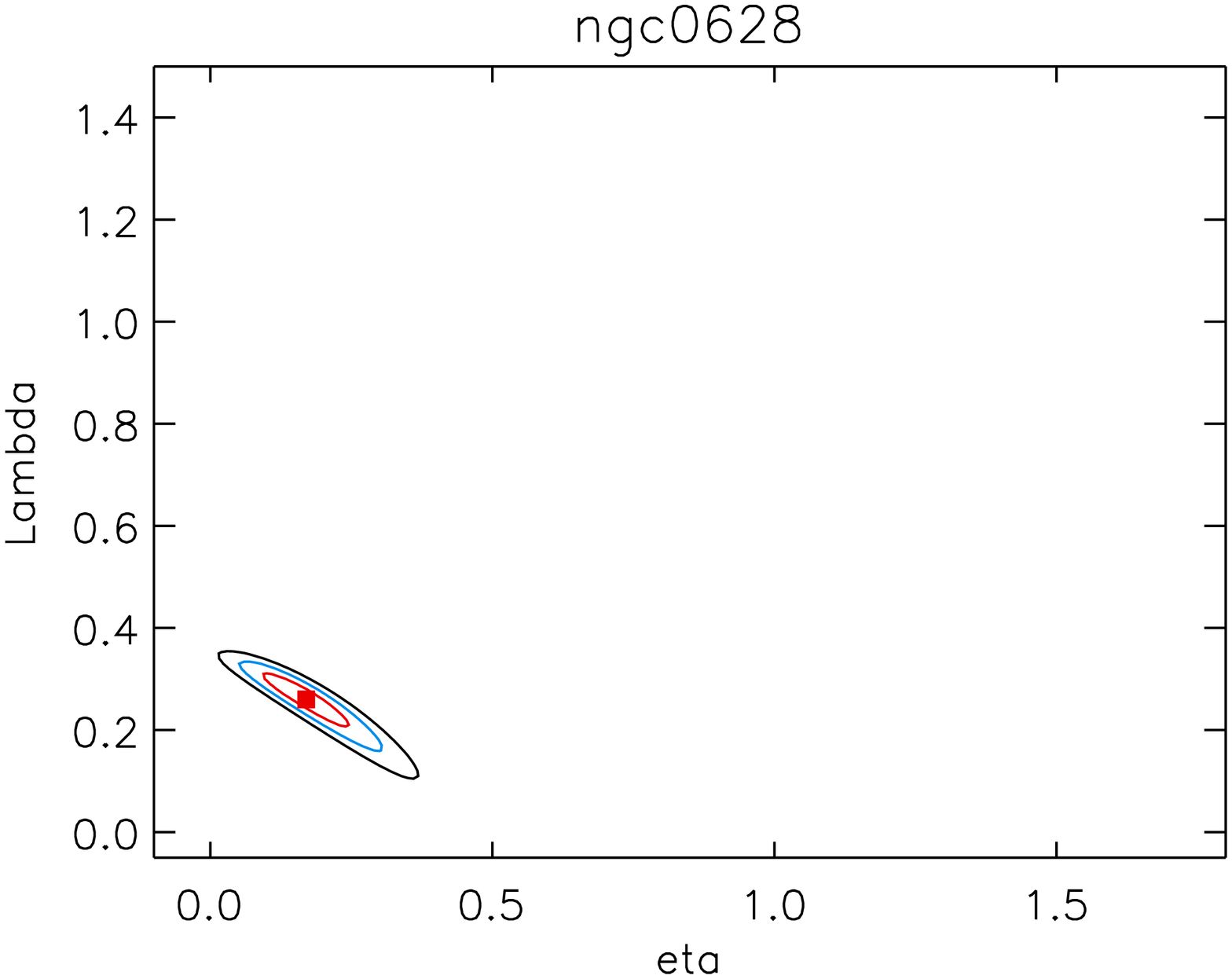}}\\
\caption{Chemical evolution fits of the galaxies NGC\,5457 (top) and 628 (bottom). The arrangement of the figures is similar to Fig.~\ref{figure7}. The observations for
galactocentric distances shorter than indicated by the dashed vertical line are not included in the $\chi^2$-fit.}\label{figure8}
\end{figure*}

\floatplacement{figure}{!t}
\begin{figure*}
\centering
\subfloat{\includegraphics[width = 6.5cm,angle=90]{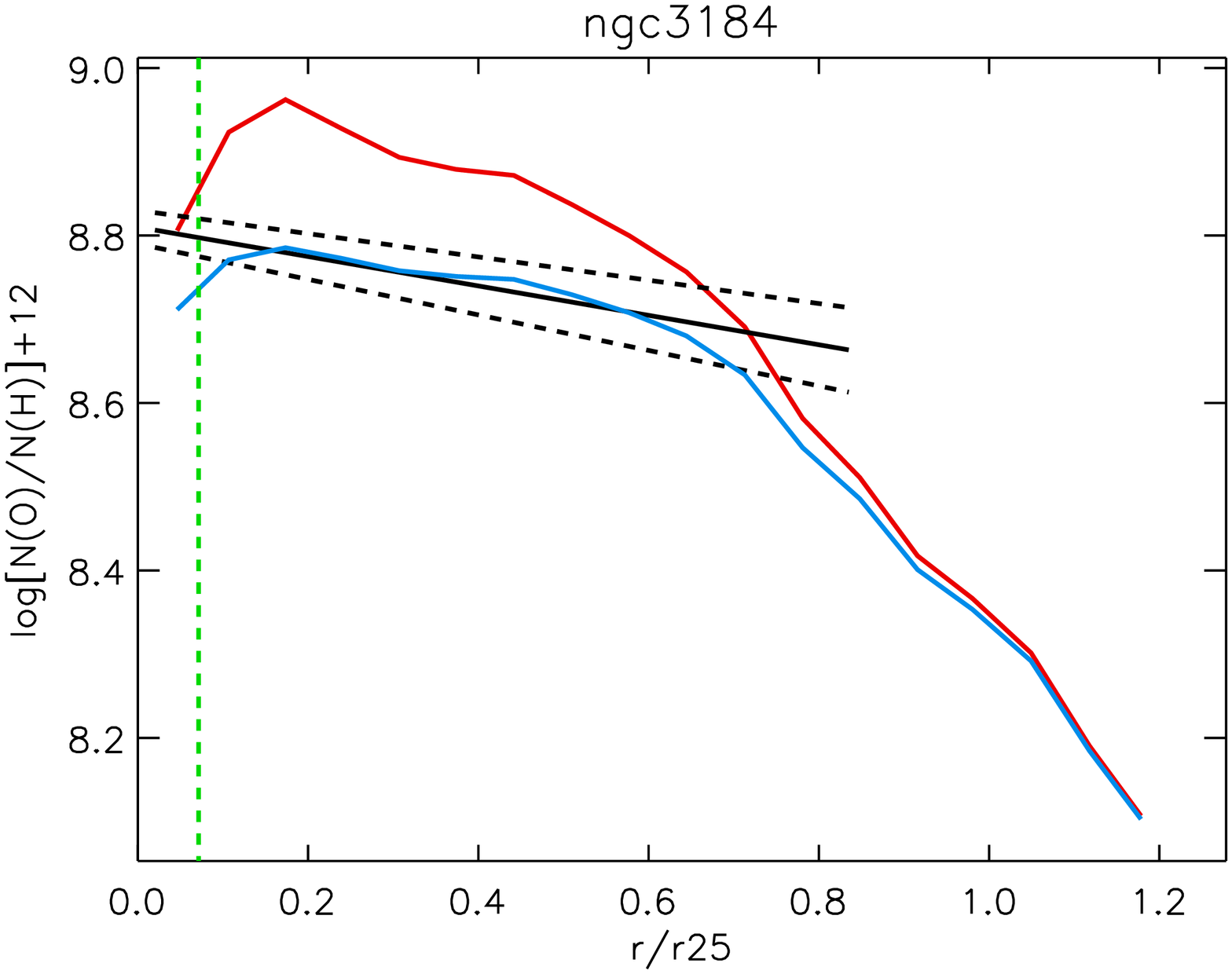}} \subfloat{\includegraphics[width = 6.5cm,angle=90]{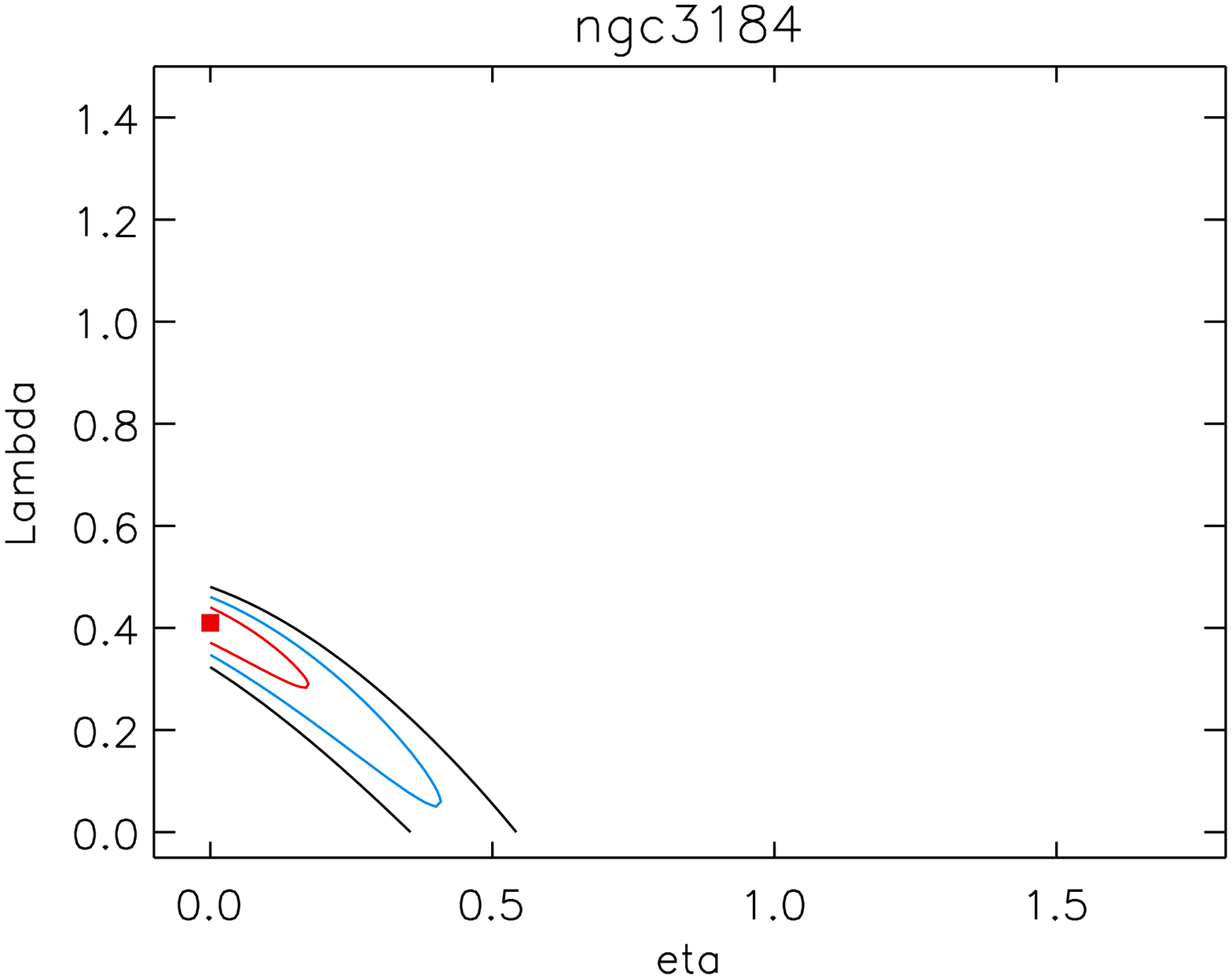}}\\
\subfloat{\includegraphics[width = 6.5cm,angle=90]{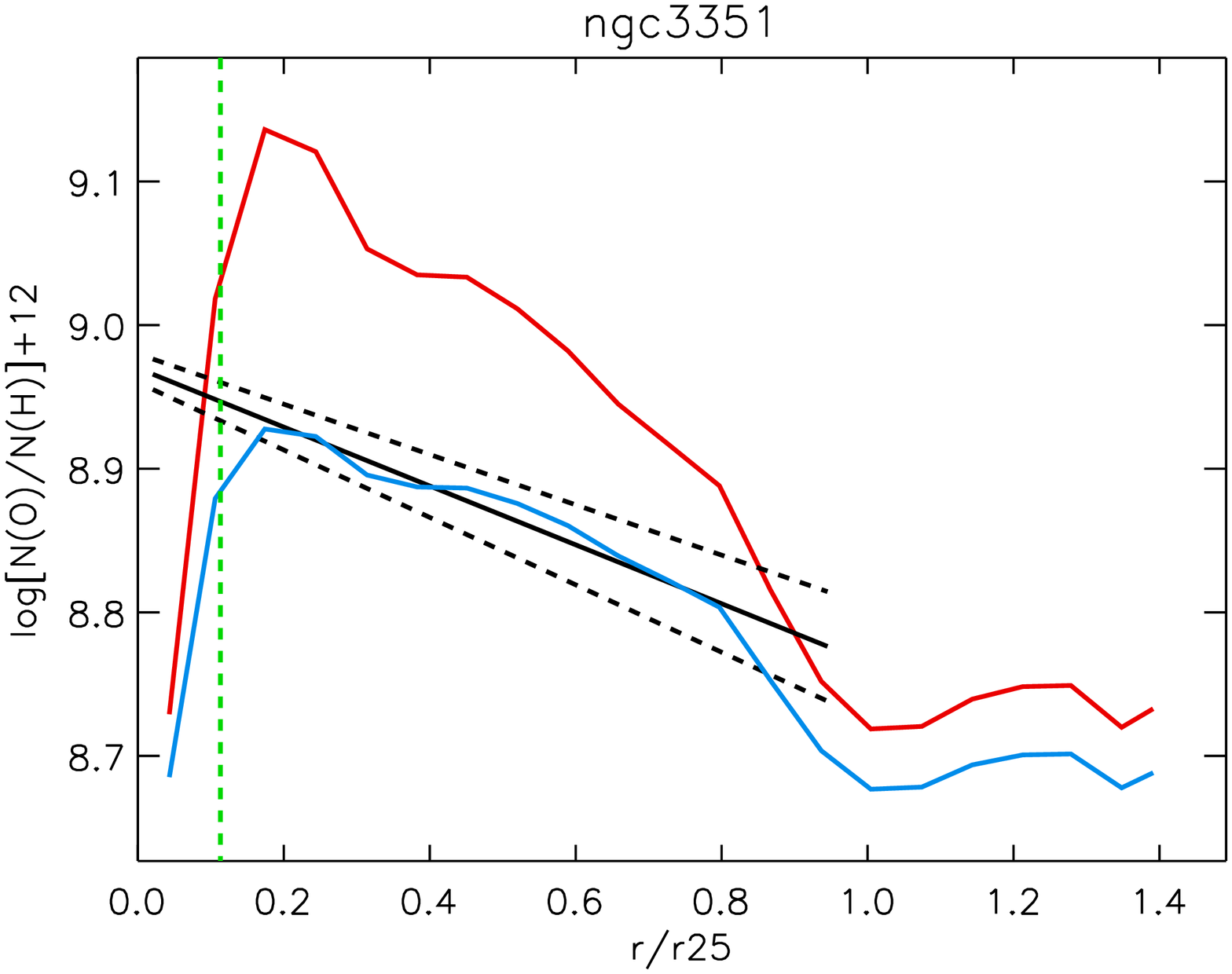}} \subfloat{\includegraphics[width = 6.5cm,angle=90]{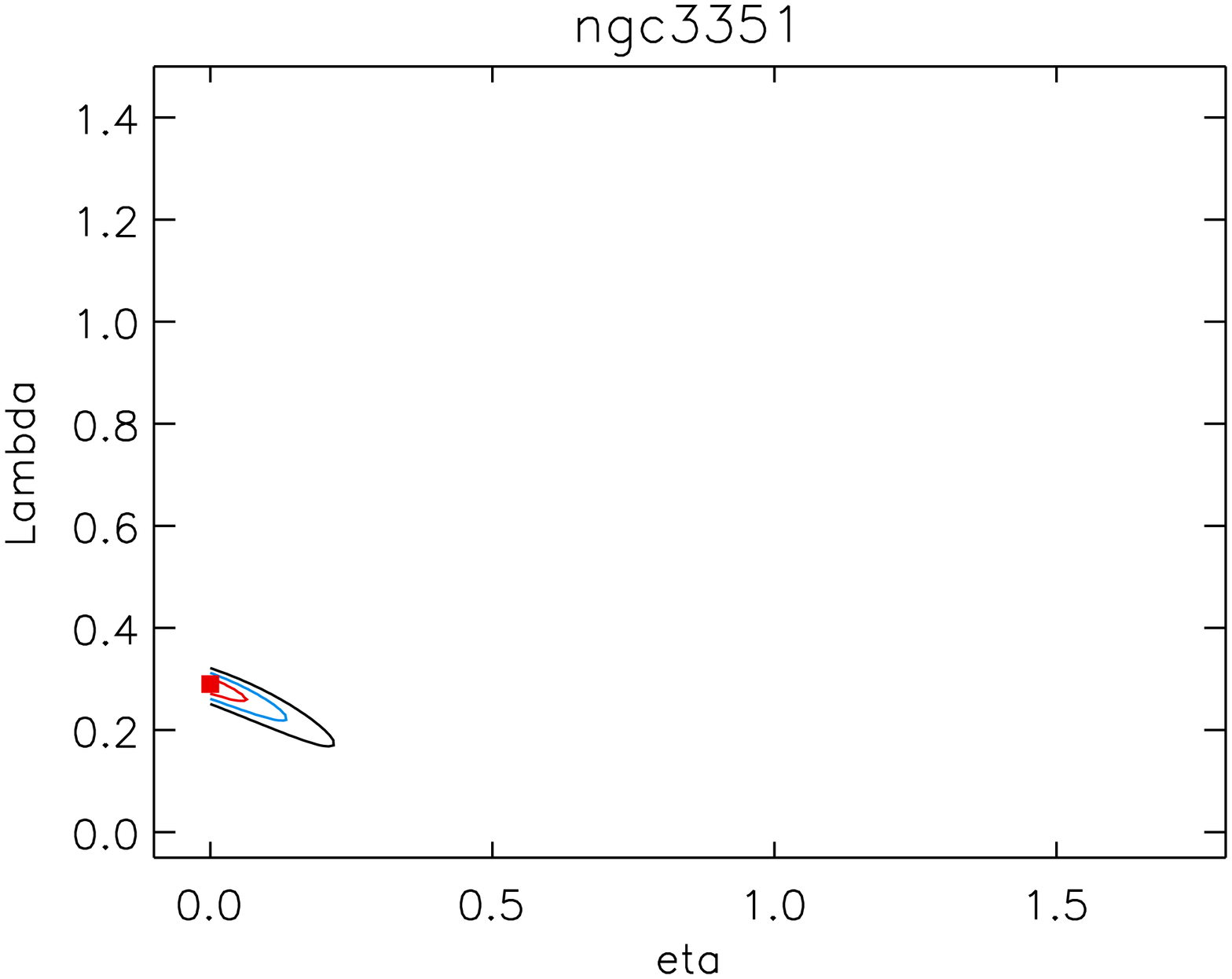}}\\
\subfloat{\includegraphics[width = 6.5cm,angle=90]{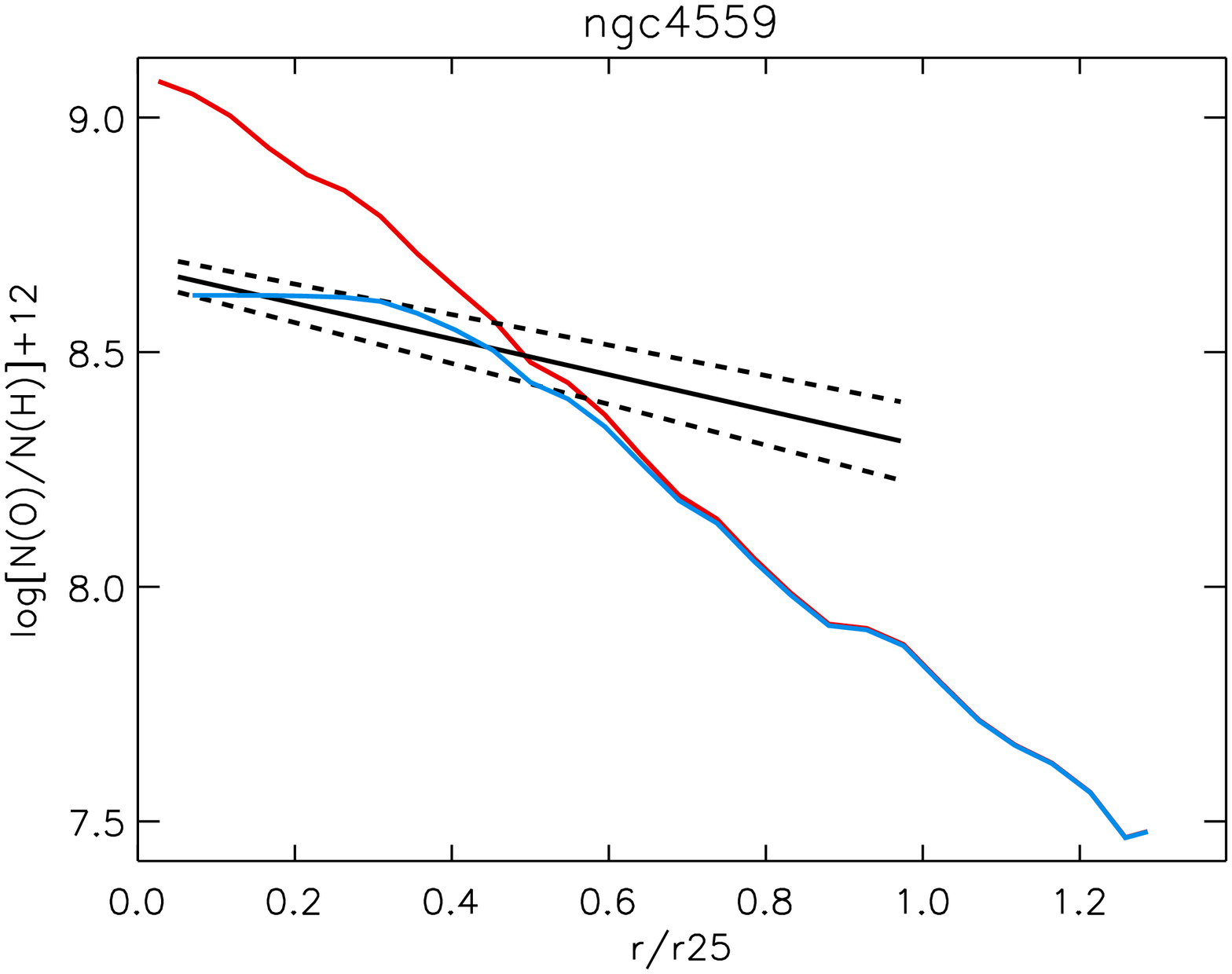}} \subfloat{\includegraphics[width = 6.5cm,angle=90]{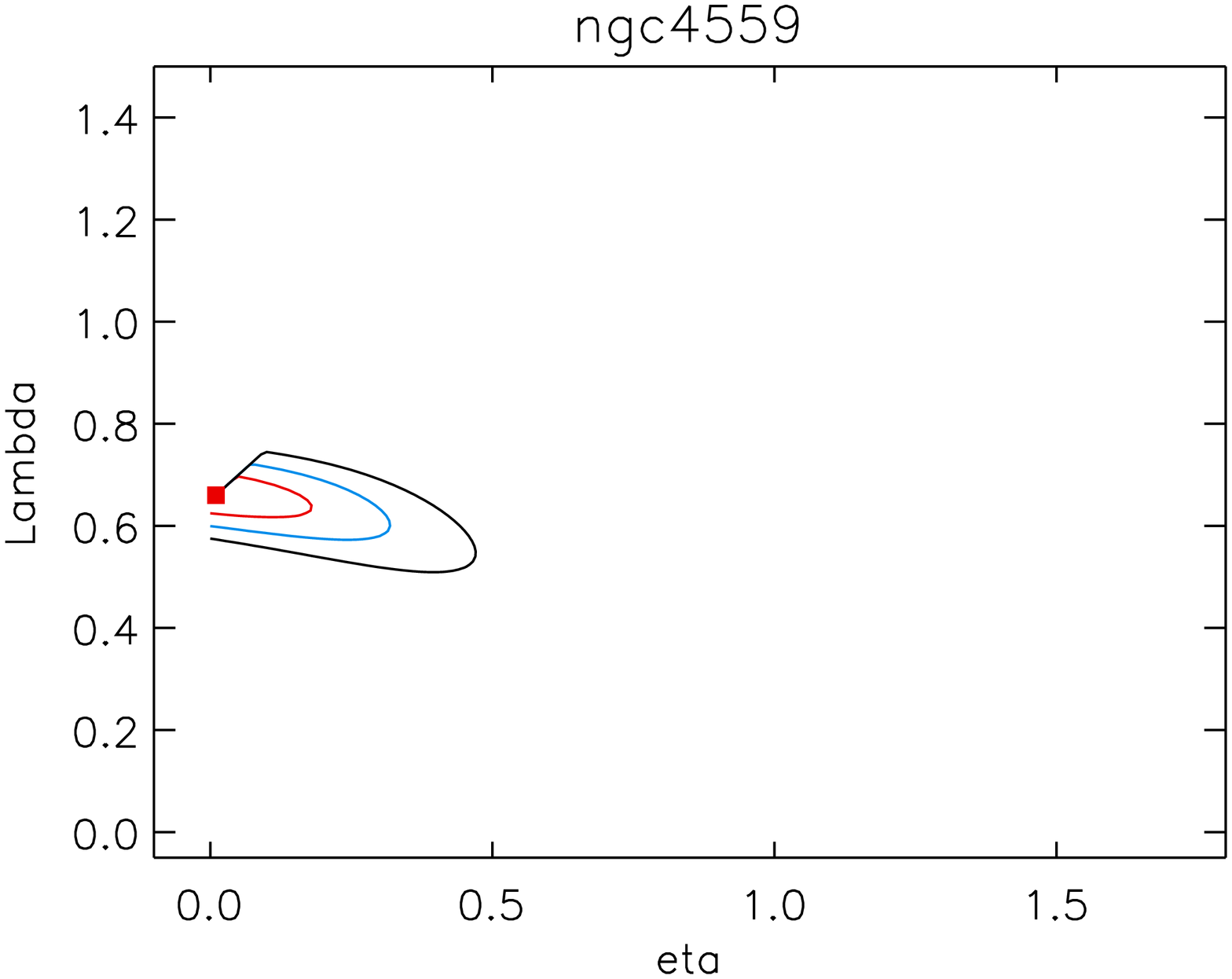}}\\
\caption{Chemical evolution fits of the galaxies NGC\,3184 (top), 3351 (middle), 4559 (bottom).}\label{figure9}
\end{figure*}

\floatplacement{figure}{!t}
\begin{figure*}
\centering
\subfloat{\includegraphics[width = 6.5cm,angle=90]{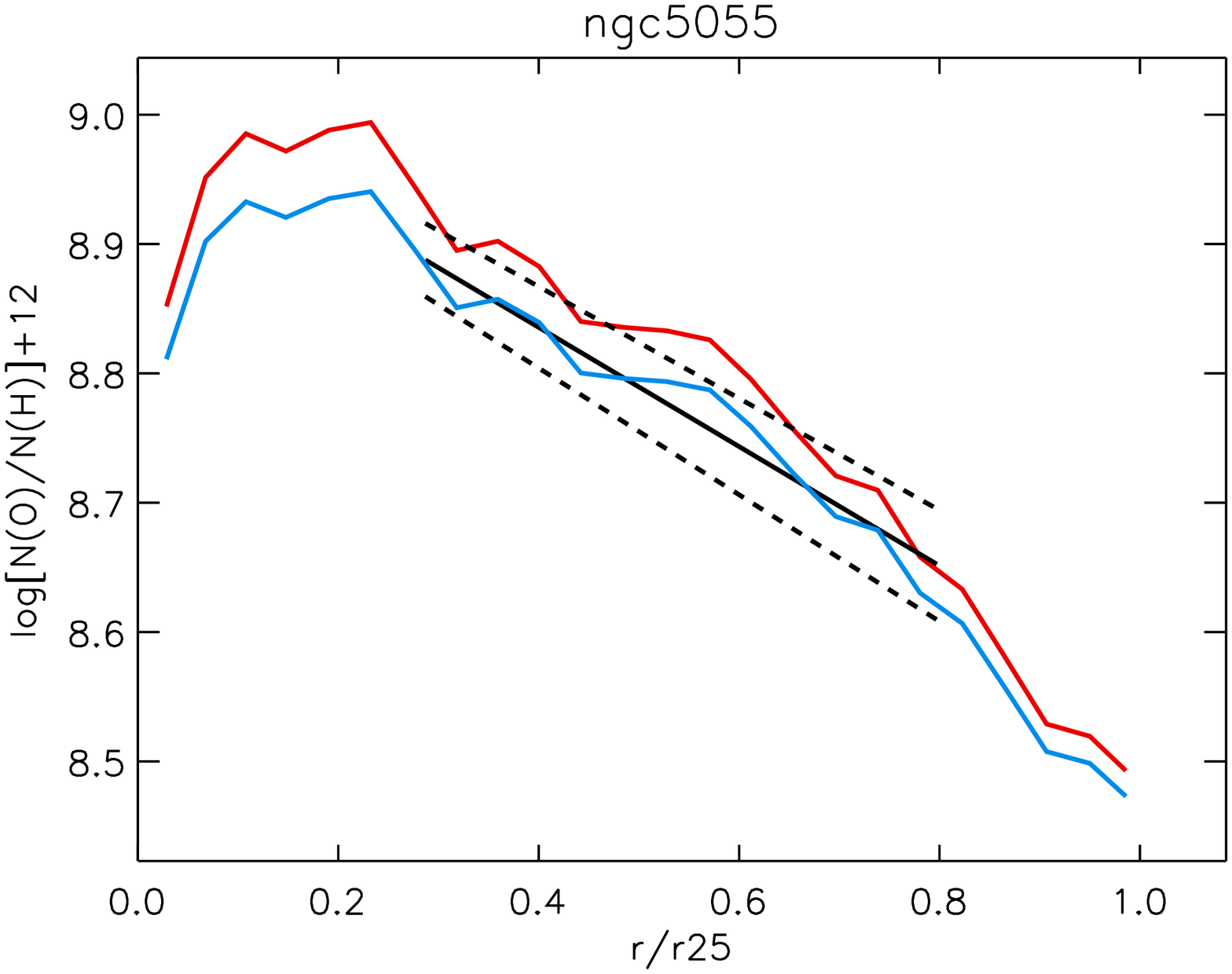}} \subfloat{\includegraphics[width = 6.5cm,angle=90]{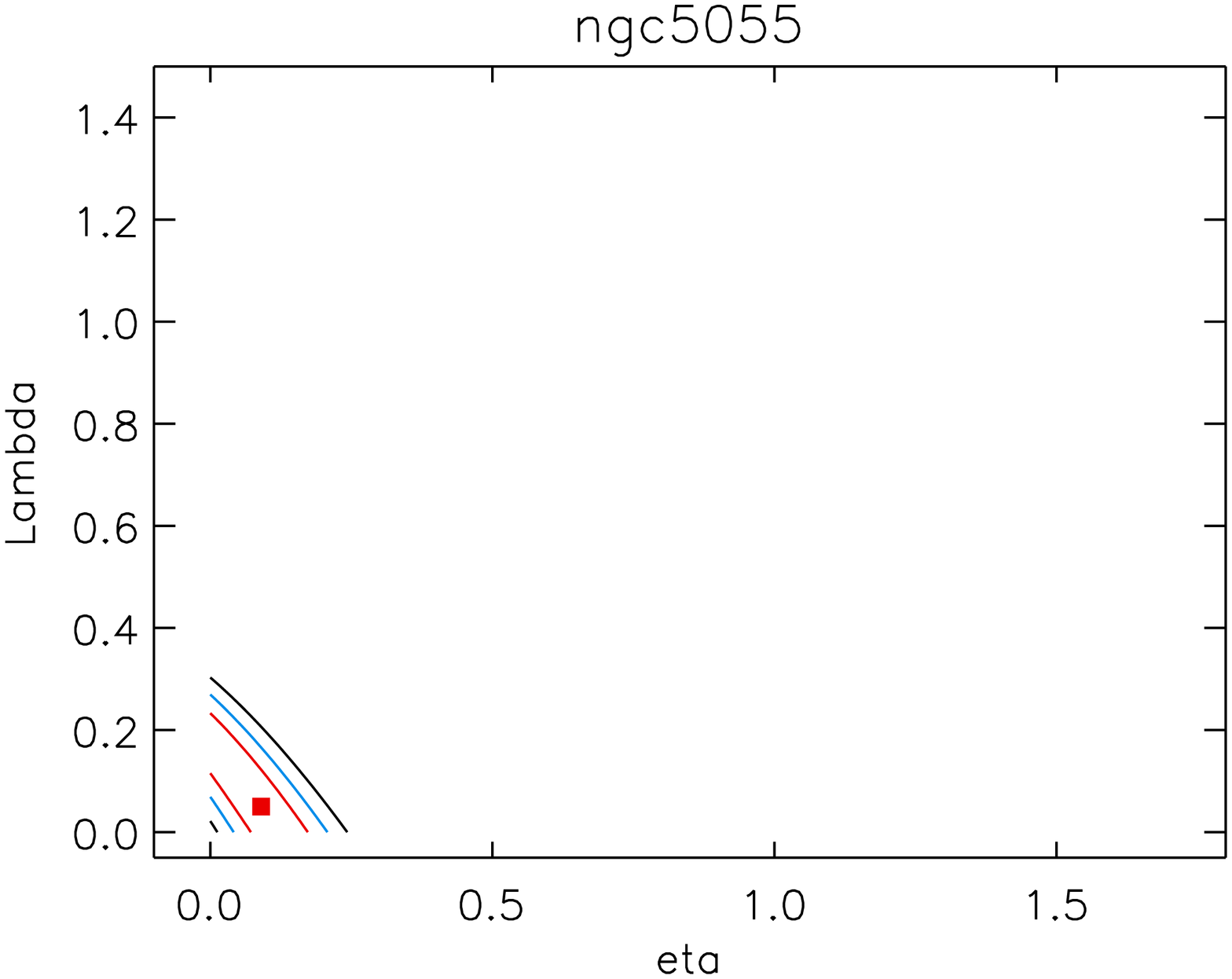}}\\
\subfloat{\includegraphics[width = 6.5cm,angle=90]{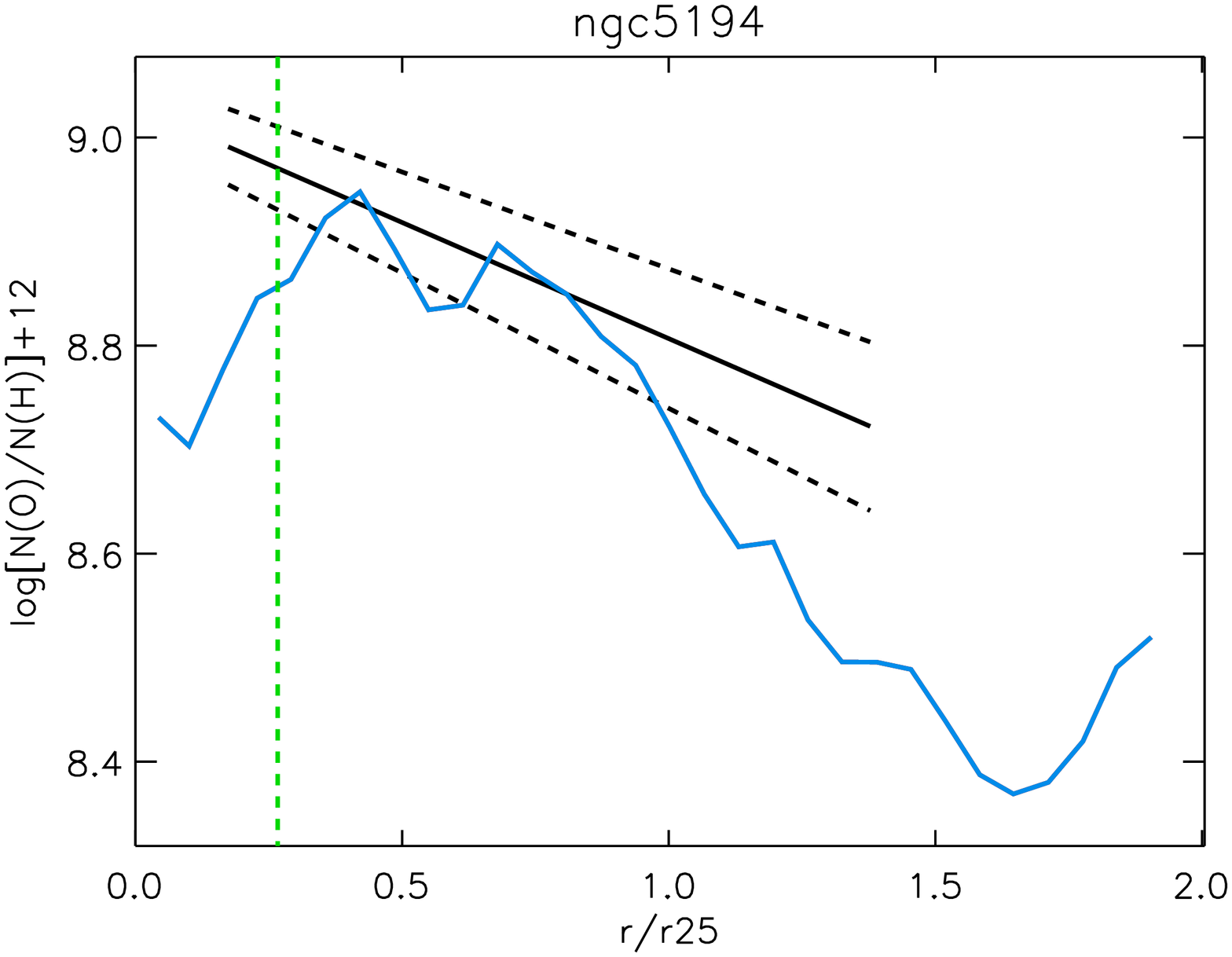}} \subfloat{\includegraphics[width = 6.5cm,angle=90]{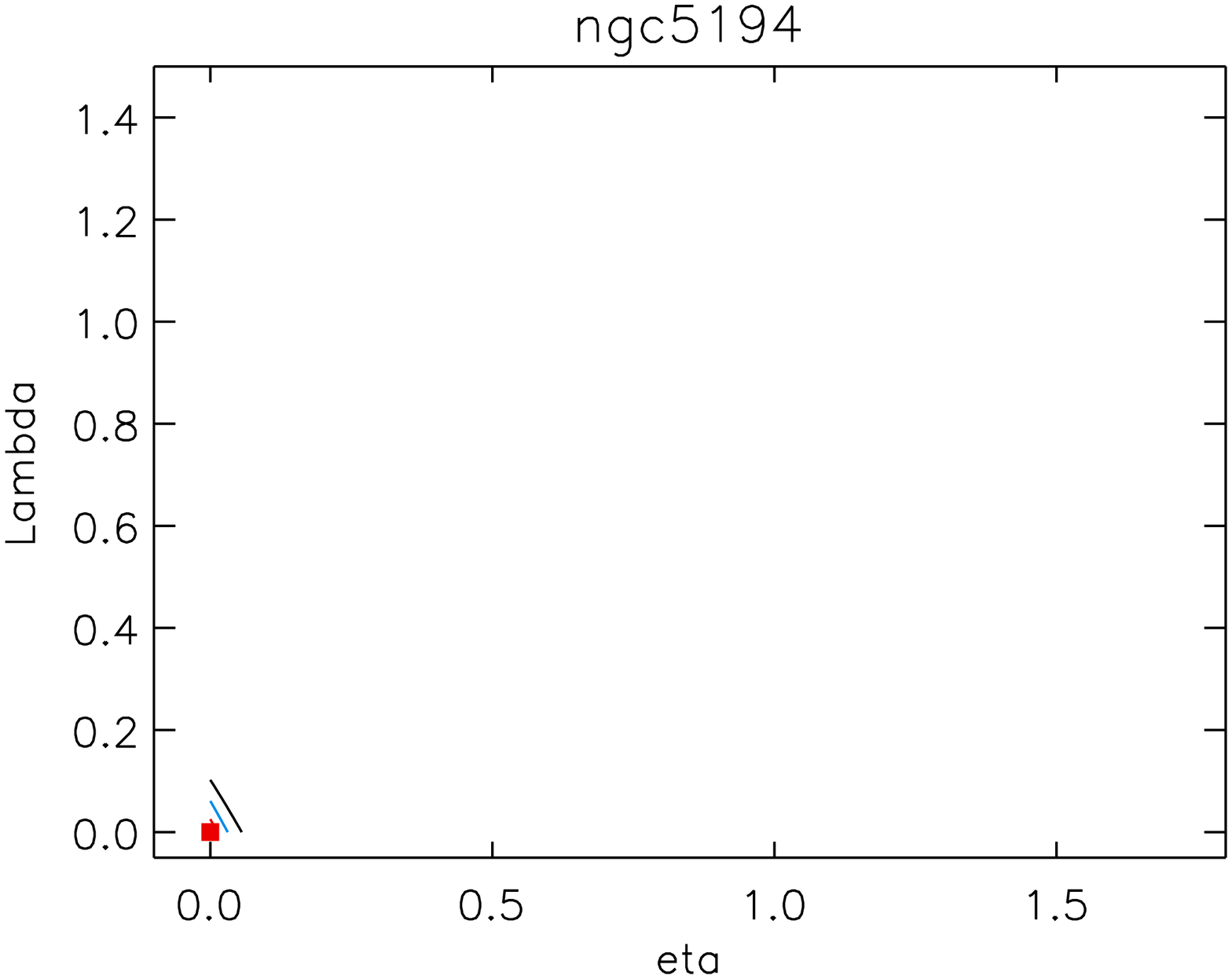}}\\
\caption{Chemical evolution fits of the galaxies NGC\,5055 (top), and 5194 (bottom).}\label{figure10}
\end{figure*}

\section{Analysis of the observed metallicity distribution of the SL-sample of galaxies}\label{sec:analysis-of-the-observed-metallicity-distribution}

With the calibration of the stellar yields obtained in the last section, we can now start to analyse the observed radial metallicity distribution of our LS-sample of galaxies. We use the observed ratios of stellar to gas mass column densities as a function of galactocentric radius and calculate model oxygen abundance profiles ${\rm (O/H)_{mod}}(r)(\eta_i,\Lambda_j)$ with $0\leq \eta_i \leq 4$ and  $0\leq \Lambda_j \leq 4$. For the calculation of ${\rm(O/H)_{mod}}(r)$, we generally use Equation~\ref{z1} except in those special cases where Equations~\ref{z3},~\ref{z5},~\ref{zcb} apply. The difference of ${\rm (O/H)_{mod}}(r)$ to the observed distribution ${\rm (O/H)_{obs}}(r)$ at discrete radial galactocentric distance points is then used to calculate a $\chi^2$-matrix $\chi^2(\eta_i,\Lambda_j)$. The minimum of this matrix then defines the best fit of the observed distribution ${\rm (O/H)_{obs}}(r)$ and provides an estimate of galactic mass-loss $\eta$ and accretion mass-gain $\Lambda$ in units of the star formation rate.

We begin the analysis with NGC\,2403, which has smooth, well observed profiles of H\textsc{i}, H$_2$ and stellar mass, and also a good coverage of metallicity measurements from H\textsc{ii} region emission lines. The result is shown in Fig.~\ref{figure7} and demonstrates that our simple chemical evolution model can reproduce the observed metallicity distribution surprisingly well. We obtain well constrained values of galactic wind mass-loss and accretion gain, $\eta = 0.81$ and $\Lambda = 0.71$. 

The galaxies NGC\,925 and 3198 are similar cases with significant amounts of galactic wind mass loss and accretion gain (see Table~\ref{table:eta-lambda}). The fit of the observed radial metallicity distribution in the range where observations exist is reasonable (for all galaxies we plot the observed metallicities exactly in the range for which \citealt{pilyugin14a} have H\textsc{ii} measurements available). The increase in the model-predicted metallicity for NGC\,925 beyond $r/R_{25} \geq 0.9$ is caused by a steep decline of H\textsc{i} gas mass column density.

Contrary to the three galaxies just discussed, we also encounter cases with either very small accretion mass-gain or galactic wind mass-loss. For instance, NGC\,5457 can be fit very well by a model with $\Lambda = 0$, as shown in Fig.~\ref{figure8}. NGC\,628 is a case with $\eta = 0.17$. Three more galaxies with very small values of $\eta$, NGC\,3184, 3351, and 4559, are displayed in Fig.\ref{figure9}. For three of the four cases with low galactic wind mass-loss rate, NGC\,628, 3184, and 4559, the model fails to describe the observed radial metallicity distribution for galactocentric radii $r/R_{25} \geq 0.8, 0.7$, and 0.6, respectively. This is caused by extended and relatively flat distributions of the ISM H\textsc{i} gas. The situation is most extreme for NGC\,4559. \citet{bresolin12} and \citet{kud14} have discussed other even more extreme cases with outer extended disks imbedded in large ISM H\textsc{i} reservoirs as examples where the simple models used here become insufficient. Our analytical models for these three galaxies start to fail when the stellar mass column density becomes smaller than the gas column density.

The galaxies NGC\,5055 and 5194 shown in Fig.\ref{figure10} are very close to the closed-box case with very small $\eta$ and $\Lambda$ values (we note that the model distribution for NGC\,5194 shows a too strong decline at larger radii similar to the cases discussed before). 

\begin{figure}
\includegraphics[width=6.5cm,angle=90]{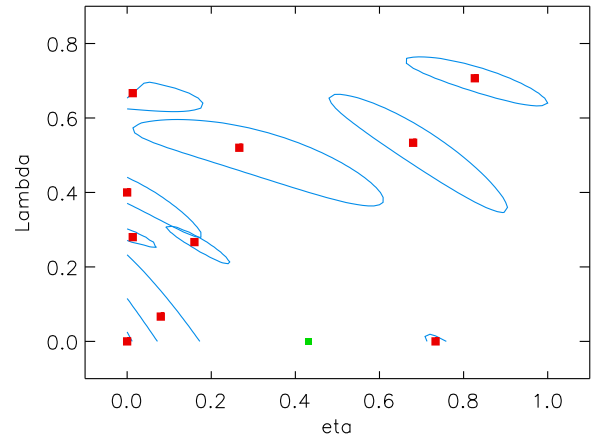}
\caption{The location of all galaxies discussed in Fig.~\ref{figure7} to~\ref{figure10} in the $(\eta,\Lambda)$-plane. The green square gives the position of the Milky Way disk
as discussed in Section 4.}\label{figure11}
\end{figure}

We summarise the results for all galaxies displayed in Fig.~\ref{figure7} to~\ref{figure10} in Fig.~\ref{figure11}. The intriguing result of this figure is the indication of the existence of three different groups in our sample: 1) galaxies with either mostly winds and almost no accretion, 2) mostly accretion and weak winds, and 3) galaxies with significant amounts of both winds and accretion, where the rates of mass-loss and mass-gain are of the same order. We also note that according to our fit the Milky Way belongs to the subgroup with a significant amount of galactic wind but with only a small accretion rate (see also Fig.~\ref{figure5}).

\begin{figure*}
\centering
\subfloat{\includegraphics[width = 6.5cm,angle=90]{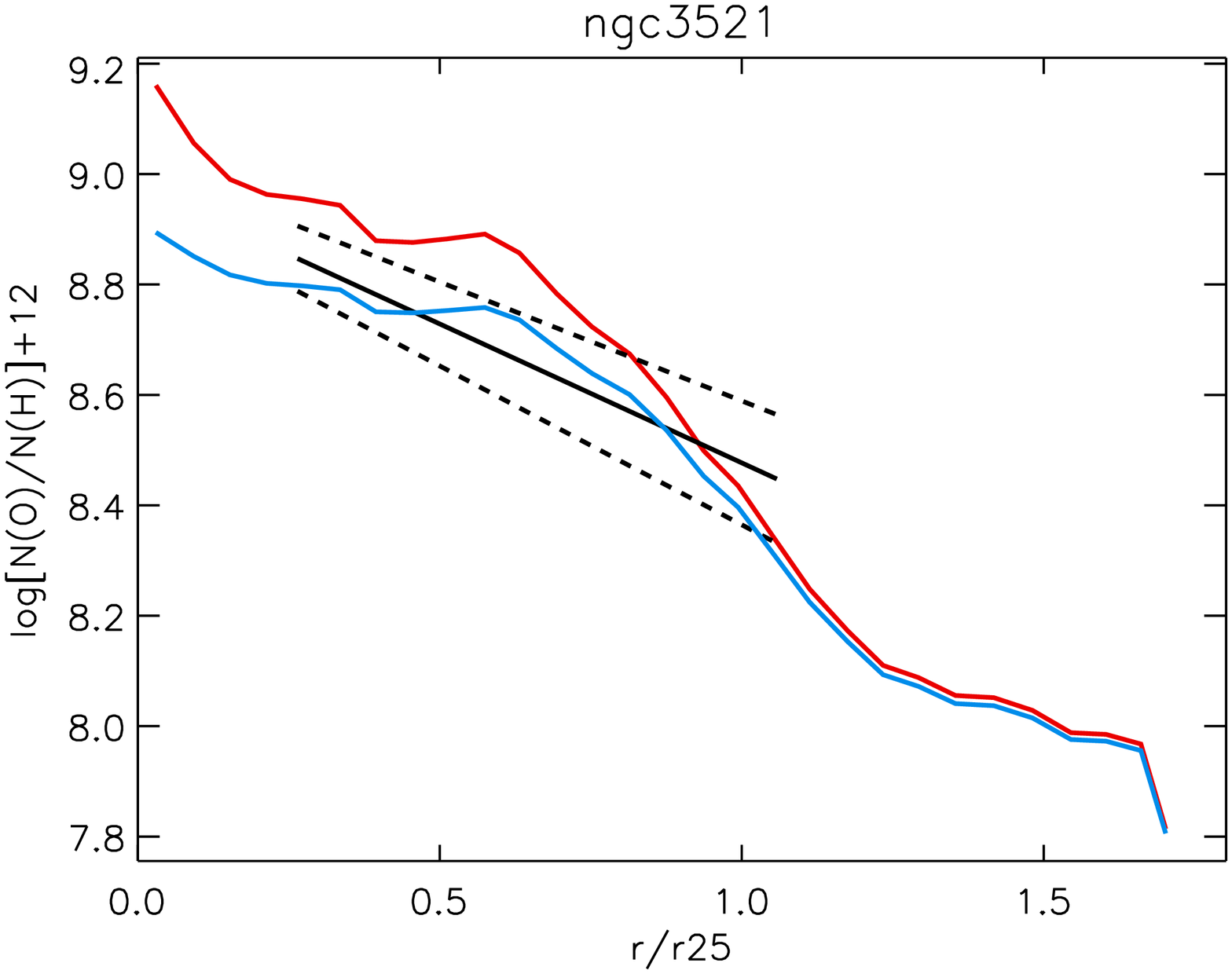}} \subfloat{\includegraphics[width = 6.5cm,angle=90]{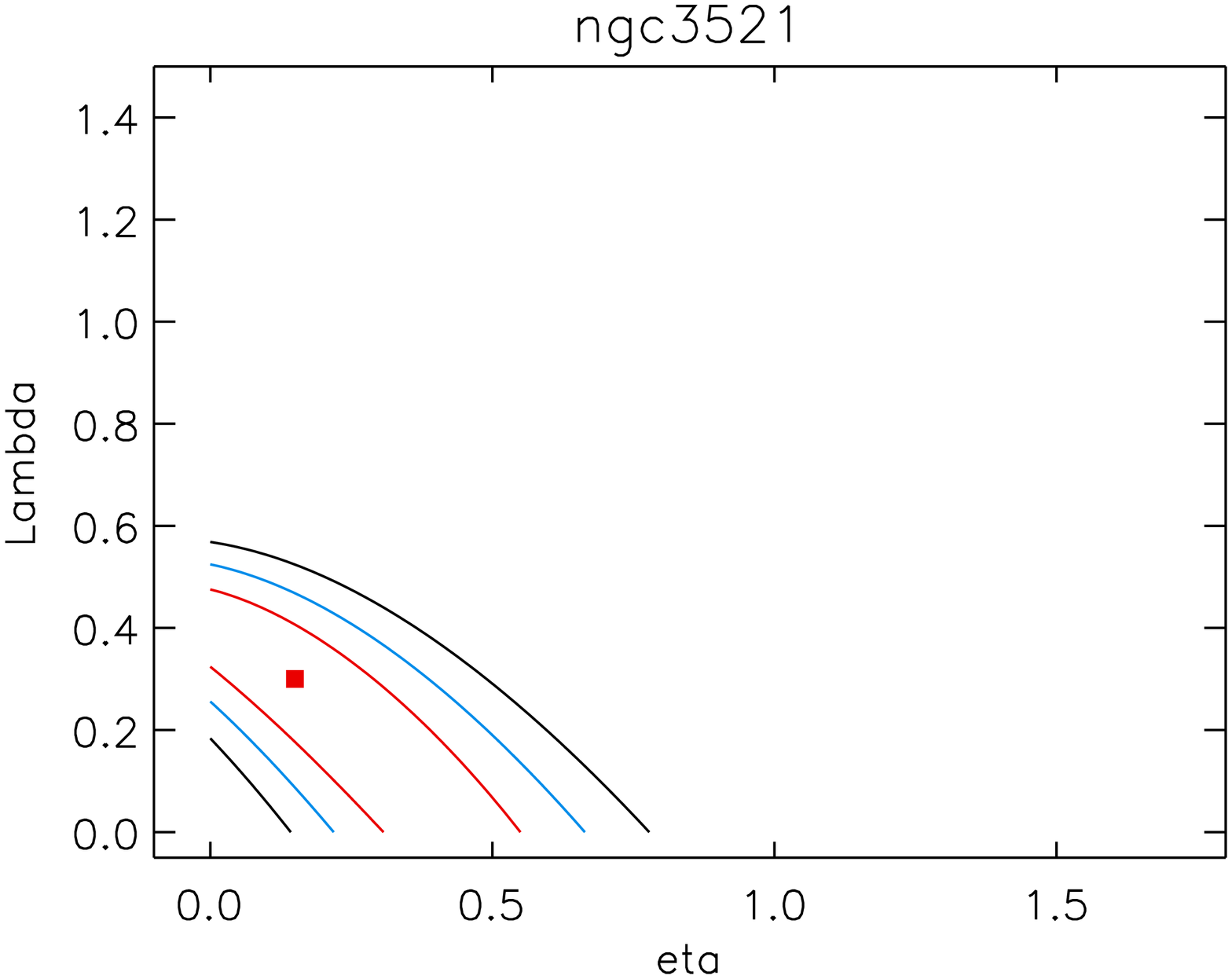}}\\
\subfloat{\includegraphics[width = 6.5cm,angle=90]{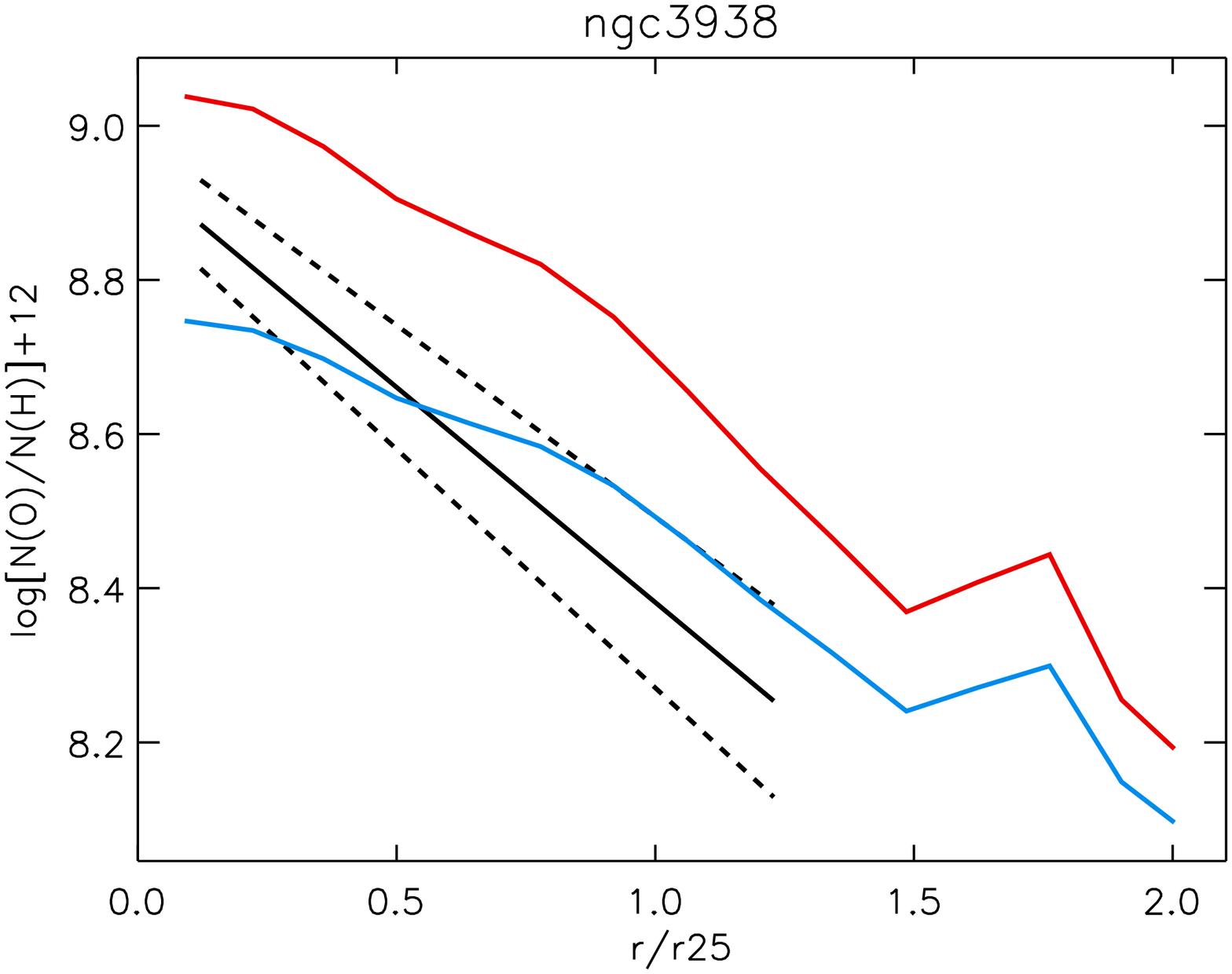}} \subfloat{\includegraphics[width = 6.5cm,angle=90]{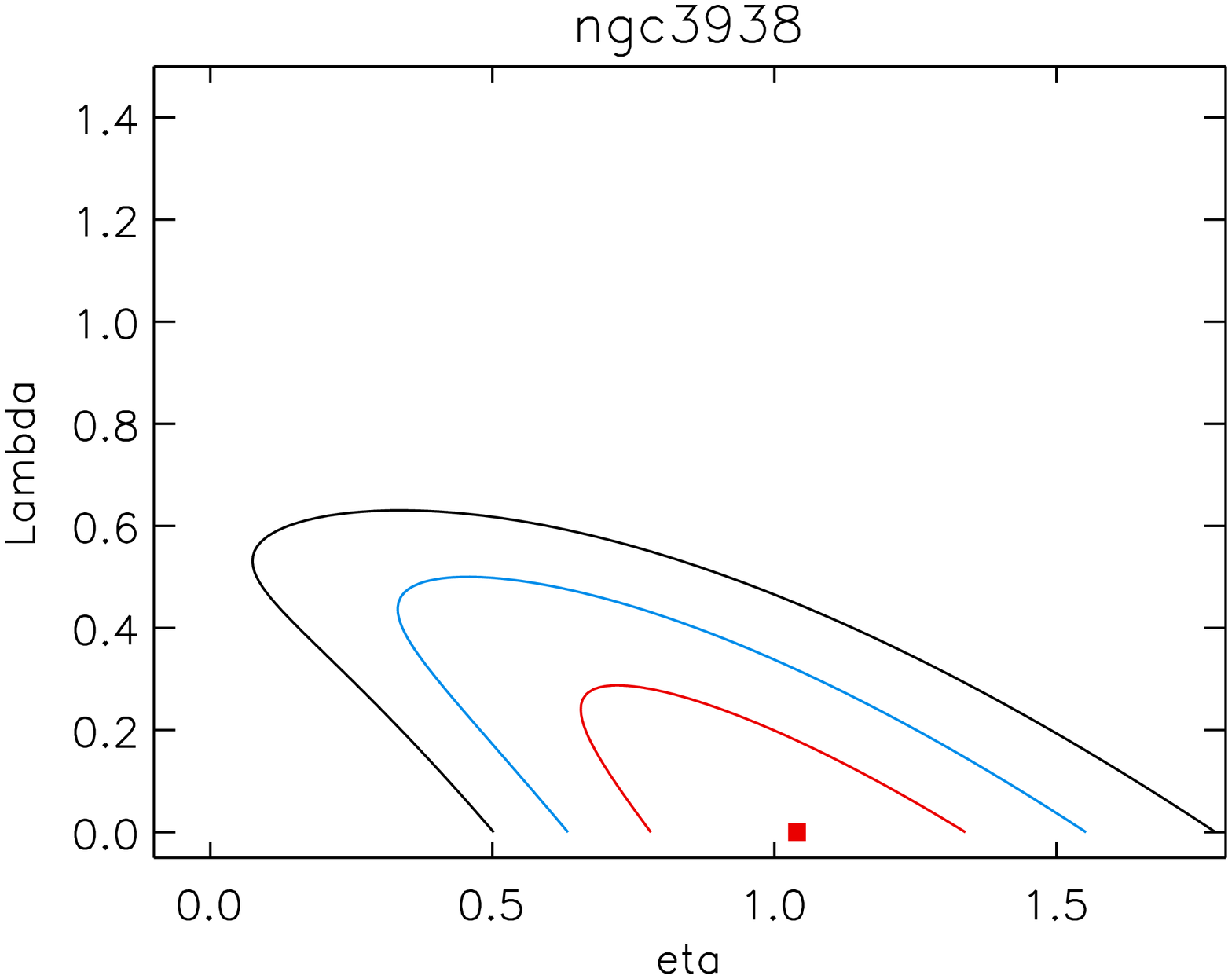}}\\
\subfloat{\includegraphics[width = 6.5cm,angle=90]{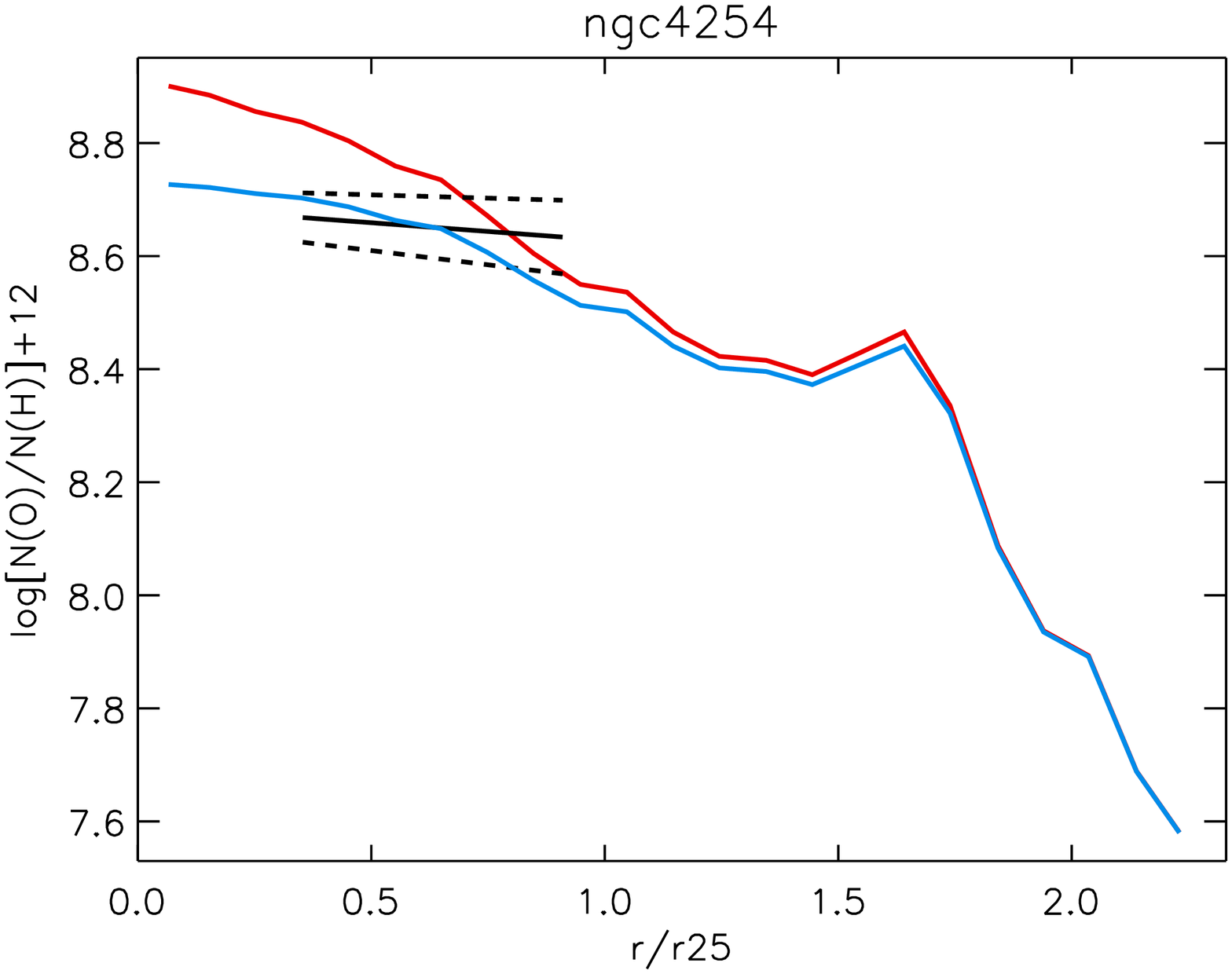}} \subfloat{\includegraphics[width = 6.5cm,angle=90]{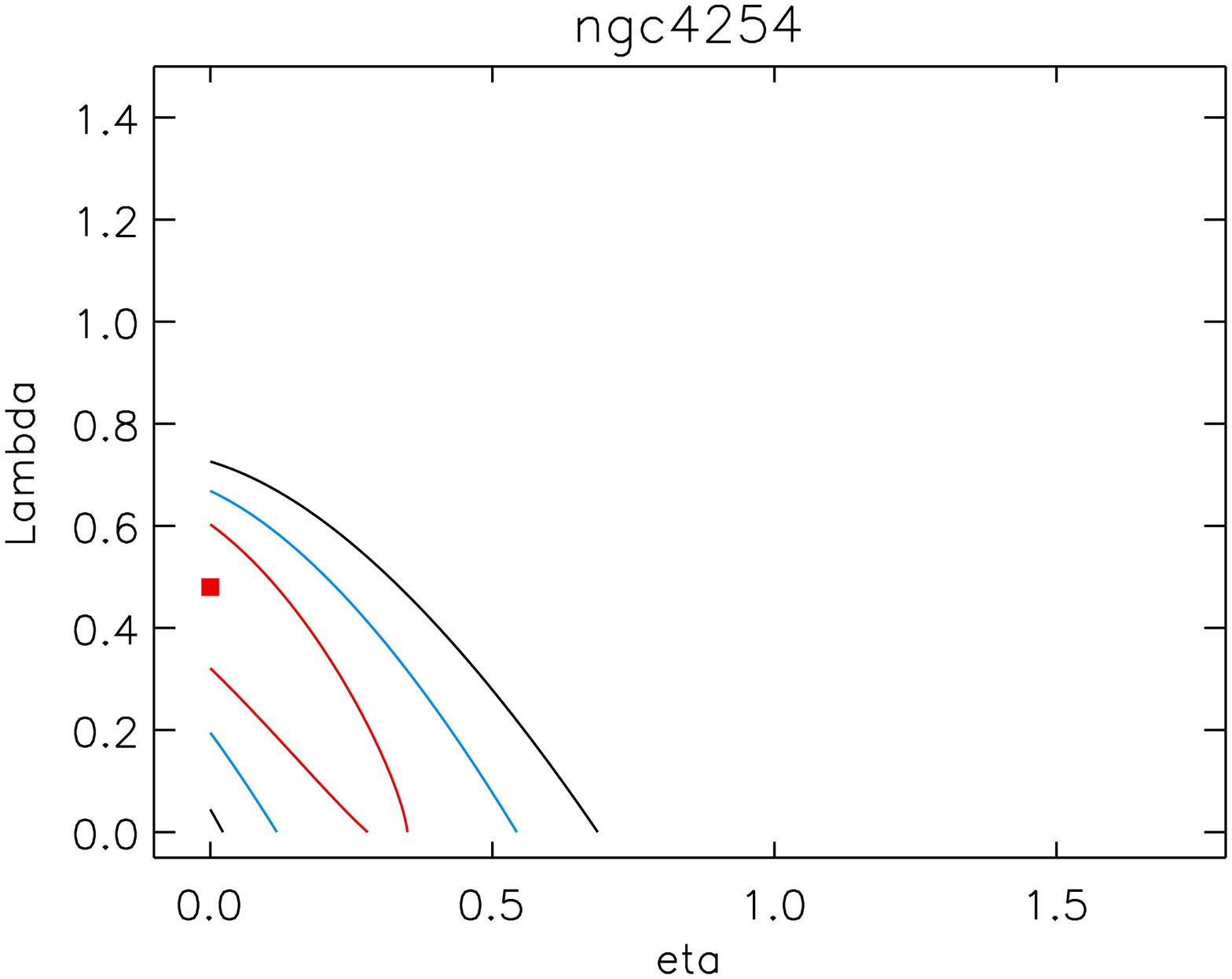}}\\
\caption{Chemical evolution fits of the galaxies NGC\,3521 (top), 3938 (middle), 4254 (bottom).}\label{figure12}
\end{figure*}

\begin{figure*}
\centering
\subfloat{\includegraphics[width = 6.5cm,angle=90]{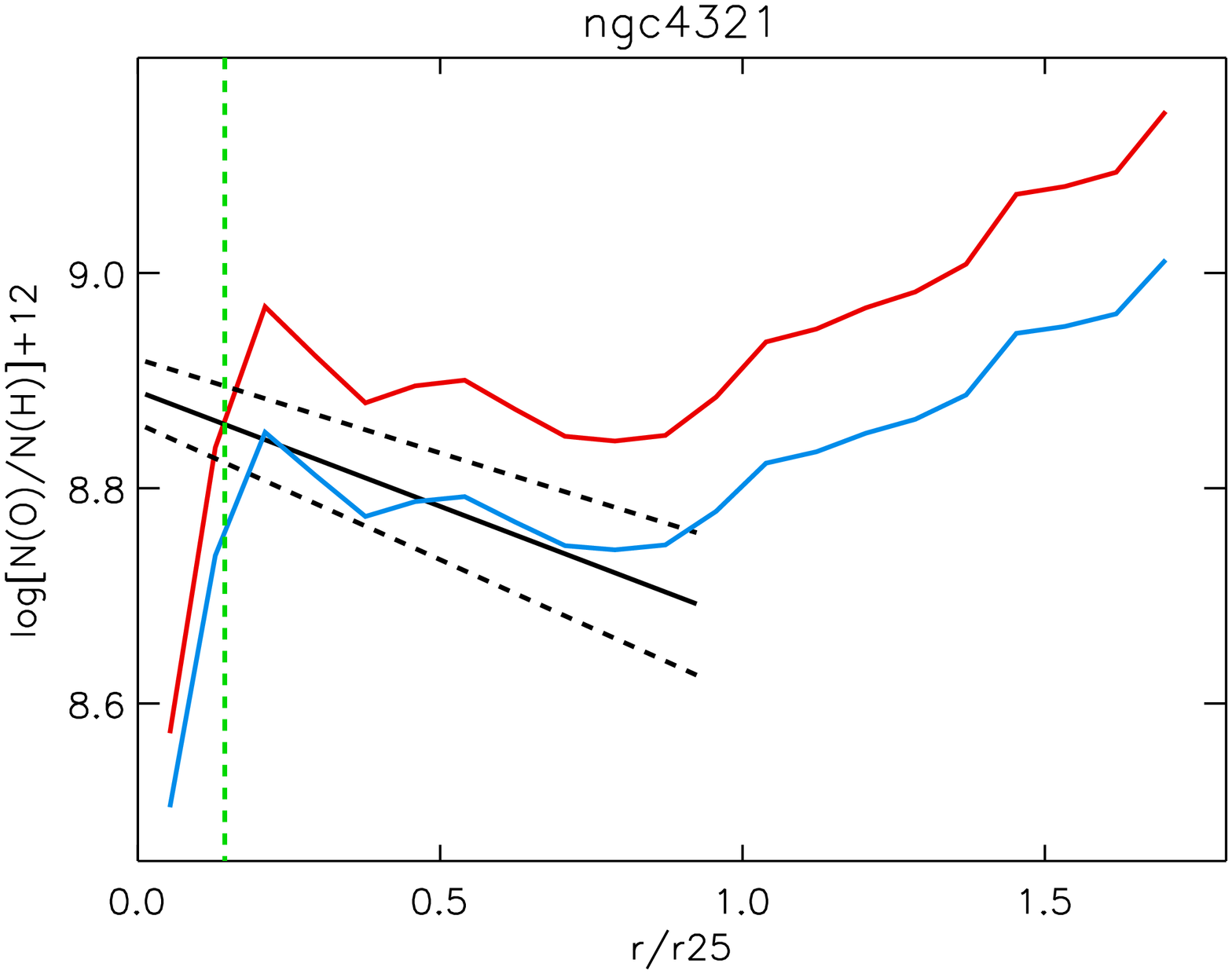}} \subfloat{\includegraphics[width = 6.5cm,angle=90]{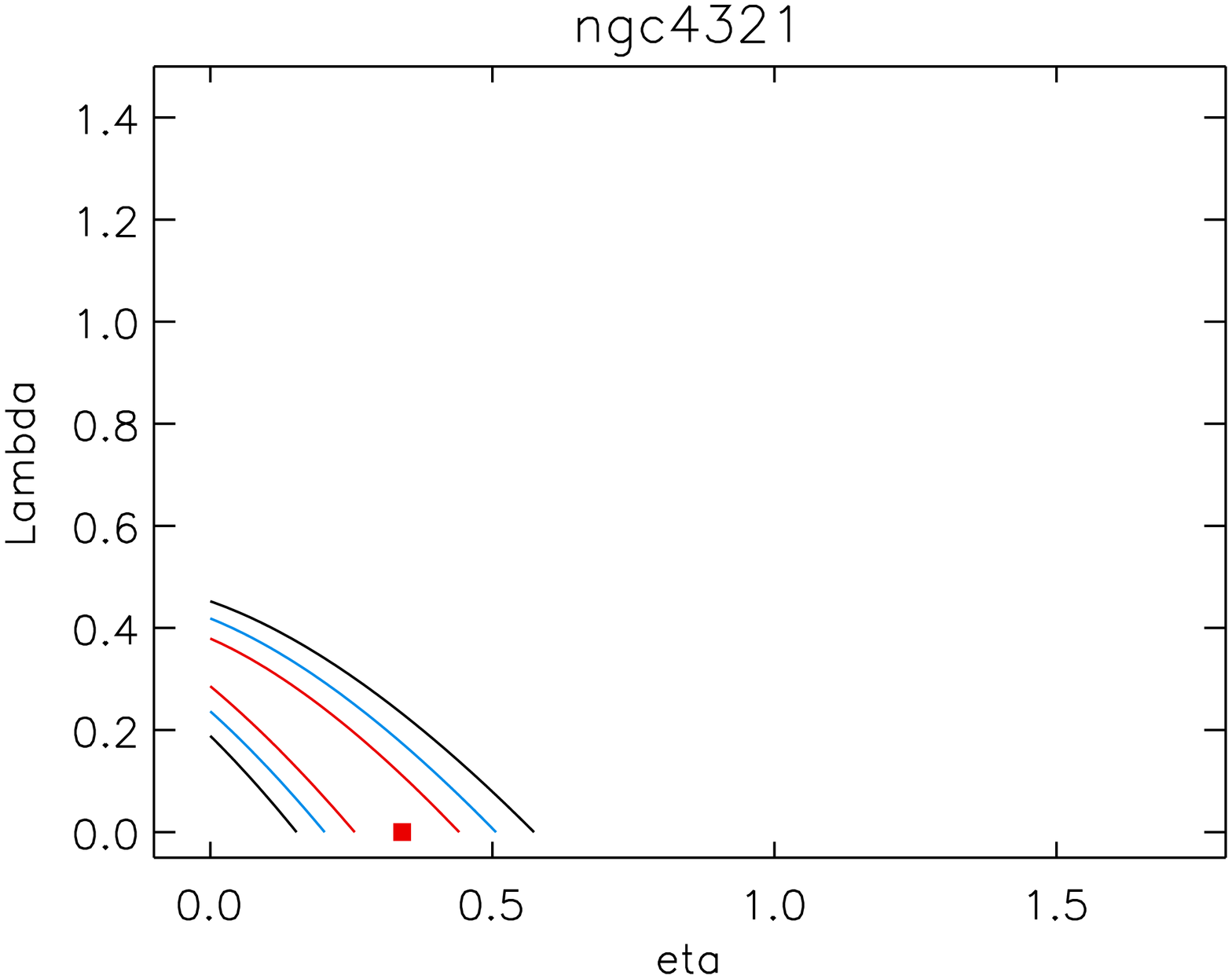}}\\
\subfloat{\includegraphics[width = 6.5cm,angle=90]{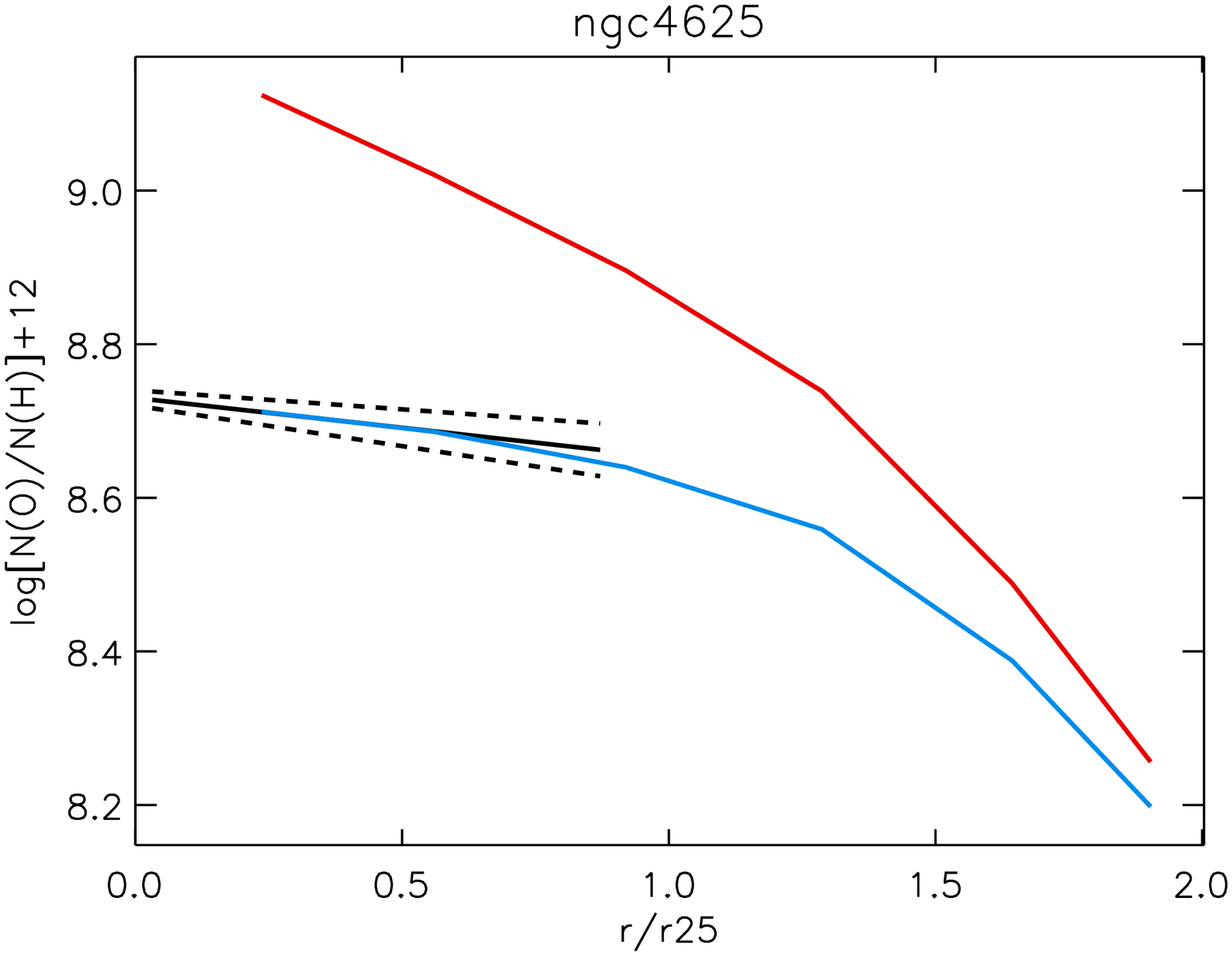}} \subfloat{\includegraphics[width = 6.5cm,angle=90]{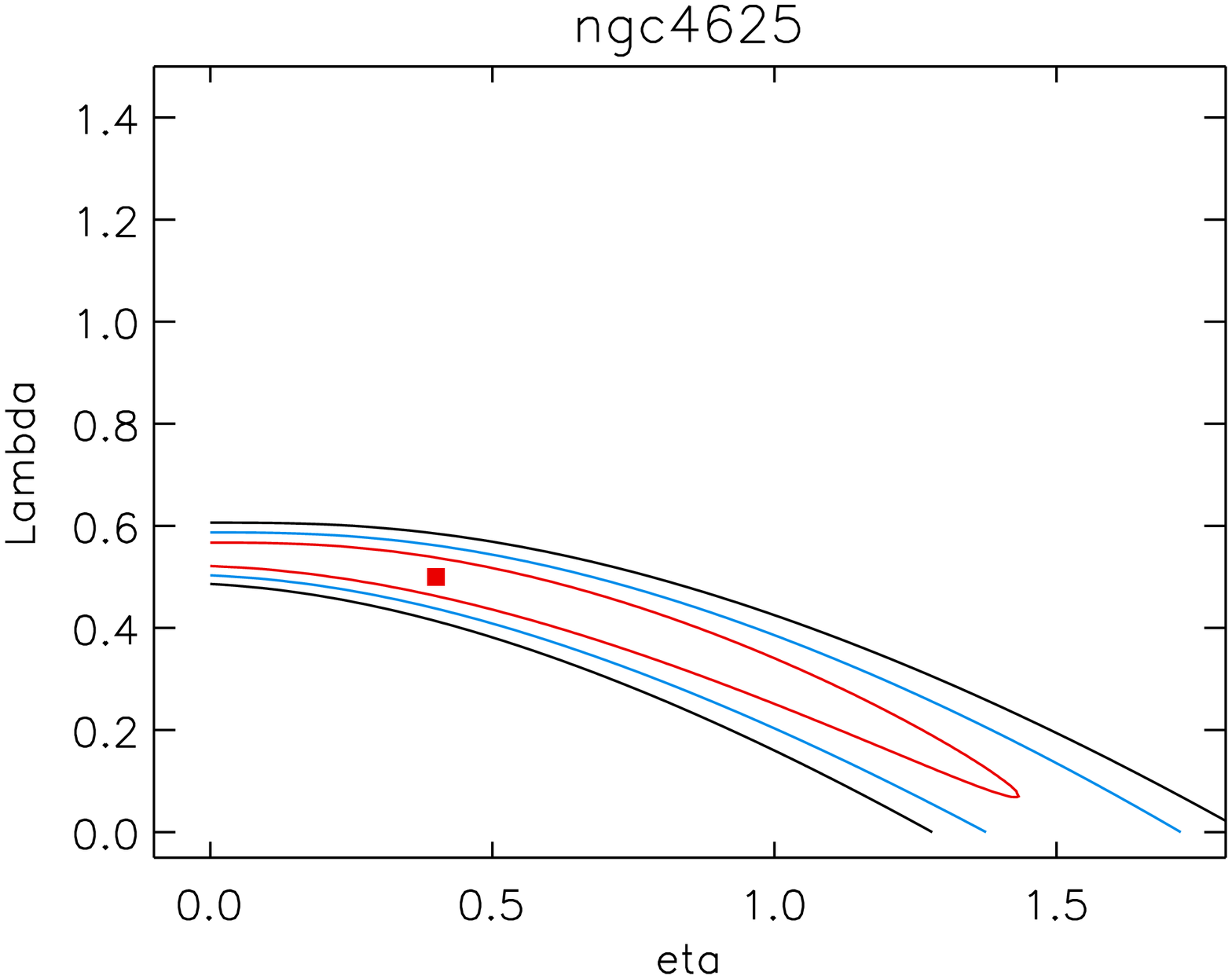}}\\
\subfloat{\includegraphics[width = 6.5cm,angle=90]{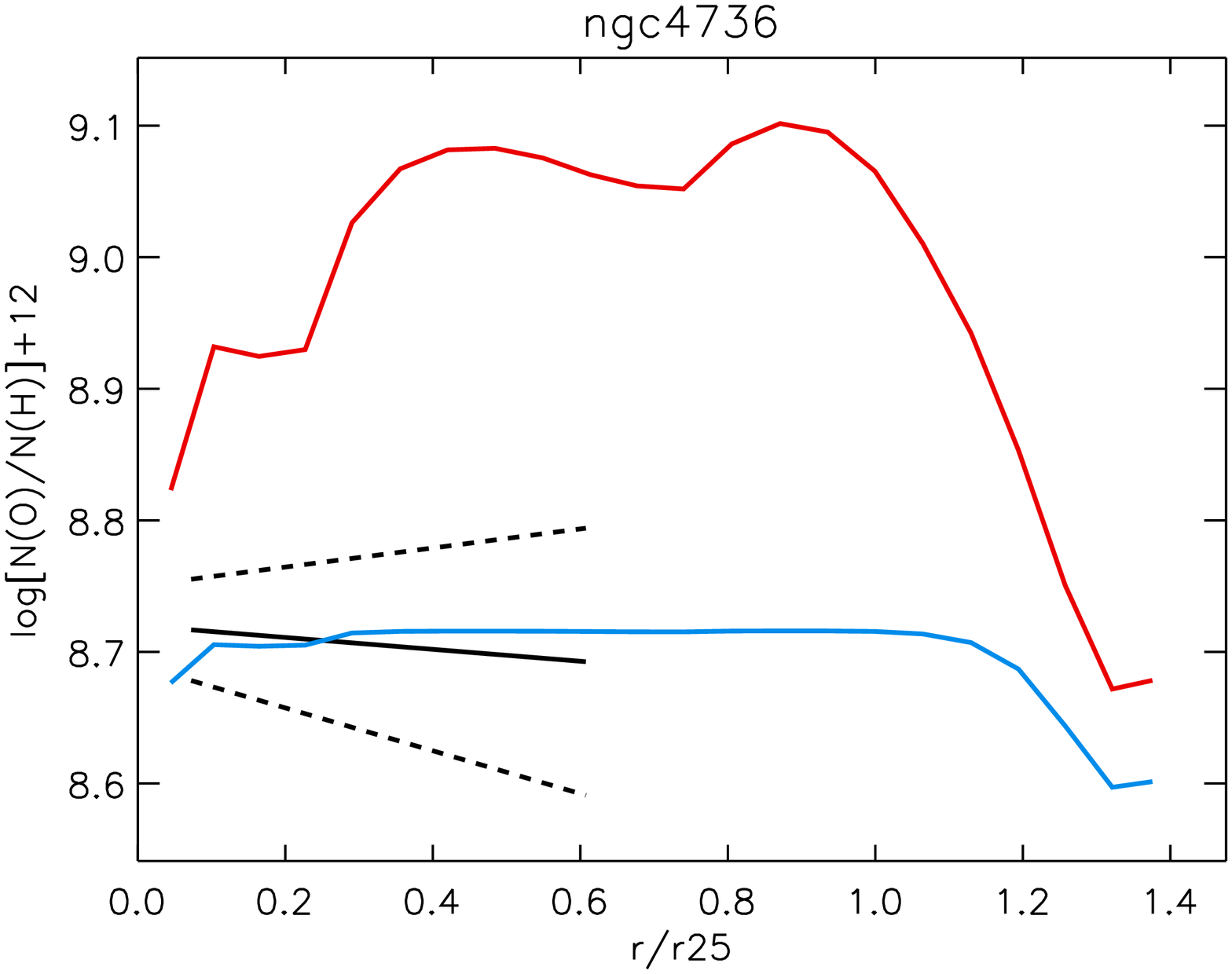}} \subfloat{\includegraphics[width = 6.5cm,angle=90]{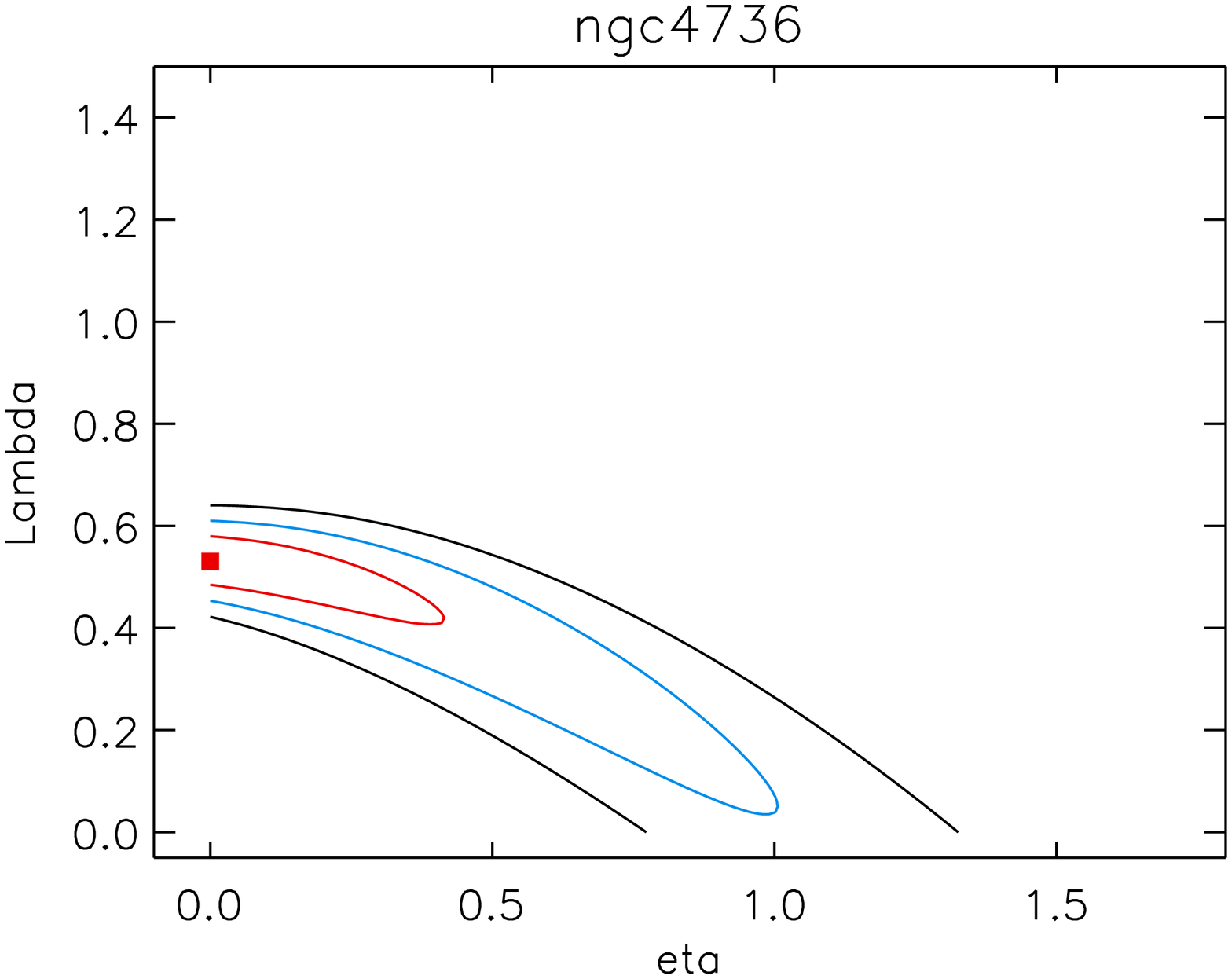}}\\
\caption{Chemical evolution fits of the galaxies NGC\,4321 (top), 4625 (middle), 4736 (bottom).}\label{figure13}
\end{figure*}

\begin{figure*}
\centering
\subfloat{\includegraphics[width = 6.5cm,angle=90]{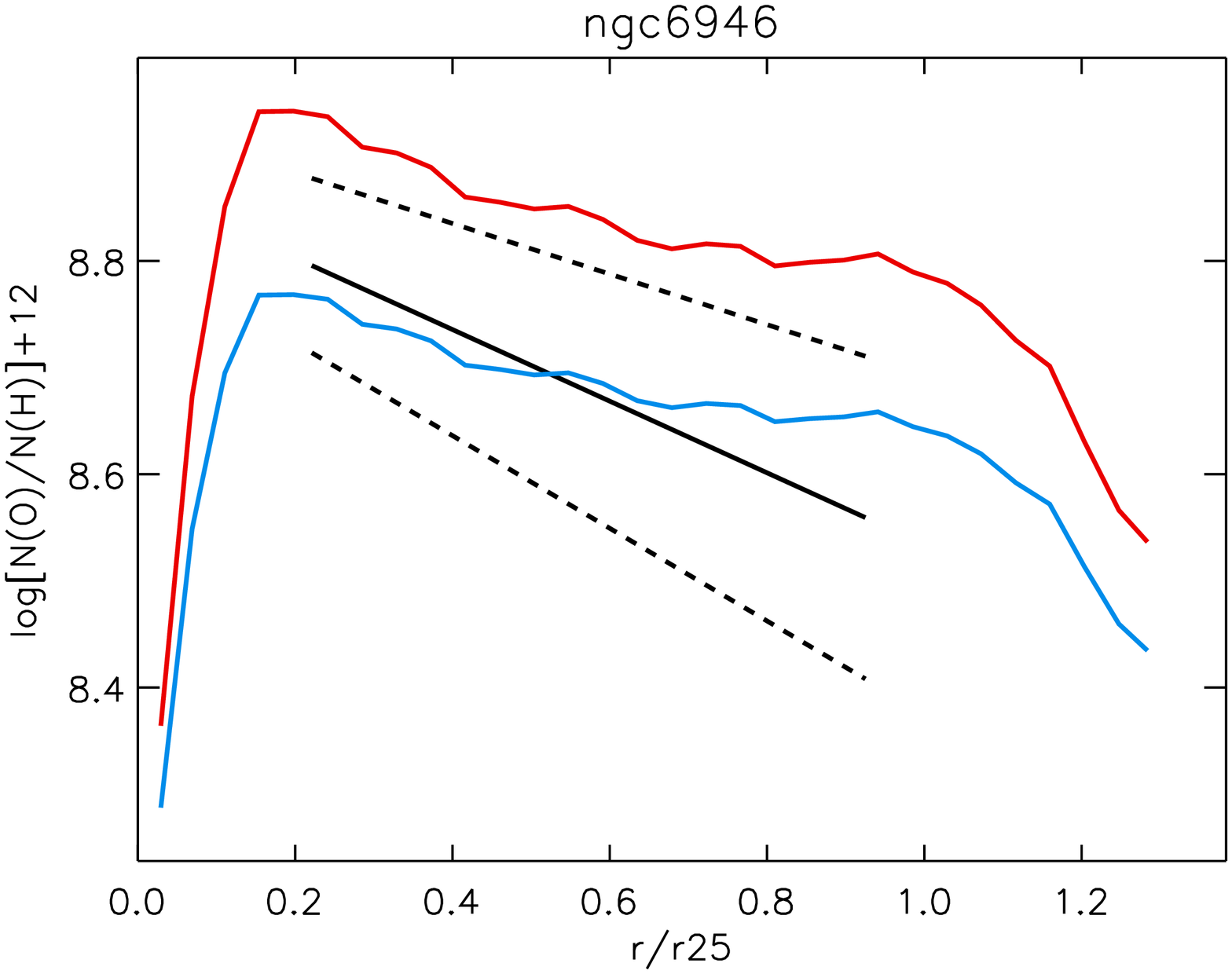}} \subfloat{\includegraphics[width = 6.5cm,angle=90]{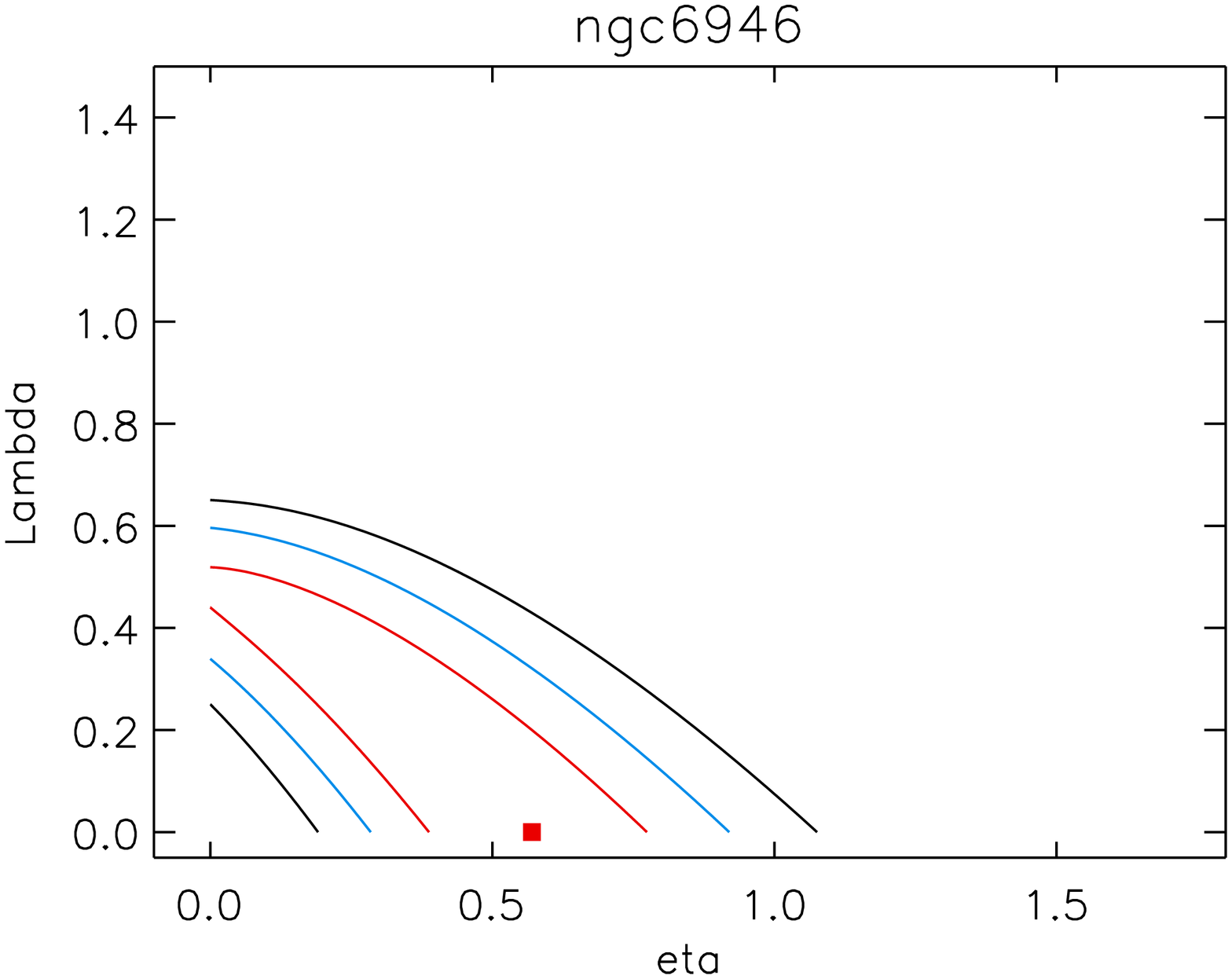}}\\
\subfloat{\includegraphics[width = 6.5cm,angle=90]{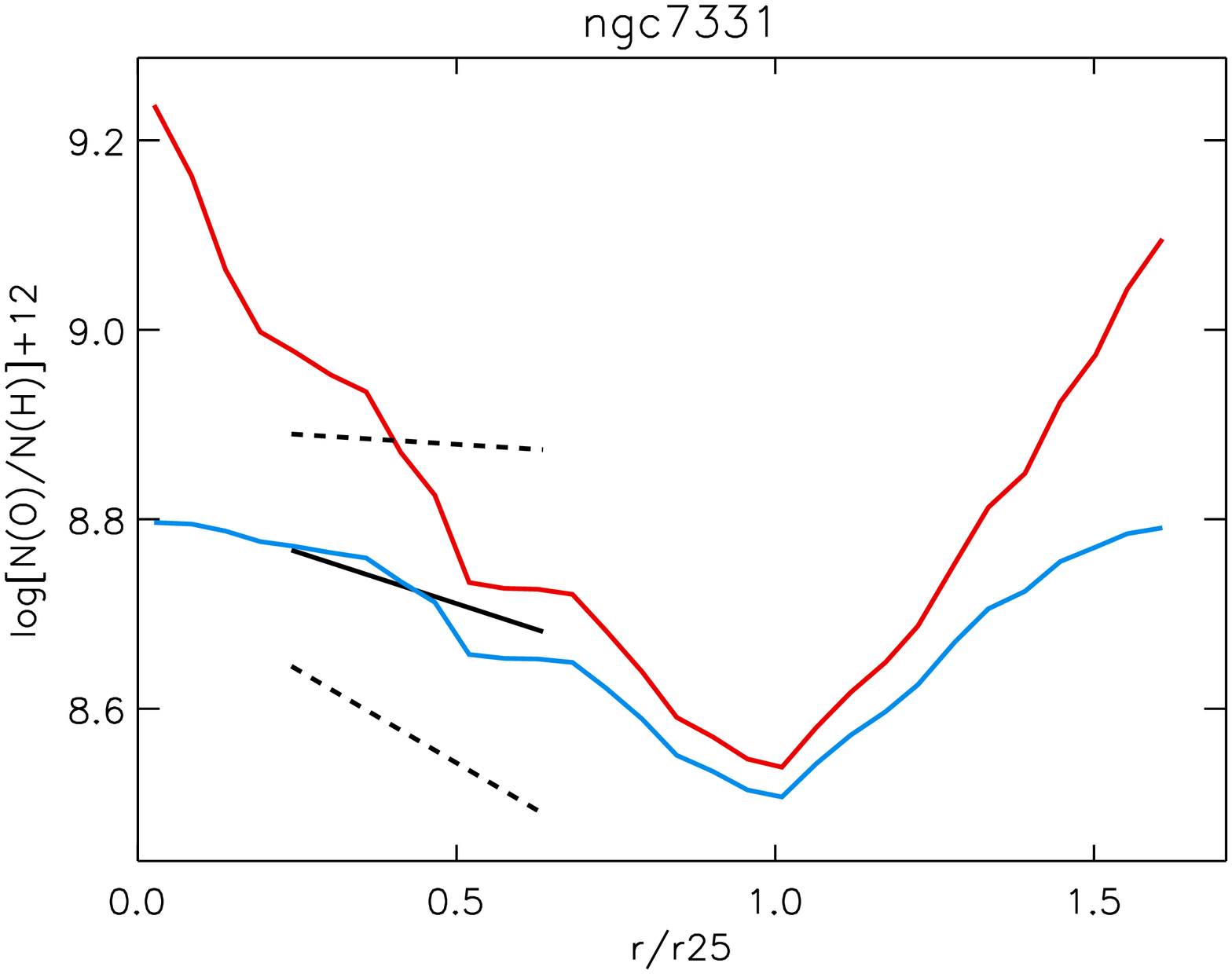}} \subfloat{\includegraphics[width = 6.5cm,angle=90]{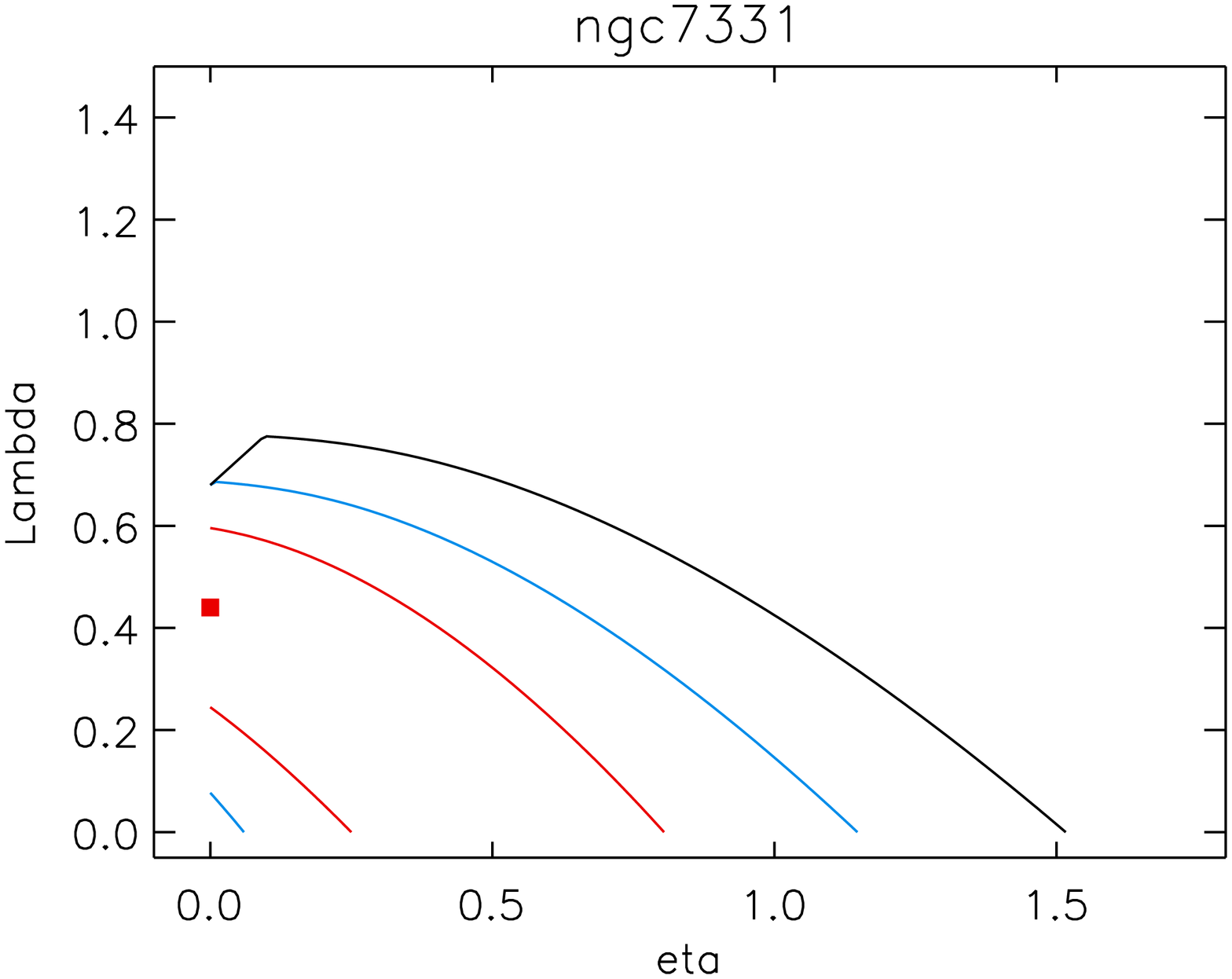}}\\
\subfloat{\includegraphics[width = 6.5cm,angle=90]{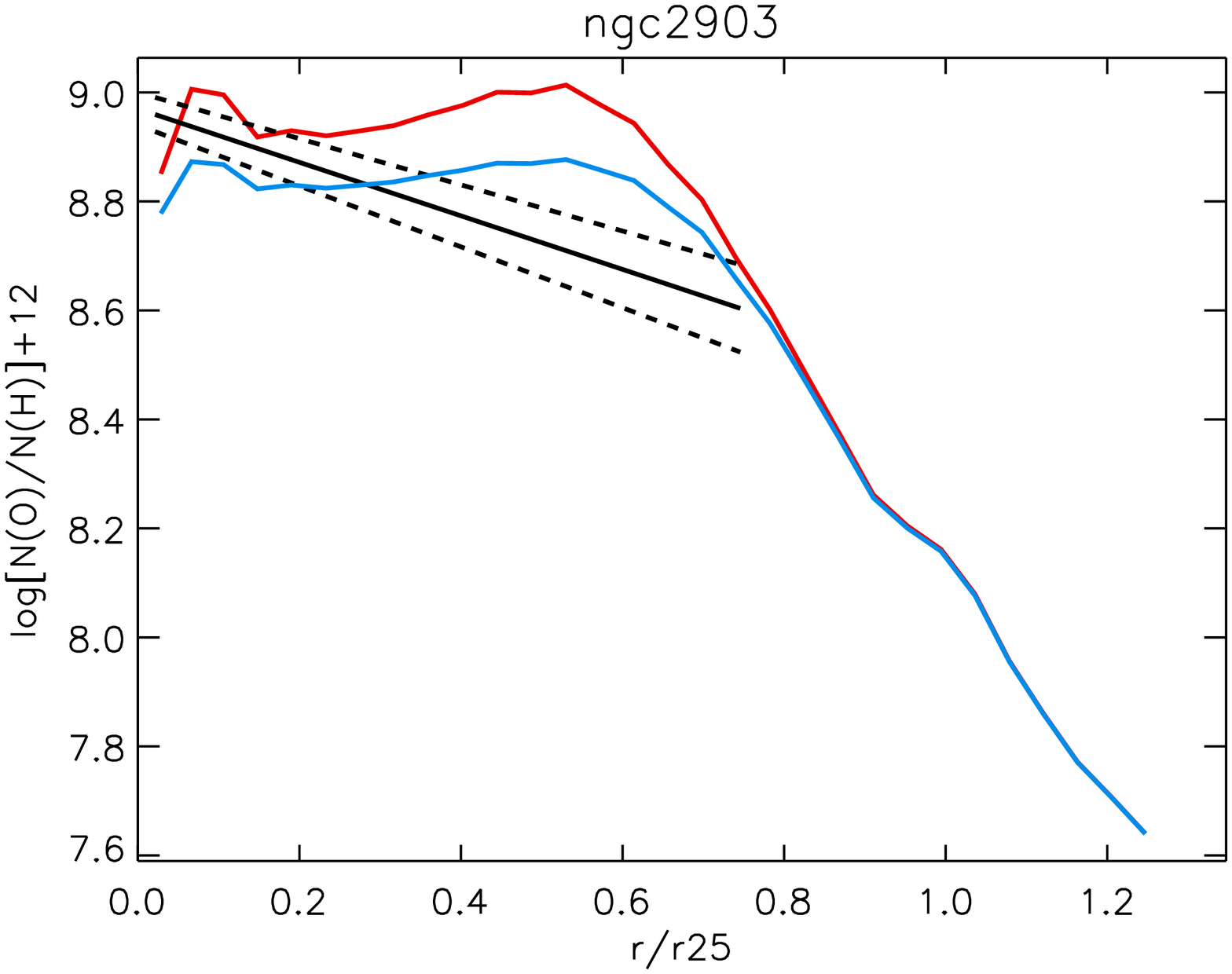}} \subfloat{\includegraphics[width = 6.5cm,angle=90]{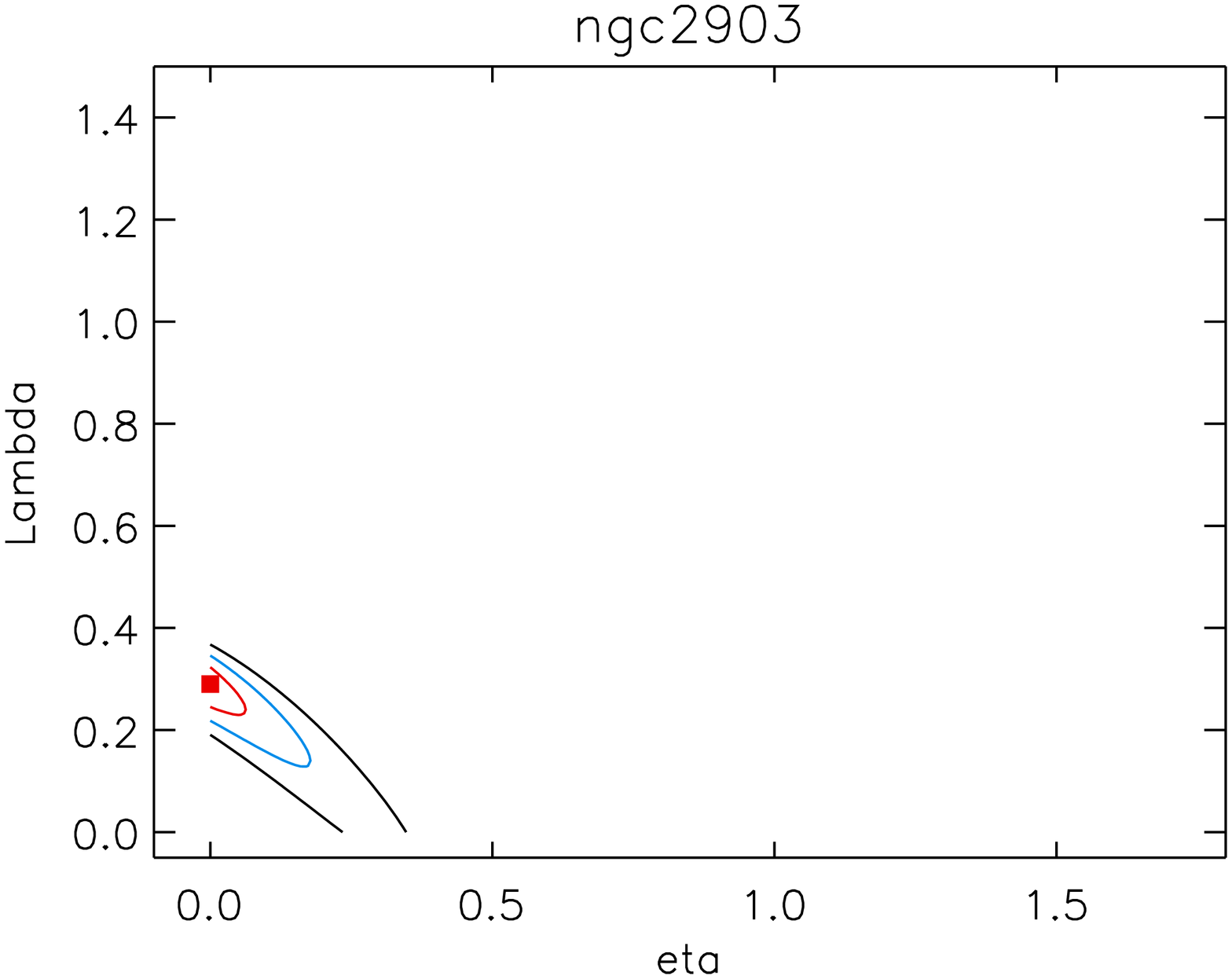}}\\
\caption{Chemical evolution fits of the galaxies NGC\,6946 (top), 7331 (middle), 2903 (bottom).}\label{figure14}
\end{figure*}

\begin{figure}
\includegraphics[width=6.5cm,angle=90]{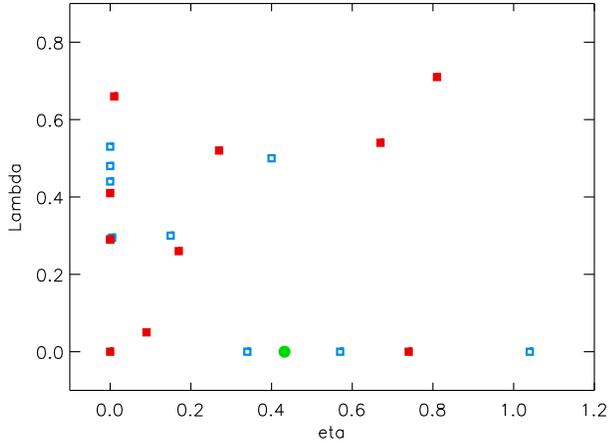}
\caption{The location of all galaxies investigated in Fig.~\ref{figure7} to~\ref{figure10} in the $(\eta,\Lambda)$-plane.
The galaxies of Fig.~\ref{figure7} to~\ref{figure10} which have more accurate values of $\eta$ and $\Lambda$ are plotted as red filled squares and the remaining galaxies as blue open squares. The filled green circle gives the position of the Milky Way disk as discussed in Section~\ref{sec:the-radial-metallicity-distribution-of-the-milky-way}.}\label{figure15}
\end{figure}

The fits of the remaining galaxies are shown in Fig.~\ref{figure12} to~\ref{figure14}. For all these galaxies, except NGC\,2903, the fits of the observed metallicity distributions are reasonable within the uncertainties of the observations and the range of galactocentric distances where observations are available. But for most of these objects the $\Delta \chi^2$ isocontour diagrams indicate that the values of $\eta$ and $\Lambda$ are not very well constrained by the fit. We add these objects to the plot of the location of galaxies in the  $(\eta,\Lambda)$-plane in Fig.~\ref{figure15} and find that we still can identify three well distinguished groups in our sample, in the same way as already discussed above. The 
galaxy NGC\,2903 has an unusual H\textsc{i} profile with a steep decline of gas mass from the centre to $0.5r/R_{25}$ and then an extended disk with constant gas mass. Our models are unable to reproduce chemical profiles for such gas distribution properly.

The observed ISM H\textsc{ii} region metallicities taken from the compilation by \citet{pilyugin14a} have been re-calibrated in Section~\ref{sec:galaxy-sample-and-observations} using metallicity determinations obtained by stellar spectroscopy. These metallicties are given in Table~\ref{table:sample} and have been used to obtain the results described so far. To investigate the systematic effects of metallicity measurements
on our method, we repeat the analysis described before with logarithmic oxygen abundances shifted by $\Delta\rm(O/H)_0 = -0.05, -0.10$, and $-0.15$, respectively, going back in small steps to the original zero point calibration by \citet{pilyugin14a}. We note that changing the zero points of the observed metallicities is equivalent to changing the yield in the chemical evolution model (see Equations~\ref{z1}, ~\ref{z3},~\ref{z5},~\ref{zcb}). Thus, the results discussed below can also be seen as a test of the influence of the uncertainty of the yield.  


\begin{table*}
 \caption{$\eta$ and $\Lambda$ determination for different metallicity zero point shifts $\Delta\rm(O/H)_0$ }
 \label{table:eta-lambda}
 \begin{tabular}{ccccccccc}
\hline
\hline
Name & $\Delta\rm(O/H)_0$& $\Delta\rm(O/H)_0$& $\Delta\rm(O/H)_0$& $\Delta\rm(O/H)_0$& $\Delta\rm(O/H)_0$& $\Delta\rm(O/H)_0$& $\Delta\rm(O/H)_0$& $\Delta\rm(O/H)_0$ \\
     &$0.0$&$0.0$&$-0.05$&$-0.05$&$-0.10$& $-0.10$&$-0.15$&$-0.15$  \\
NGC  & $\eta$ & $\Lambda$ &  $\eta$ & $\Lambda$ & $\eta$ & $\Lambda$ & $\eta$ & $\Lambda$ \\
\hline
0628& 0.17& 0.26& 0.39& 0.25& 0.70& 0.20& 1.07& 0.13 \\
0925& 0.67& 0.54& 1.10& 0.49& 1.65& 0.39& 2.38& 0.21 \\
2403& 0.81& 0.71& 1.20& 0.75& 1.64& 0.80& 2.19& 0.83 \\
2903& 0.00& 0.29& 0.00& 0.38& 0.00& 0.47& 0.00& 0.55 \\
3184& 0.00& 0.41& 0.07& 0.58& 0.25& 0.52& 0.46& 0.56 \\
3198& 0.27& 0.52& 0.50& 0.56& 0.76& 0.61& 1.10& 0.65 \\
3351& 0.00& 0.29& 0.08& 0.33& 0.23& 0.36& 0.42& 0.39 \\
3521& 0.15& 0.30& 0.40& 0.26& 0.74& 0.18& 1.18& 0.05 \\
3938& 1.04& 0.00& 1.36& 0.00& 1.74& 0.00& 2.18& 0.00 \\
4254& 0.00& 0.48& 0.00& 0.64& 0.00& 0.75& 0.06& 0.86 \\
4321& 0.34& 0.00& 0.55& 0.00& 0.78& 0.00& 1.06& 0.00 \\
4559& 0.01& 0.66& 0.17& 0.74& 0.41& 0.82& 0.67& 0.91 \\
4625& 0.40& 0.50& 0.57& 0.56& 0.79& 0.62& 1.04& 0.69 \\
4736& 0.00& 0.53& 0.00& 0.60& 0.04& 0.68& 0.12& 0.76 \\
5055& 0.09& 0.05& 0.30& 0.00& 0.51& 0.00& 0.76& 0.00 \\
5194& 0.00& 0.00& 0.00& 0.04& 0.00& 0.23& 0.00& 0.38 \\
5457& 0.74& 0.00& 1.00& 0.00& 1.31& 0.00& 1.67& 0.00 \\
6946& 0.57& 0.00& 0.82& 0.00& 1.10& 0.00& 1.44& 0.00 \\
7331& 0.00& 0.44& 0.00& 0.53& 0.11& 0.60& 0.27& 0.67 \\
\hline
\end{tabular}
\end{table*}

\begin{figure}
\includegraphics[width=6.5cm,angle=90]{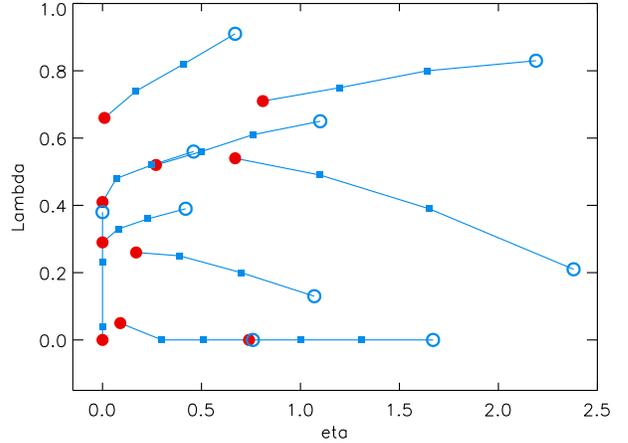}
\caption{The shift of the location of the galaxies of Fig.~\ref{figure7} to~\ref{figure10} 
in the $(\eta,\Lambda)$-plane with different assumptions for the zero point of the metallicity determination from H\textsc{ii} regions. Red filled circles refer to the metallicities given in Table~\ref{table:sample}, which were also used for the previous plots. Blue open circles are the results when a zero point shift of $\Delta\rm(O/H) = -0.15$ is applied to the logarithmic oxygen abundances. Blue filled squares represent shifts of $\Delta\rm(O/H) = -0.05$ and $-0.10$, respectively.}\label{figure16}
\end{figure}

The results for the different metallicity zero points are given in Table~\ref{table:eta-lambda}. Fig.~\ref{figure16} displays the shifts
in the ($\eta$,$\Lambda$)-plane caused by these systematic changes of the metallicity zero points. The changes obtained in this way are relatively large in the sense that systematically decreasing the metallicty leads to larger values of $\eta$ and $\Lambda$, which is easily understood with the help of Fig.~\ref{figure1} and~\ref{figure2}. This changes the conclusions somewhat. While galaxies with very small values of $\Lambda$ at $\Delta\rm(O/H)_0 = 0.0$ remain low accretion cases, most of the weak galactic wind objects are now fit with higher mass-loss rates. This increases the number of objects with both wind and accretion, at least for the largest zero point shift $\Delta\rm(O/H)_0 = -0.15$.

\begin{figure}
\includegraphics[width=6.5cm,angle=90]{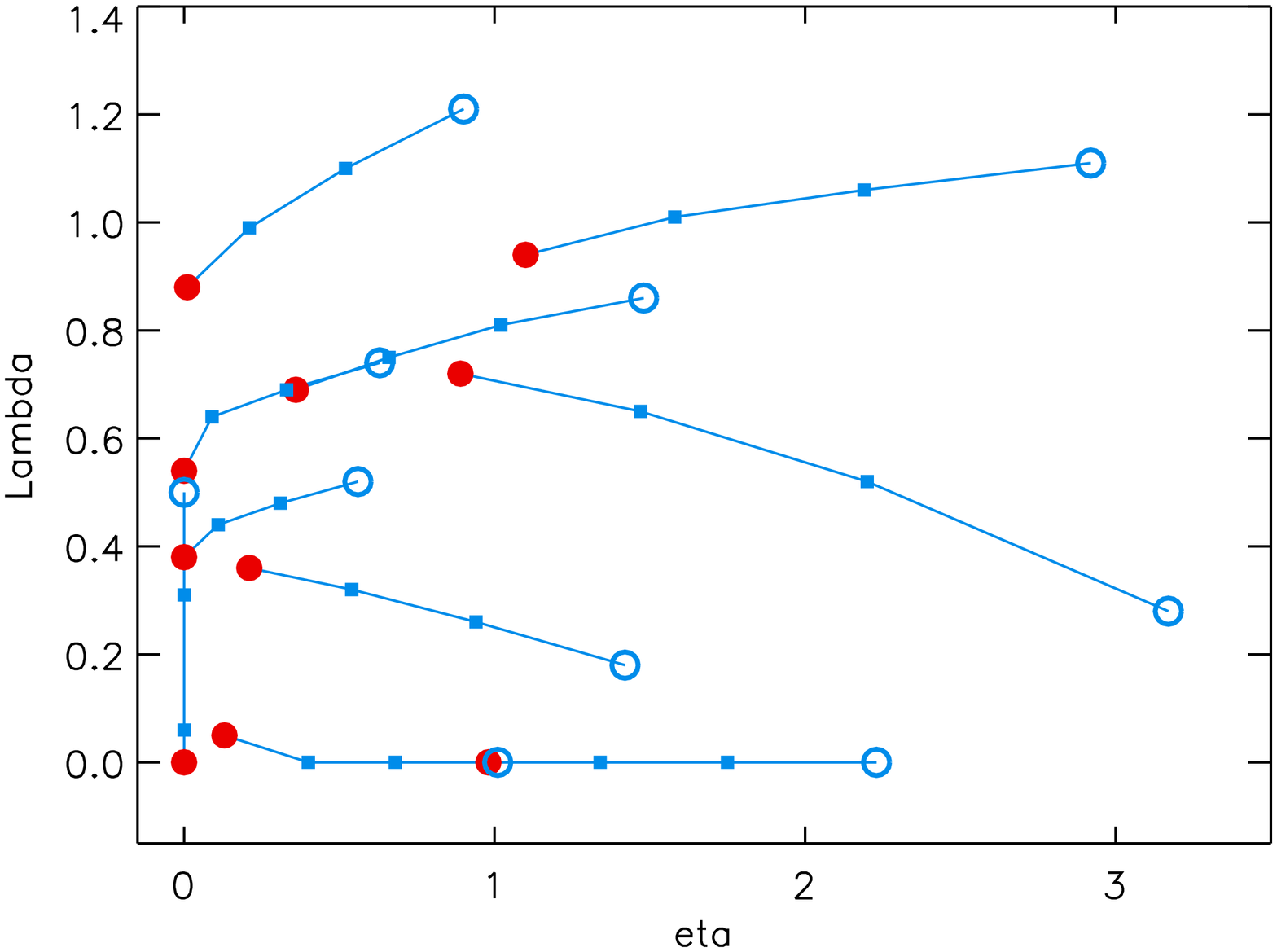}
\caption{Same as Fig.~\ref{figure16} but for for the stellar mass return fraction $R = 0.2$ (see text).}\label{figure17}
\end{figure}

Finally, we test the influence of the stellar mass return factor. While we argued in Section~\ref{sec:the-chemical evolution-model} that $R = 0.4$ is the preferable choice, $R = 0.2$ cannot be completely excluded. In consequence, we repeat our analysis for $R = 0.2$ and the corresponding yield (see Section~\ref{sec:the-radial-metallicity-distribution-of-the-milky-way}). The results are shown in Fig.~\ref{figure17}. While for $R = 0.2$ the $\eta$ and $\Lambda$ values for the fits obtained are somewhat larger, the figure clearly indicates that none of the conclusions made before are affected.

\floatplacement{figure}{!t}
\floatplacement{figure}{!t}
\floatplacement{figure}{!t}
\floatplacement{figure}{!t}

\floatplacement{figure}{!t}

\floatplacement{figure}{!t}

\section{Discussion and Summary}\label{sec:discussion-and-summary}

In the previous sections, we have developed and tested a new method to use the radial metallicity distributions of the ISM and the young stellar population in connection with the radial profiles of stellar mass and ISM gas mass column densities to constrain the strengths of galactic wind and accretion processes. The method is based on the application of an analytical chemical evolution model, which assumes constant ratios of galactic mass-loss and mass-gain to the star formation rate and which then predicts radial metallicity profiles as a function of the ratio of stellar mass to ISM gas mass. Generally, this relatively simple model reproduces the observed metallicies and their radial gradients very well. In some cases, however, the model fails at outer galactocentric distances mostly for galaxies with extended disks of high H\textsc{i} gas column densities, but it still remains useful in the inner parts.

We have applied the model to the observations of the Milky Way disk and empirically determined the metallicity yield $y_Z$. We obtained $y_Z/(1-R) = 0.0112\pm17\% = 0.8~Z_{\rm B-stars}$, where $Z_{\rm B-stars} = 0.014$ is the metallicity mass fraction of young B-stars in the solar neighbourhood \citep{nieva12}. $R$ is the fraction of stellar mass returned to the ISM and has been adopted as R = 0.4 in our model calculation (other values of $R$ were also tested but the influence of varying $R$ was found to be small). Using a B-star mass fraction of oxygen to total metallicy $O_m/Z_{\rm B-stars} = 0.467$, we determined an oxygen yield of $y_O$ = 0.00313. With a galactic wind mass-loading factor 
$\eta \equiv \dot{M}_{loss} / \psi = 0.432\pm{0.12}$ and a very low accretion gain $\Lambda \equiv \dot{M}_{accr} / \psi = 0.0\pm_{0.00}^{0.10}$, the model reproduces the observed metallicity gradient very accurately. We conclude that based on our model fit the Milky Way disk is characterised by a low accretion rate and a moderate galactic wind mass loading factor. We note, however, that this result is in disagreement with the estimate of the Milky Way infall rate from ultraviolet absorption line studies of halo high velocity clouds in the sightline of halo stars from which \citet{lehner11} estimate an infall rate of 0.8 to 1.4~\msun~yr$^{-1}$. With a Milky way star formation rate of $1.9\pm{0.4}$~\msun~yr$^{-1}$ \citep{chomiuk11,davies11}, this would lead to a value of $\Lambda$ between 0.35 to 0.93 much larger than our very good fit indicates.  However, we also note that \citet{chomiuk11} do not rule out star formation rates a factor of two higher for alternative choices of the initial mass function and that the determination of the infall rate depends on a variety of parameters which are not well constrained as discussed by \citet{lehner11}. 

We also tested the alternative chemical evolution models developed recently by \citet{ascasibar14} to reproduce the observed global metallicities of galaxies by the use of the global galaxy gas and stellar mass. While this model has been successfully applied to explain the observed global mass-metallicity relationship of galaxies, it fails in the spatially resolved case of the Milky Way to reproduce the observed metallicity gradient. 

About $\approx 30\%$ of our sample of galaxies studied showed similarly low values of accretion rates $\Lambda$ as the Milky Way disk with simultaneously a range of mass loading factors between $0 \leq \eta \leq 1$.  However, we also find a group with very low mass loading factor $\eta$ and  $0.3 \leq \Lambda \leq 0.7$. This group comprises $\approx 40\%$ of our sample. The remaining galaxies have roughly equal rates of accretion and mass-loss in a range $0.2 \leq \Lambda \approx \eta \leq 0.8$.

The range of galactic mass-loss found in our study is in agreement with spectroscopic studies searching for outflows in the local universe \citep[e.g.,][]{rupke05, veilleux05,bouche12,zahid12,yabe15}. \citet{bigiel14} identify clear inflow signatures form their study of H\textsc{i} velocity fields for NGC\,2403, 3198, 2903, 7331, for which we also determine significant rates of mass infall (see also \citealt{deblok14}). \citet{bigiel14} also detect infall for NGC\,6946. We note that while our fit for NGC\,6946 finds a solution with $\Lambda =0$ as the best value, the shape of the isocontours in Fig.~\ref{figure14} does not rule out a value of $\Lambda$ as high as 0.4. For NGC\,925, \citet{bigiel14} find no signs of inflows but instead significant outflows. With $\eta=0.67$ and $\Lambda=0.54$, we detect in NGC\,925 both outflows and inflows, but the inflow rate within 3$\sigma$ can still be consistent with zero (Fig.~\ref{figure7}).

The division into three groups with either mostly winds and weak accretion or mostly accretion and weak winds or winds and accretion roughly equal is an intriguing result. In particular, the existence of a significant group of objects disconnected from the cosmic web with  $\Lambda \approx 0$ is somewhat surprising, although cosmological simulations do not rule out such cases in the local universe (\citealt{keres05}; \citealt{bouche10}; see \citealt{yabe15}). Our results do not confirm the observational finding by \citet{yabe15} based on an investigation of global metallicities, masses and star formation rates that galaxies evolve along the ``equilibrium relationship'' $\Lambda$ = $\eta$ + (1-R) by \citet{dave11a}. It is interesting that \citet{yabe15} applied the same chemical evolution model as this study; however, an important difference is that in our investigation the spatially resolved information about metallicity, gas and stellar masses is used and the gas mass profiles are obtained directly from radio observations, whereas in \citet{yabe15} the information about gas masses is obtained indirectly from the Kennicutt-Schmidt law and H$\alpha$ observations. In addition, \citet{yabe15} do not disentangle the effects of galactic bulges by a bulge decomposition of the stellar mass profiles. 

We find that the members of the three galaxy groups have distinct properties of their gas disks. NGC\,2403, 925, 3198, 4625 which comprise the group with relatively large and roughly equal $\eta$ and $\Lambda$ values in the diagonal of Fig.~\ref{figure15} are all characterised by relatively metal-poor H\textsc{i} dominated gas disks with much lower H$_2$ column densities and an almost flat radial profile of the total gas mass column densities ($\Delta \log \Sigma_g \lesssim 0.3~{\rm dex}~R_{25}^{-1}$; see figures in \citealt{schruba11}). Of the remaining galaxies only NGC\,4559 has similar gas disk properties. We note that  NGC\,4559 has also a high $\Lambda$ value (0.66) but a very small value of $\eta$. The gas disks of all other galaxies are dominated by molecular hydrogen with column densities larger than the one of H\textsc{i} in the inner disk region ($r \leq 0.5$ to 0.7 $R_{25}$) and show a clear exponential decline ($\Delta \log \Sigma_g \gtrsim 0.6~{\rm dex}~R_{25}^{-1}$) of their total gas mass column densities over the galactocentric radius range used for our metallicity fit. It seems that among these galaxies the ones with low accretion rates have on average a steeper exponential decline. For the low $\Lambda$ cases NGC\,5194, 5055, 4321, 6946, 5457 we find $\Delta \log \Sigma_g = 1.4, 1.2 1.3, 1.2, 1.1~{\rm dex}~R_{25}^{-1}$, respectively (NGC\,3938, another low accretion case but with a very high $\eta = 1.04$, is an exception with $\Delta \log \Sigma_g = 0.7~{\rm dex}~R_{25}^{-1}$). On the other hand, most of the galaxies for which we determine low outflow rates, have mostly flatter profiles. NGC\,628, 3184, 3351, 3521, 7331 show $\Delta \log \Sigma_g = 0.6$ to $0.7~{\rm dex}~R_{25}^{-1}$. The exceptions are NGC\,4736 and 2903 with a steeper decline. NGC\,4254 also shows a high value of $\Delta \log \Sigma_g = 1.3$, but the isocontour in Fig.~\ref{figure12} does not rule out this galaxy as a potential low accretion case.

For the Milky Way, we have also determined a relatively low accretion rate (Figs.~\ref{figure5} and \ref{figure15}), however, its ISM gas properties are different from the rest of the low accretion group. In the range of galactocentric distance from 4 to 9 kpc, which we used for the fit in Fig.~\ref{figure4}, the Milky Way gas disk is dominated by atomic hydrogen and has a relatively flat radial profile. The stellar mass column density is much higher than the gas mass column density. Comparing the Milky Way to the group in the upper diagonal of Fig.~\ref{figure15}, for which the gas disks are also flat and dominated by H\textsc{i}, the metallicity of the Milky Way is significantly higher.

As any other study of this kind using metallicities obtained from H\textsc{ii} region emission lines, our results depend on the calibration of the strong-line diagnostics adopted. In our work, we have re-calibrated the \citet{pilyugin14a} oxygen abundances by applying a shift $\Delta \rm (O/H)_0 = 0.15~\rm dex$ based on the quantitative spectroscopy of blue supergiants in spiral galaxies \citep{kud08,kud12,kud13,u09}. If we use the original \citet{pilyugin14a} calibration, we obtain larger values for $\eta$ but the qualitative conclusions remain the same as before. In particular, the fraction of galaxies with very small accretion rates $\Lambda$ remains unchanged. For future work, we regard it as crucial to improve the calibration of the H\textsc{ii} region strong line methods based on a larger sample of stellar spectroscopy of blue and also red supergiants \citep{gazak14}.

In summary, we conclude that the use of spatially resolved information of observed metallicity, stellar mass and gas mass is a powerful tool to constrain the effects of galactic winds and accretion on the chemical evolution of star forming galaxies.

\section*{Acknowledgments}

RPK and FB acknowledge support by the National Science Foundation under grants AST-1008798 and AST-1108906 and by the Munich Institute for Astro- and Particle Physics (MIAPP) of the DFG cluster of excellence ``Origin and Structure of the Universe'', where part of this work was carried out. We wish to thank our referee, Dr. Yago Ascasibar, for his helpful and very constructive review.

This research used NASA's Astrophysics Data System Bibliographic Services and the NASA/IPAC Extragalactic Database (NED). 
We acknowledge the usage of the HyperLeda database (\href{http://leda.univ-lyon1.fr}{http://leda.univ-lyon1.fr}).


\begin{thebibliography}{0}
\providecommand{\natexlab}[1]{#1}

\end{thebibliography}


\begin{thebibliography}{}

\bibitem[Ascasibar et al.(2014)]{ascasibar14} Ascasibar, Y., Gavilan, M., Pinto, N. et al.\ 2014, {\tt arXiv:1406.6397 [astro-ph.GA]}
\bibitem[Asplund et al.(2009)]{asplund09} Asplund, M., Grevesse, N., Sauval, A.~J., \& Scott, P.\ 2009, ARA\&A, 47, 481 
\bibitem[Bigiel et al.(2010)]{bigiel10} Bigiel, F., Leroy, A., Walter, F., et al.\ 2010, \aj, 140, 1194
\bibitem[Bigiel et al.(2014)]{bigiel14} Bigiel, F., Cornier, P., \& Schmid, T.\ 2014, \an, 335, 470
\bibitem[Burkert et al.(1992)]{burkert92} Burkert, A., Truran, J.~W., \& Hensler, G.\ 1992, \apj, 391, 651
\bibitem[Bovy \& Rix(2013)]{bovy13} Bovy, J., Rix, H.~W. \ 2013, \apj, 779, 115
\bibitem[Bouche et al.(2010)]{bouche10} Bouche, N., Dekel, A., Genzel, R., et al.\ 2010, \apj, 718, 1001
\bibitem[Bouche et al.(2012)]{bouche12} Bouche, N., et al.\ 2012, \mnras, 426, 801
\bibitem[Bresolin et al.(2009a)]{bresolin09a} Bresolin, F., Gieren, W., Kudritzki, R.-P., et al.\ 2009, \apj, 700, 309 
\bibitem[Bresolin et al.(2009b)]{bresolin09b} Bresolin, F., Ryan-Weber, E., Kennicutt, R.~C., \& Goddard, Q. \ 2009, \apj, 695, 580
\bibitem[Bresolin et al.(2012)]{bresolin12} Bresolin, F., Kennicutt, R.~C., \& Ryan-Weber, E. \ 2012, \apj, 750, 122
\bibitem[Bresolin (2011)]{bresolin11} Bresolin, F.\ 2011, \apj, 729, 56 
\bibitem[Chiappini et al.(2001)]{chiappini01} Chiappini, C., Matteucci, F., Romano, D.\ 2001, \apj, 554, 1044
\bibitem[Chiappini et al.(2001)]{chiappini01} Chiappini, C., Matteucci, F., \& Romano, D.\ 2001, \apj, 554, 1044 
\bibitem[Chomiuk \& Povich(2011)]{chomiuk11} Chomiuk, L., Povich, M.~S.\ 2011 \aj, 142, 197
\bibitem[Dalcanton(2007)]{dalcanton07} Dalcanton, J.~J.\ 2007, \apj, 658, 941 
\bibitem[Dav{\'e} et al.(2011a)]{dave11a} Dav{\'e}, R., Oppenheimer, B.~D., \& Finlator, K.\ 2011, \mnras, 415, 11 (a) 
\bibitem[Dav{\'e} et al.(2011b)]{dave11b} Dav{\'e}, R., Finlator, K., \& Oppenheimer, B.~D.\ 2011, \mnras, 416, 1354 (b)
\bibitem[Dav{\'e} et al.(2012)]{dave12} Dav{\'e}, R., Finlator, K., \& Oppenheimer, B.~D.\ 2012, \mnras, 421, 98
\bibitem[Davies et al.(2011)]{davies11} Davies, B., Hoare, M.~G., Lumsden, S.~L. et al.,\ 2011, \mnras, 416, 972
\bibitem[Dayal et al., 2013]{dayal13} Dayal, P., Ferrara, A. \& Dunlop, J.~F.\ 2013, \mnras, 430, 2891
\bibitem[de Blok et al.(2014)]{deblok14} de Blok, W.~J.~G., Keating, K.~M., Pisano, D.~J., et al.\ 2014, \aap, 569, A68 
\bibitem[Edmunds(1990)]{edmunds90} Edmunds, M.~G.\ 1990, \mnras, 246, 678
\bibitem[Edmunds \& Greenhow(1995)]{edmunds95} Edmunds, M.~G., Greenhow, R.~M.\ 1995, \mnras, 272, 241
\bibitem[Fu et al.(2009)]{fu09} Fu, J., Hou, J.~H., Yin, J. et al.\ 2009\, \apj, 696, 668
\bibitem[Garnett \& Shields(1987)]{garnett87} Garnett, D.~R., \& Shields, G.~A.\ 1987, \apj, 317, 82 
\bibitem[Garnett et al.(1997)]{garnett97} Garnett, D.~R., Shields, G.~A., Skillman, E.~D., Sagan, S.~P., \& Dufour, R.~J.\ 1997, \apj, 489, 63 
\bibitem[Gazak et al.(2014)]{gazak14} Gazak, J.~Z., Davies, B., Bastian, N., Kudritzki, R.~P., Bergemann, M., Plez, B., Evans, C., Pattrick, L., Bresolin, F., Schinnerer, E. \ 2014, \apj, 787, 142
\bibitem[Genovali et al.(2014)]{genovali14} Genovali, K., Lemasle, B., Bono, G. et al.\ 2014, \aap, 566, A37
\bibitem[Graham(2001)]{graham01} Graham, A.~W.\ 2001, \aj, 121, 820
\bibitem[Ho et al.(2015)]{ho15} Ho, I.-T., Kudritzki, R.~P., Kewley, L.~J. et al.,\ 2015, \mnras, 448, 2030
\bibitem[Jarret et al.(2003)]{jarrett03} Jarrett, T.~H., Chester, T., Cutri, R., Schneider, S.~E., \& Huchra, J.~P.\ 2003, \aj, 125, 525
\bibitem[Kennicutt et al.(2003)]{kennicutt03} Kennicutt, R.~C., Jr., et al.\ 2003, \pasp, 115, 928 
\bibitem[Keres et al.(2005)]{keres05} Keres, D., Katz, N., Weinberg, D.~H., \& Dave, R.\ 2005, \mnras, 363, 2 
\bibitem[Kewley \& Ellison(2008)]{kewley08} Kewley, L.~J., \& Ellison, S.~L.\ 2008, \apj, 681, 1183
\bibitem[Koribalski et al.(2004)]{koribalski04} Koribalski, B.~S., Stavely-Smith, L., Kilborn, V.~A., et al. \ 2004, \aj, 128, 16
\bibitem[K{\"o}ppen \& Edmunds(1999)]{koeppen99} K{\"o}ppen, Edmunds, M.~G.\ 2007, \mnras, 306, 317
\bibitem[Kudritzki et al.(2014)]{kud14} Kudritzki, R.~P., Urbaneja, M.~A., Bresolin, F., et al.\ 2014, \apj, 788, 56
\bibitem[Kudritzki et al.(2013)]{kud13} Kudritzki, R.~P., Urbaneja, M.~A., Gazak, Z., et al.\ 2013, \apjl, 779, L20 
\bibitem[Kudritzki et al.(2012)]{kud12} Kudritzki, R.~P., Urbaneja, M.~A., Gazak, Z., et al.\ 2012, \apj, 747, 15 
\bibitem[Kudritzki et al.(2008)]{kud08} Kudritzki, R.~P., Urbaneja, M.~A., Bresolin, F., et al.\ 2008, \apj, 681, 269 
\bibitem[Leitner \& Kravtsov(2011)]{leitner11} Leitner, S.~N., Kravtsov, A.~V. \ 2011, \apj, 734, 48
\bibitem[Lequeux et al.(1979)]{lequeux79} Lequeux, J., Peimbert, M., Rayo, J.~F., Serrano, A., \& Torres-Peimbert, S.\ 1979, \aap, 80, 155 
\bibitem[Lehner \& Howk(2011)]{lehner11} Lehner, N., Howk, J.~C.\ 2011, Science, 334, 955
\bibitem[Leroy et al.(2008)]{leroy08} Leroy, A.~K., Walter, F., Brinks, E. et al. \ 2008, \aj, 136, 2782 
\bibitem[Leroy et al.(2009)]{leroy09} Leroy, A.~K. et al. \ 2009, \aj, 137, 4670
\bibitem[Lilly et al.(2013)]{lilly13} Lilly, S.~J., Carollo, C.~M., Pipino, A., et al.\ 2013, \apj, 772, 119
\bibitem[Maeder(1992)]{maeder92} Maeder, A.\ 1992, \aap, 264, 105
\bibitem[Martin et al.(2002)]{martin02} Martin, C.~l., Kobulnicky, H.~A., \& Heckman, T.~M.\ 2002, \apj, 574, 663  
\bibitem[Matteucci(2012)]{matteucci12} Matteucci, F.\ 2012, ``Chemical Evolution of Galaxies'', Astronomy and Astrophysics Library, Springer-Verlag Berlin Heidelberg 
\bibitem[Meynet \& Maeder\/(2005)]{meynet05} Meynet, G. \& Maeder, A. \ 2005, \aap, 429, 581
\bibitem[Minchev et al.(2013)]{minchev13} Minchev, I., Chiappini, C., Martig, M.\ 2013, \aap, 558, A9
\bibitem[Molla et al.(1997)]{molla97} Molla, M., Ferrini, F., Diaz, A.~I.\ 1997, \apj 475, 519
\bibitem[Mott et al.(2013)]{mott13} Mott, A., Spitoni, E., Matteucci, F.\ 2013, \mnras, 435, 2918
\bibitem[Nieva \& Przybilla(2012)]{nieva12} Nieva, M.~F., Przybilla, N.\ 2012, \aap, 539, A143
\bibitem[Paturel et al.(2002)]{paturel02} Paturel, G., Teerikorpi, P., Theureau, G., Fouque, P., Musella, I., Terry, J.~N.\ 2002, \aap, 389, 19 
\bibitem[Pagel \& Patchett(1975)]{pagel75} Pagel, B.~E.~J., Patchett\ 1975, \mnras, 172, 13
\bibitem[Pagel (2009)]{pagel09} Pagel, B.~E.~J.\ 2009, \ {\em Nucleosynthesis and Chemical Evolution of Galaxies}, second edition, Cambridge University Press
\bibitem[Pilkington et al.(2012)]{pilkington12} Pilkington, K. et al.\ 2012, \aap, 540, A56
\bibitem[Pilyugin et al.(2012)]{pilyugin12} Pilyugin, L.~S., Grebel, E.~K., \& Mattsson, L.\ 2012, \mnras, 424, 2316
\bibitem[Pilyugin et al.(2014)]{pilyugin14a} Pilyugin, L.~S., Grebel, E.~K., \& Kniazev, A.~Y.\ 2014, \aj, 147, 131 
\bibitem[Pipino et al.(2014)]{pipino14} Pipino, A., Lilly, S.~J., \% Carollo, C.~M.\ 2014, \mnras, 441, 1444
\bibitem[Prantzos \& Boissier(2000)]{prantzos00} Prantzos, N., \& Boissier, S.\ 2000, \mnras, 313, 338 
\bibitem[Przybilla et al.(2006)]{przybilla06} Przybilla, N., Butler, K., Becker, S.~R., \& Kudritzki, R.~P.\ 2006, \aap, 445, 1099 
\bibitem[Recchi et al.(2008)]{recchi08} Recchi, S., Spitoni, E., Matteucci, F., \% Lanfranchi, G.~A.\ 2008, \aap, 489, 555
\bibitem[Rupke et al.(2005)]{rupke05} Rupke, D.~S., Veilleux, S., Sanders, D.~B.\ 2015, \apjs, 160, 115
\bibitem[Rupke et al.(2013)]{rupke13} Rupke, D.~S., Veilleux, S.\ 2013, \apj, 768, 75
\bibitem[Schoenrich \& Binney(2009a)]{schoenrich09a} Schoenrich, R., Binney, J.\ 2009a, \mnras, 396, 203
\bibitem[Schoenrich \& Binney(2009b)]{schoenrich09b} Schoenrich, R., Binney, J.\ 2009b, \mnras, 399, 1145
\bibitem[Schruba et al.(2011)]{schruba11} Schruba, A., Leroy, A.~K., Walter, F. et al.\ 2011, \aj, 142, 37
\bibitem[Searle(1971)]{searle71} Searle, L.\ 1971, \apj, 168, 322
\bibitem[Searle \& Sargent(1972)]{searle72} Searle, L. \& Sargent, W.~L.~W\ 1972, \apj, 173, 25
\bibitem[Sellwood \& Binney(2002)]{sellwood02} Sellwood, J.~A., Binney, J.\ 2002, \mnras, 336, 785
\bibitem[Skillman(1998)]{skillman98} Skillman, E.~D.\ 1998, Stellar astrophysics for the local group: VIII Canary Islands Winter School 
of Astrophysics, 457 
\bibitem[Spitoni et al.(2010)]{spitoni10} Spitoni, E., Calura, F., Matteucci, F., \& Recchi, S.\ 2010, \aap, 514, A73 
\bibitem[Tremonti et al.(2004)]{tremonti04} Tremonti, C.~A., Heckman, T.~M., Kauffmann, G., et al.\ 2004, \apj, 613, 898 
\bibitem[U et al.(2009)]{u09} U, V., Urbaneja, M.~A., Kudritzki, R.-P., et al.\ 2009, \apj, 704, 1120 
\bibitem[Veilleux et al.(2005)]{veilleux05} Veilleux, S., Cecil, g., Bland-Hawthorn, J.\ 2005, \araa, 43, 769
\bibitem[Vila-Costas \& Edmunds(1992)]{vilacostas92} Vila-Costas, M.~B., Edmunds, M.~G.\ 1992, \mnras, 259, 121
\bibitem[Walter et al.(2008)]{walter08} Walter, F., Brinks, E., de Blok, W.~J., et al.\ 2008, \aj, 136, 2563
\bibitem[Wolfire et al.(2003)]{wolfire03} Wolfire, M.~G., McKee, C.~F., Hollenbach, D., \& Tielens, A.~G.~G.~M. \ 2003, \apj, 587, 278
\bibitem[Werk et al.(2011)]{werk11} Werk, J.~K., Putman, M.~E., Meurer, G.~R., \& Santiago-Figueroa, N. \ 2011, \apj, 735, 71 
\bibitem[Wright et al.(2010)]{wright10} Wright, E.~L., Eisenhardt, P.~R., Mainzer, A.~K., et al.\ 2010, \aj, 140, 1868
\bibitem[Yabe et al.(2015)]{yabe15} Yabe, K., Ohta, K., Akiyama, M., Iwamuro, F., Tamura, N., Yuma, S., Dalton, G., \& Lewis, I.\ 2015, \apj, 798, 45 
\bibitem[Yates et al. (2012)]{yates12} Yates, R.~M., Kauffmann, G., \& Guo, Q.\ 2012, \mnras, 422, 215
\bibitem[Zahid et al.(2012)]{zahid12} Zahid, H.~J., Dima, G.~I., Kewley, L.~J. et al.\ 2012, \apj, 757, 54
\bibitem[Zahid et al.(2013)]{zahid13} Zahid, H.~J., Geller, M.~J., Kewley, L.~J. et al.\ 2013, \apj, 771, L19
\bibitem[Zahid et al.(2014)]{zahid14} Zahid, H.~J., Dima, G.~I., Kudritzki, R.~P. et al.\ 2014, \apj, 791, 130
\bibitem[Zaritsky et al.(1994)]{zaritsky94} Zaritsky, D., Kennicutt, R.~C., Jr., \& Huchra, J.~P.\ 1994, \apj, 420, 87 
\bibitem[Zoccali et al.(2003)]{zoccali03} Zoccali, M., Renzini, A., Ortolani, S. et al.\ 2003, \aap, 399, 931

\end{thebibliography}
\end{document}